\documentclass[superscriptaddress,twocolumn,10pt,pra,aps,reprint,showpacs,longbibliography]{revtex4-1}
\usepackage{float}
\floatstyle{boxed}
\usepackage{wrapfig}
\usepackage{array}
\usepackage{graphicx}
\usepackage{amsbsy}
\usepackage{xcolor}
\usepackage[utf8x]{inputenc}
\usepackage{epstopdf}
\usepackage{amsmath,amssymb,amsfonts}
\usepackage{indentfirst}
\usepackage{soul}
\usepackage[T1]{fontenc}

\renewcommand{\eqref}[1]{\mbox{Eq.~(\ref{#1})}}

\newcommand{\be}{\begin{equation}}
\newcommand{\ee}{\end{equation}}
\newcommand{\bea}{\begin{eqnarray}}
\newcommand{\eea}{\end{eqnarray}}
\newcommand{\Hef}{\hat H_{\rm eff}}
\newcommand{\dline}[1]{\boldsymbol{#1}}

\usepackage{url}
\usepackage[colorlinks]{hyperref}
\hypersetup{
    plainpages=true,
    breaklinks=true,
    hypertexnames=false,
    pageanchor=true,
    colorlinks=true,
    linkcolor={blue},
    citecolor={red},
    urlcolor={blue},
    anchorcolor={black}
}

\usepackage[normalem]{ulem} 
\newcommand\redline{\bgroup\markoverwith
    {\textcolor{red}{\rule[.5ex]{2pt}{1pt}}}\ULon}

\begin{document}

\title{Generating high-order quantum exceptional points in synthetic dimensions}

\author{Ievgen I. Arkhipov}
\email{ievgen.arkhipov@upol.cz} \affiliation{Joint Laboratory of
Optics of Palack\'y University and Institute of Physics of CAS,
Faculty of Science, Palack\'y University, 17. listopadu 12, 771 46
Olomouc, Czech Republic}

\author{Fabrizio Minganti}
\email{fabrizio.minganti@riken.jp} \affiliation{Theoretical
Quantum Physics Laboratory, RIKEN Cluster for Pioneering Research,
Wako-shi, Saitama 351-0198, Japan}

\author{Adam Miranowicz}
\email{miran@amu.edu.pl}
\affiliation{Theoretical Quantum Physics Laboratory, RIKEN Cluster
for Pioneering Research, Wako-shi, Saitama 351-0198, Japan}
\affiliation{Institute of Spintronics and Quantum Information,
Faculty of Physics, Adam Mickiewicz University, 61-614 Pozna\'n, Poland}

\author{Franco Nori}
\email{fnori@riken.jp} 
\affiliation{Theoretical Quantum Physics
Laboratory, RIKEN Cluster for Pioneering Research, Wako-shi,
Saitama 351-0198, Japan}
\affiliation{RIKEN Center for Quantum Computing, 2-1 Hirosawa, Wako-shi, Saitama 351-0198, Japan}
\affiliation{Physics Department, The
University of Michigan, Ann Arbor, Michigan 48109-1040, USA}

\begin{abstract}
Recently, there has been intense research in  proposing and
developing various methods for constructing high-order exceptional
points (EPs) in dissipative systems. These EPs can possess a
number of intriguing properties related to,  e.g., chiral
transport and enhanced sensitivity.
Previous proposals to realize non-Hermitian Hamiltonians (NHHs) with high-order EPs have been mainly based on either direct construction of spatial networks of coupled modes
or utilization of synthetic dimensions, e.g., of mapping spatial lattices to time or photon-number space.
Both methods rely on the construction of effective NHHs describing classical or postselected quantum fields, which neglect the effects of quantum jumps, 
and which, thus, suffer from a scalability problem in the {\it quantum regime}, when the probability of quantum jumps increases with the number of excitations and dissipation rate.
Here, by considering the full quantum dynamics of a quadratic
Liouvillian superoperator,  we introduce a simple and effective
method for engineering NHHs with high-order quantum EPs, derived
from evolution matrices of system operators moments.
That is, by quantizing higher-order moments of system operators, e.g., of a quadratic two-mode system,
the resulting evolution matrices can be interpreted as alternative NHHs
describing, e.g., a spatial lattice of coupled resonators, where spatial sites are represented by high-order field moments in the synthetic space of field moments.
Notably, such a mapping allows correct reproduction of the results
of the Liouvillian dynamics, including quantum jumps.
As an example, we consider a $U(1)$-symmetric quadratic
Liouvillian describing a {\it bimodal} cavity with incoherent mode
coupling, which can also possess anti-$\cal PT$-symmetry, whose field moment dynamics can be mapped to an NHH governing a spatial {\it network} of coupled resonators with high-order EPs.
\end{abstract}

\date{\today}

\maketitle


\section{Introduction}
Recently, the field of open quantum systems has attracted much interest.
While dissipation is often seen as detrimental, there exists a whole class of processes which can never take place in Hermitian (i.e., non-dissipative) systems.
In these systems, the existence of exotic spectral degenaracies called exceptional points (EPs) has attracted much attention \cite{Ozdemir2019,Miri2019,Ashida2020}. At an EP, two or more eigenvalues, along with their eigenstates, coalesce. 
Since the eigenstates of a Hermitian operator are always orthogonal, EPs require non-Hermitian operators. Historically, EPs were first investigated in the context of non-Hermitian Hamiltonians (NHHs), primarily in the framework of parity-time ($\cal PT$)-symmetric systems, i.e., those for which an NHH commutes with the $\cal PT$ operator~\cite{Bender1998}. 
Note that non-Hermitian Hamiltonians do not lead to the violation of
no-go theorems as explicitly demonstrated in Ref.~\cite{Ju2019}.
The
existence of EPs has been further generalized to any NHH exhibiting pseudo-Hermiticity~\cite{Mostafazadeh2002}, for which the $\cal PT$-symmetry is a particular case. 

Beyond linear optical systems (see Refs.~\cite{Ozdemir2019,Miri2019} and references therein), EPs have been realized in various experimental platforms, e.g., in nonlinear optics~\cite{Roy2020,Roy2021,Jahani2021},
electronics~\cite{Schindler2011}, optomechanics
\cite{Jing2014,Jing2015,Harris2016,Jing2017}, acoustics
\cite{Zhu2014,Alu2015}, plasmonics~\cite{Benisty2011},
metamaterials~\cite{Kang2013}, and ion trapped systems~\cite{Ding2021}. 

Many interesting and nontrivial effects are associated with the presence of EPs~\cite{Heiss2001,Dembowski2003,Lin2011,Regensburger2012,Feng2014,Hodaei2014,Peng2014,Chang2014,Arkhipov2019b,Huang2020,Brands2014a,Peng2014a,Perina2019b,Lange2020,Kuo2020}. 
One of these is enhanced system sensitivity to external perturbations in the vicinity of EPs
\cite{Wiersig2014,Zhang2015,Wiersig2016,Ren2017,Chen2017,
Hodaei2017,ChenNat2018,Liu2016,Mortensen2018,Wiersig2020}.
If $n$ eigenstates coalesce (so the order of an EP is $n$), the response of a system to a perturbation of intensity $\epsilon$ scales as $\sqrt[n]{\epsilon}$. 
Although some recent studies (both theoretical and experimental) have questioned the presence of enhanced sensing at EPs~\cite{Wiersig2020b,Wang2020,Chen2019,Lau2018,Zhang2019,Langbein2018},
Refs.~\cite{Lau2018,Zhang2019,Yu2020} have argued that EPs lead to enhanced sensitivity.

The interesting properties of high-order EPs ignited the search for methods which enable one to construct higher-order EPs~\cite{Teimourpour2014,Nada2017,Wu2018,Zhang2020c,Sahin2020}.
For NHHs, the proposed techniques require the realization of complex networks of coherently coupled resonators.
A major drawback is that one has to finely tune the system parameters, due to the incommensurate mode couplings arising from the form of the mode coupling in NHHs~\cite{Teimourpour2014,Nada2017,Zhang2020c}. In a recent work~\cite{Sahin2020}, the authors proposed a novel approach for constructing tight-binding networks with higher-order EPs, based on chiral-mode coupling instead. 

On the other hand, to avoid the experimentally demanding construction of complex spatial networks of coupled modes, one can also turn for help to synthetic dimensions, where a spatial dimension is mapped to a more abstract synthetic one~\cite{Tschernig18,Tschernig20,Quiroz19}. That is, one assigns to the internal degrees of freedom of the modes the role of spatially arranged modes. For instance, considering a Fock state synthetic dimension~\cite{Tschernig20,Quiroz19}, each spatial optical mode is mapped to a photon number state, thus recovering all the relevant physics of a lattice system in the photon-number synthetic dimensions.

Alternatively, one can just map space to time and realize multimode NHHs in  synthetic temporal lattices~\cite{Regensburger2012}.
Nonetheless, these methods rely on the construction of effective NHHs describing classical or postselected quantum fields. The latter removes the effects of quantum jumps, 
and, thus, suffers from a scalability problem in the {\it quantum} regime, because the probability of quantum jumps increases with the number of excitations and dissipative rates in a system. 

Apart from the previously-detailed experimental difficulties in realizing such NHHs in the quantum domain, another problem arises concerning the interpretation of NHHs.
While the wave functions of an Hermitian Hamiltonian have a clear physical meaning, the state vectors and inner product of Hilbert space spanned by such NHHs require finding a nontrivial metric in order to be compatible with conventional quantum mechanics~\cite{MOSTAFAZADEH_2010}.

Non-Hermiticity naturally emerges in the context of open quantum systems.
Indeed, the Lindblad master equation of a Markovian open system, although Hermiticity preserving, has a well-defined arrow of time. Therefore, the Liouvillian superoperator associated with the master equation is non-Hermitian.
With respect to an NHH, the Liouvillian also accounts for the presence of quantum jumps and has a well-defined inner product for its eigenstates.
The extension of EPs of NHHs to those based on Liouvillians~\cite{Minganti2019} has shown that quantum jumps can play a crucial role in the properties of EPs~\cite{Prosen2012,Arkhipov2020,Minganti2020,Jaramillo2020,Huber2020,Nakanishi2021,Wiersig2020,Arkhipov2020b}. 
Furthermore, the evolution of a density matrix of an open quantum system is described by a completely-positive and trace-preserving (CPTP) linear map. 
As such, an NHH cannot describe the evolution of an arbitrary quantum system.

Despite the fact that an NHH may not describe the time evolution of a Lindblad master equation (i.e., the eigenstates of a NHH do not reproduce those of the Liouvillian), the dynamics of some operators
can be derived in terms of the action of an NHH \cite{Arkhipov2020b}.
This apparent contradiction results from the fact that the operators in the Heisenberg picture do not evolve under a CPTP map.
In other words, the NHH can describe the evolution of operators even in the quantum limit.

In this article, 
we propose a method to properly define a new class of effective NHHs for quadratic Liouvillian systems. These NHHs are derived from evolution matrices governing \textit{the moments of system operators} and as such are called  \textit{moments-based NHHs}. Most importantly, these moments-based NHHs can reveal higher-order \textit{quantum} EPs, residing in the Liouvillian eigenspace~\cite{Arkhipov2020b}.
That is, by quantizing system operators moments by mapping the corresponding evolution matrices to moments-based NHHs, one can engineer quantum systems with high-order EPs in the field moments space.

Such constructed moments-based NHHs with high-order EPs substantially differ from {\it standard} NHHs. Whereas the latter are constructed by expanding the \textit{Hilbert space} of system operators~\cite{Teimourpour2014,Nada2017,Zhang2020c,Quiroz19,Tschernig20}, the former are constructed by rather expanding the space of system-operator {\it moments}.
Indeed, if one uses the eigenstates of the standard NHH for low-order operator moments to express higher-order ones, the result would be unphysical and give different results compared to the full Liouvillian dynamics~\cite{Arkhipov2020b}. 
However, the moments-based NHHs are derived via the quantization of the system operator moments obtained from the Liouvillian, and, thus, correctly capture the dynamics of high-order system operators without approximations (i.e., including the effects of quantum jumps~\cite{Minganti2019,Arkhipov2020}).

Note that the knowledge of operator moments allows one, in principle, to fully reconstruct a system state $\hat\rho$~\cite{Wunsche1996}.
The dynamics of such moments-based NHH can describe a spatial lattice, e.g., a spatial network of coupled resonators.
In other words, by starting from a quadratic Liouvillian, describing, e.g., a two-mode system,
the resulting evolution matrices, governing the higher-order field moments, can be cast
to the new NHHs, which can describe a lattice of coupled cavities instead.

These newly obtained moments-based NHHs can reveal the presence of arbitrarily-high order quantum EPs in the Liouvillian dynamics. 
Physically speaking, one can witness the presence of high-order quantum EPs by means of the coherence and spectral  functions~\cite{Arkhipov2020b} or by properly initializing the system.  
To put it another way, instead of considering spatial \textit{networks} of $n$ resonators (where high-order EPs 
can be engineered), by considering higher-order moments of, e.g., 
\textit{two coupled} resonators, one can obtain the \textit{same} spectral degeneracies.
Apart from a theoretical interest in defining the NHHs in the quantum limit, the advantage of the use of moments-based NHHs with respect to standard NHHs lies in its manifesting simplicity and in the possibility of preserving the commensurate character of modes coupling.

As an example, we implement our method for the $U(1)$-symmetric quadratic Liouvillian that describes a two-mode optical cavity with {\it incoherent} mode coupling. This model is also characterized by the
anti-$\cal PT$-symmetry~\cite{Arkhipov2020b,Zhang2020,Zhang_2020,Fan2020,Choi2018,Nair2020,Peng2016}, as defined in Eq.~(\ref{APT}). 
We show how the Liouvillian eigenspace of such a \textit{two-mode} system, expressed via field moments, can be mapped to the eigenspace of an effective NHH of a \textit{multimode} system. 
The benefit of considering anti-$\cal PT$-symmetric systems, compared to their $\cal PT$-symmetric counterparts with exclusively {\it coherent} mode coupling, is the absence of any active elements, and their scalability, which is seemingly easier to realize. That is, one does not need to build up complex networks to achieve higher-order EPs, but only to \textit{excite additional modes} in the anti-$\cal PT$-symmetric cavity, at the same time ensuring the incoherent character of the coupling between the newly excited modes. Recent studies show that such systems, with incoherent mode interactions, can be realized via incoherent mode backscattering in 
waveguide networks~\cite{Mukherjee2017} and cavity-based photonic devices~\cite{Metelmann2015,Kullig2018,Kullig2019,Ding2019,Qin2021}.
Moreover, the dissipative couplings can play a prominent role in the experimental realizations of photonic and quantum computing in time-multiplexed optical systems~\cite{Marandi2014,Inagaki2016}.
 
 As a byproduct of our method,  we also highlight the rich structure of  quadratic Liouvillians. Indeed, the correspondence between multimode systems and higher-order moments is a peculiarity of the Liouvillian space structure \cite{Wiersig2020}. We argue that, although the correspondence between the Liouvillian evolution of higher-order correlation functions and that of lower-order correlation functions of more complex multimode systems is exact only for quadratic Liouvillians (i.e., describing Gaussian states), such a procedure should be valid also in the presence of weak nonlinearity, where a Gaussian state approximation can still be valid.
 
We note that our method can be implemented irrespective of the knowledge of the specific details of a given quadratic Liouvillian.
That is, in order to realize a physical system with high-order EPs,
initially one only has to have some physically realizable NHH (which of course can again be related to some quadratic Liouvillian), whose matrix mode representation reveals an EP of at least order 2. 
Then, by increasing the order of the EP, according to the method described here, one obtains a non-Hermitian model, which can be a guide to realize extended lattice systems whose NHH
has a higher-order EP. 

This paper is organized as follows. In Sec.~\ref{II}, we introduce a general model of quadratic Liouvillians. In Sec.~\ref{III}, we analyze the dynamics of higher-order moments of system operators in the model, expressed via the corresponding evolution matrices. Then, by performing a second quantization of the operator moments, we introduce a map between moments evolution matrices and the new class of NHHs, called moments-based NHHs, which can genuinely capture the quantum effects in a system.
That is, the evolution matrices of the field moments can be assigned to the new NHHs, which can govern, e.g., the dynamics of spatial networks of coupled resonators. The latter procedure allows to map a spatial lattice to a synthetic space of field moments, where spatial sites are represented by high-order field moments.
In Sec.~\ref{IV},  we explicitly demonstrate how the described method enables to engineer higher-order EPs determined from such moments-based NHHs. 
In Sec.~\ref{V}, we implement the proposed scheme on the example of the $U(1)$ and anti-${\cal PT}$-symmetric cavity with incoherent mode coupling. The discussion of the proposed method and its comparison to existing methods along with conclusions are given in Sec.~\ref{VI}. 


\section{General Model of a system described by a quadratic Liouvillian}\label{II}

The evolution of a density matrix $\hat\rho(t)$ is described by the master equation
\begin{equation} \label{ME} 
\frac{{\rm d}}{{\rm d}t}\hat\rho(t)={\cal L}\hat\rho(t),
\end{equation}
which for a quadratic Liouvillian superoperator $\cal L$ in the Gorini-Kossakowski-Sudarshan-Lindblad form
 reads ($\hbar=1$)
\begin{equation}\label{L} 
{\cal L}\hat\rho(t)=-i\left(\hat H_{\rm eff}\hat\rho(t)-\hat\rho(t)\hat H^{\dagger}_{\rm eff}\right)+2 \sum\limits_{jkl}\Gamma^{l}_{jk}\hat s^{(l)}_{j}\hat\rho(t) \left(\hat s^{(l)}_{k}\right)^\dagger,
\end{equation}
where $\Hef$ is the effective NHH given by:
\begin{equation}\label{Heffg} 
\Hef=\sum\kappa_{jk}^{lm}\hat s_j^l\hat s_k^m-{i}\sum\Gamma_{jk}^{l}\hat
s_j^{(l)} \left(\hat s^{(l)}_{k}\right)^\dagger.
\end{equation}
In Eqs.~(\ref{L}) and (\ref{Heffg}), the indices $\{j,\,k\}$ indicate the sites (modes) of a system, and $\{l,\,m\}=\{1,\,2\}$ are such that $\hat s^{(1,2)}_q=\{\hat a_q,\hat a_q^{\dagger}\}$, where $\hat a_q$ ($\hat a_q^\dagger$) is the annihilation (creation) operator of a particle at the $q$th site (e.g., a photon).
The  coefficients $\kappa_{jk}^{lm},\Gamma_{jk}^{lm}\in{\mathbb R}$ describe the coherent and incoherent parts of the system evolution, respectively. 
Such a quadratic Liouvillian describes the dissipative dynamics of Gaussian states that, in optics, describes, e.g., linearly coupled waveguides (cavities) or nonlinear parametric processes~\cite{Perina1991Book}.

Henceforth, we describe $\Hef$ in \eqref{Heffg} as the \textit{effective NHH} to distinguish it from the \textit{moments-based NHH}, which is associated with system operators moments of higher order and which we introduce in Sec.~\ref{Sec:moments_NHH}.

According to the quantum trajectory theory, the Liouvillian in Eq.~(\ref{L}) can be divided into two parts, namely, in a {\it continuous} nonunitary evolution described by the effective NHH $\Hef$, and in the action of {\it discrete} random changes expressed by quantum jumps \cite{Dalibard92,Molmer93}.
The effective NHH is especially useful in the semiclassical limit, where the jump action can be neglected. It can also be used to describe systems where it is possible to determine whether a quantum jump took place, e.g., cavities with a small photon number and with a very high finesse \cite{HarocheBook} or postselected systems \cite{Naghiloo19}.
In other cases, the last term in Eq.~(\ref{L}), describing the quantum jump effects (in this case, the sudden creation or annihilation of a particle) is essential to faithfully capture the system dynamics at the quantum level. Indeed, for the case studied here of open {\it linear} systems, the effective NHH $\Hef$ does not reflect the non-conservative character of  dissipation \cite{Minganti2020}.


\section{dynamics of the moments of system operators and the non-Hermitian Hamiltonian}\label{III}

In the following, we assume that the dissipators in Eq.~(\ref{L}) induce no incoherent amplification. Otherwise, the dynamics of some moments is
affected by an additional noise vector~\cite{AgarwalBook,CarmichaelBook}, which is of no relevance here.

\subsection{Field moments evolution for a quadratic $U(1)$-symmetric two-mode Liouvillian}

We begin by considering the simplest case, i.e., that of a quadratic and $U(1)$ symmetric system describing two cavities (its generalization is provided in Sec.~\ref{IIIB}). 
We refer to the corresponding $\mathcal{L}$ as a \textit{linear} Liouvillian because it emerges in the context of dissipative coupled linear systems, e.g., coupled waveguides or cavities. 

Any Liouvillian for coupled bosonic systems is said to be $U(1)$-symmetric if it commutes with a phase rotation operator $\cal U$, defined as
\begin{equation} 
    {\cal U}\hat\rho=\exp\!\!\left({i\phi \sum_{j} \hat a_j^{\dagger}\hat a_j}\right)\hat\rho \exp\!\!\left({-i\phi \sum_{j} \hat a_j^{\dagger}\hat a_j}\right).
\end{equation}
In other words, the master equation must be invariant under a simultaneous arbitrary phase shift $\hat a_j\to\hat a_j e^{i\phi}$.

For two coupled cavities, i.e., $j, \, k= 1,\,2$ in \eqref{L}, $\cal U$ constraints the rate equations for the  \textit{field moments}
\begin{equation*}
\langle\hat a_1^{\dagger m}\hat a_2^{\dagger n}\hat a_1^p\hat a_2^q\rangle,\quad \text{for}\quad m,n,p,q=0,1,\dots.
\end{equation*}
Normally, the dynamics of the field moments relates all the possible combinations of $\{m,n,p,q\}$. However, in the presence of the $\mathcal{U}$ symmetry only moments that have the same order $(m+n-p-q)$ can be coupled.
Since the considered  Liouvillians here are also quadratic, one obtains $p+q=m+n$. Thus, the dynamics of the moments is captured by \textit{closed sets of coupled equations}. For example, the first-order moment $\langle\hat a_1\rangle$ ($\langle\hat a_1^{\dagger}\rangle$) would be coupled only to the moment $\langle\hat a_2\rangle$ ($\langle\hat a_2^{\dagger}\rangle$).

Given the closed structure of the rate equations for the first-order moments we have:
\begin{equation}\label{RES} 
\frac{{\rm d}}{{\rm d}t}\langle\vec{A}\rangle=\boldsymbol{M}_A \langle\vec A\rangle,
\end{equation}
where $\vec A=[\hat a_1,\hat
a_2]^T$, and $\boldsymbol{M}_A$ is a $2\times2$ evolution matrix. 

The matrix $\boldsymbol{M}_A$ is the building block to obtain the evolution matrix for higher-order field moments by constructing  various Kronecker products of the vectors $\vec A$ and $\vec A^{\dagger} \equiv [\hat a_1^\dagger,\hat
a_2^\dagger] $. For instance, given the second-order moments
\begin{equation}\label{B} 
    \langle\vec B\rangle =\langle\vec A\otimes\vec A \rangle= \langle[\hat a_1^2, \hat a_2\hat a_1, \hat a_1\hat a_2, \hat a_2^2]^T\rangle,
\end{equation}
the evolution matrix $\boldsymbol{M}_B $, such that $\partial_t\langle\vec B\rangle=\boldsymbol{M}_B \langle\vec B\rangle$, is the Kronecker sum of the same two matrices $\boldsymbol{M}_A$,
\begin{equation}\label{MB} 
    \boldsymbol{M}_B =\boldsymbol{M}_A\oplus\boldsymbol{M}_A =\boldsymbol{M}_A\otimes\dline I_2+\dline I_2\otimes\boldsymbol{M}_A.
\end{equation}
Here, the symbol $\oplus$ denotes a Kronecker sum, $\otimes$ is the Kronecker product, and $\dline I_2$ is the $2\times2$ identity matrix.
Note that we keep the order between the products of the operators $\hat a_1$ and $\hat a_2$  in Eq.~(\ref{B}). Although, in this case, it is not relevant because $[\hat a_1,\hat a_2]=0$, but in general the order should be preserved. 
We remark that a Kronecker sum naturally appears in problems when solving Lyapunov and/or Sylvester equations~\cite{Lototsky2015}.

 The same procedure can  recursively be applied to obtain the evolution matrices for higher-order moments. The generalization of \eqref{MB} is given by:
 \begin{eqnarray}\label{Mgamma} 
&\langle\vec\gamma\rangle=\langle\vec\alpha\otimes\vec\beta\rangle,& \nonumber \\
&\boldsymbol{M}_\gamma=\boldsymbol{M}_\alpha\oplus\boldsymbol{M}_\beta=\boldsymbol{M}_\alpha\otimes\dline I_\beta+\dline I_\alpha\otimes\boldsymbol{M}_\beta,&
 \end{eqnarray}
 where the vectors of operators, $\vec\alpha$ and $\vec\beta$, are the Kronecker products of the initial vectors $\vec A$ and/or $\vec A^{\dagger}$, while $\boldsymbol{M}_{\alpha,\beta}$ are the corresponding evolution matrices.
 The resulting evolution matrix $\boldsymbol{M}_\gamma$ of higher-order field moments is the recursive Kronecker sum of the matrices $\boldsymbol{M}_A$.
The dimension $N_\gamma$ of the matrix $\boldsymbol{M}_\gamma$ is the product of the dimensions of the matrices $\boldsymbol{M}_\alpha$ and $\boldsymbol{M}_\beta$, i.e., $N_\gamma=N_\alpha N_\beta$. Moreover, due to the standard properties of the Kronecker sum, the symmetry of the matrix $\boldsymbol{M}_A$  is retained by all the matrices $\boldsymbol{M}_\gamma$.

\subsection{Evolution of field moments  for a generic quadratic Liouvillian}\label{IIIB}

The previous derivation for a $U(1)$ quadratic two-mode system can be extended to generic quadratic Liouvillians, even if the form of $\boldsymbol{M}_\gamma$ is slightly more involved.

For an arbitrary quadratic Liouvillian (i.e., not necessarily $U(1)$-symmetric),  describing an $n$-mode system, all  field moments are generated  by the tensor product of the $2n$-dimensional vector
\begin{equation}\label{Agen} 
    \vec A =[\hat a_1,\hat a_1^{\dagger},\dots,\hat a_n,\hat a_n^{\dagger}]^T.
\end{equation}
The time evolution of $ \langle\vec A\rangle$ is given by $\partial_t \langle\vec{A}\rangle = \boldsymbol{M}_A  \langle\vec A\rangle$, which is the same as in Eq.~(\ref{RES}).

Given the building block $\dline M_A$, the previously detailed procedure, to obtain the evolution matrices for higher-order moments, remains valid.
Similarly to the $U(1)$ case, the evolution matrices $\boldsymbol{M}_\gamma$, which determine the dynamics of various higher-order moments, are obtained by taking a recursive Kronecker sum of the corresponding $2n\times2n$ matrix $\boldsymbol{M}_A$. 
That is,  Eq.~(\ref{Mgamma}) is valid for any quadratic Liouvillian.  
Also, as it was stressed earlier, the operators order should be kept when constructing Kronecker tensors out of the vector of operators $\vec A$ in Eq.~(\ref{Agen}). 
In other words, no permutations are allowed in the obtained products of the operators.

\subsection{Second quantization of field moments and a definition of the moments-based non-Hermitian Hamiltonian}
\label{Sec:moments_NHH}

\subsubsection{Moments-based non-Hermitian Hamiltonian of $U(1)$ quadratic Liouvillians}\label{IIIC1}

In the case of $U(1)$ quadratic Liouvillians, the evolution matrix $\dline M_A$ for the first-order field moments becomes equivalent (up to the imaginary factor) to the matrix form of the 
corresponding NHH, i.e.,: 
\begin{equation}\label{MH} 
\dline M_A=-i \dline H_{\rm eff},  
\end{equation}
where the matrix form of the NHH is defined as follows 
\begin{eqnarray}
\Hef\equiv \left(\vec{A}\right)^\dagger \dline H_{\rm eff} \vec{A},
\end{eqnarray}
and vector of operators $\vec{A}$, for a two-mode case, is given in Eq.~(\ref{RES}).
In other words, there is a one-to-one correspondence between the evolution of the operator $\vec{A}$ and its first-order moments $\langle\vec{A}\rangle$ for the $U(1)$ systems. 
This correspondence is quite intriguing, since it provides a clear physical meaning to the effective NHH of  two coupled bosonic systems via the introduction of its first-order moments, and it has been intensively exploited in a number of previous works on {\it quantum} EPs~\cite{Zueco2018,Downing2020,Arkhipov2020,Purkayastha2020}. 

The above described  correspondence, however, cannot be simply extended to higher-order field moments, since the evolution matrices governing the dynamics of the operators and their moments would in general be different. This stems from the fact that the dynamics of the field moments of any order is determined by the Liouvillian which includes quantum jump effects, whereas the effective NHH applied to the same moments (in Heisenberg picture) in general fails to incorporate them~\cite{Arkhipov2020b}.

\begin{figure*}
    \centering
    \includegraphics[width=0.95\linewidth]{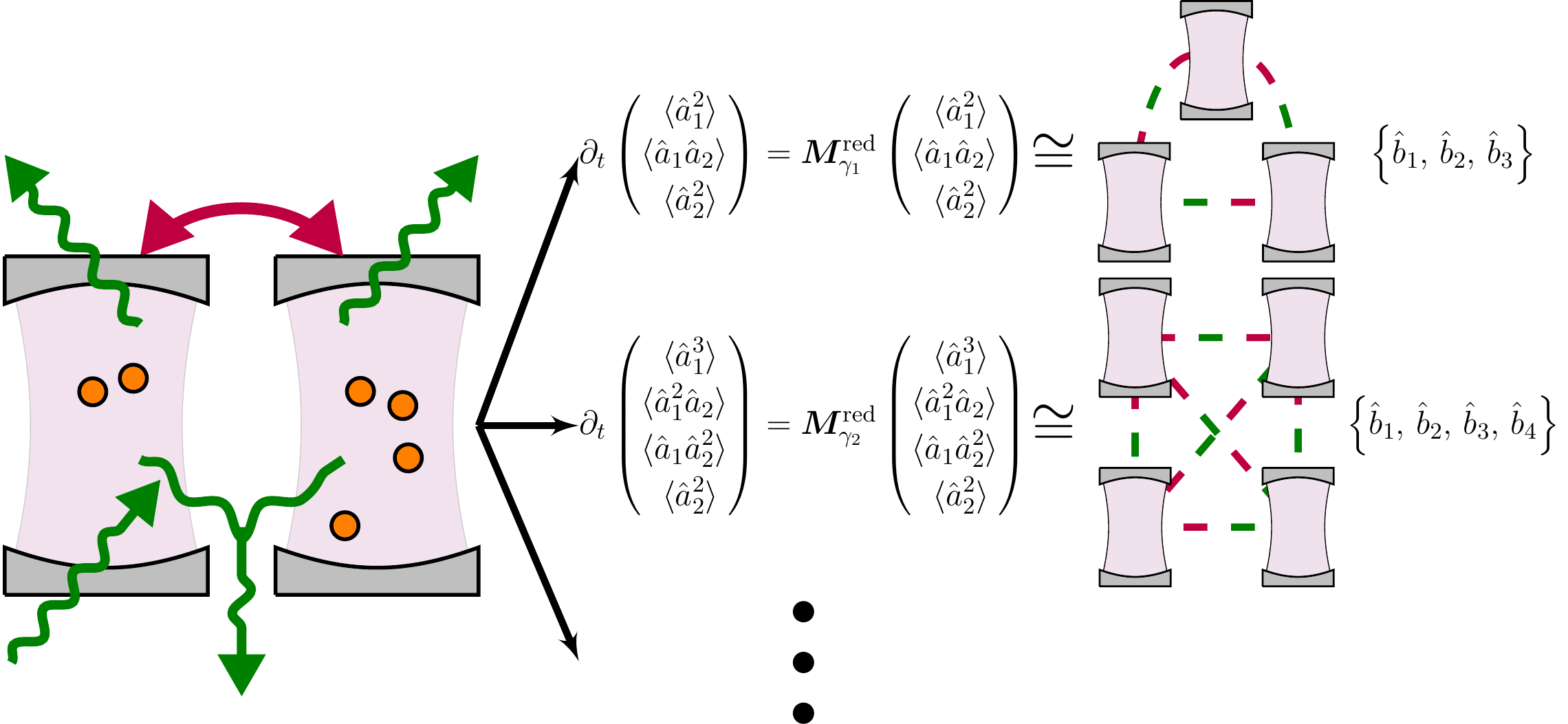}
    \caption{Schematic representation of the procedure to define a moments-based NHH and its relation to a spatial lattice of coupled  resonators in the $U(1)$ symmetric case. A quadratic Liouvillian system (e.g., the two cavities on the left) is characterized by coherent interactions
    (red double arrows), dissipation (green arrows pointing outwards), and amplification channels (green arrows pointing inwards) that compete in determining the photonic field inside the cavity (orange circles).
    The equations of motion for the various moments $\langle\hat a_1^{\dagger m}\hat a_2^{\dagger n}\hat a_1^p\hat a_2^q\rangle$ of such a system (middle portion) form a finite set, which can be described by the reduced evolution matrices $\dline M_\gamma^{\rm red}$ (see Sec.~\ref{IIIC1} for details). By quantizing the moments, such matrices $\dline M_\gamma^{\rm red}$ can be interpreted as an effective NHH of a more complex lattice of resonators (right portion), according to Eq.~(\ref{Hb}), defined in the synthetic space of field moments. Indeed, each moment can be mathematically treated as a separate bosonic field ($\hat b_i$), i.e., a  driven dissipative quadratic bosonic system, as shown at the right with cavities. Each of these fields interact with the others via dissipative or coherent interactions (represented by dashed green and red lines).
    This procedure is general, and explores a direct correspondence between the evolution of higher-order moments of a Liouvillian system and that of lower-order moments of a larger system. Interestingly, this correspondence can be exploited to observe higher-order EPs in simple lattices by considering higher-order correlation functions and, vice versa, as a guideline to engineer lattices which display higher-order EPs.}
    \label{fig:lattice_equivalence}
\end{figure*}

Nevertheless, one can assign to any evolution matrix $\dline M_\gamma$  a new NHH in analogy to Eq.~(\ref{MH}), which we thus call a moments-based NHH, by quantizing the corresponding higher-order field moments $\langle\vec{\gamma}\rangle$ (see also Fig.~\ref{fig:lattice_equivalence}).
The determination of such moments-based NHHs, however, requires an intermediate passage.
Namely, in general, $\vec \gamma$ contains terms which are identical [e.g., $\hat a_1\hat a_2=\hat a_2\hat a_1$  in Eq.~(\ref{B})], and it cannot be straightforwardly quantized.
The degeneracy of $\boldsymbol{M}_\gamma$ can be eliminated by introducing the \textit{reduced} matrix $\dline M^{\rm red}_\gamma$. 
For instance, $\boldsymbol{M}_\gamma$ of any non-Hermitian moment of the vector $\vec A$, i.e., $\vec\gamma=\bigotimes\limits_{i=1}^{m}\vec A$, has an ``initial'' dimension $N_\gamma=2^m$. 
By collecting identical terms, the resulting matrix $\boldsymbol{M}^{\rm red}_\gamma$ has the dimension $N^{\rm eff}_\gamma=m+1$. 
The price paid for this reduction is, in general, the loss of some initial symmetry of the matrix $\dline M_\gamma$. For instance, if $\dline M_\gamma= \dline M_\gamma^T$ then, in general, $\dline M^{\rm red}_\gamma\neq \left(\dline M^{\rm red}_\gamma\right)^T$. That is, if the mode coupling in the matrix $\dline M_\gamma$ is symmetric, then that coupling in the effective evolution matrix $\dline M^{\rm eff}_\gamma$ is, in general, asymmetric. 

Having eliminated the redundant variables, any $N$ dimensional vector $\hat\gamma$ in Eq.~(\ref{Mgamma}) can be quantized as $\vec \gamma\to \vec\gamma'$, where
$\vec\gamma'=[\hat b_1,\dots,\hat b_{N}]$ is the vector of the boson annihilation operators  $\hat b_j$, which describe new fields, as shown in Fig.~\ref{fig:lattice_equivalence}. The emerging physics is that a \textit{new linear dissipative system} is constructed.
This new system is described by $N$ coupled fields $\hat b_j$ and it evolves under a moments-based NHH $\hat H_{\gamma}^{\rm mb}$ given by [c.f. Eq.~(\ref{MH})]:
\begin{equation}\label{Hb} 
 \dline H_{\gamma}^{\rm mb} =i\dline M_\gamma^{\rm red},
\end{equation}
where $ \dline H_{\gamma}^{\rm mb}$ is, as before, a matrix form of the moments-based NHH $\hat H_{\gamma}^{\rm mb}=(\vec\gamma')^{\dagger} \dline H_{\gamma}^{\rm mb}\vec{\gamma'}$.

This procedure is easily and recursively extended to any set of higher-order moments.
Thus, moments-based NHHs representing large systems can be constructed by considering an initial $U(1)$ two-mode system with an evolution matrix $\dline M_A$ for the first-order field moments.
Vice versa, one can obtain higher-order EPs by using the moments-based NHHs as a guideline to realize dissipative lattices of coupled resonator, whose effective NHH will have higher-order EPs (see Fig.~\ref{fig:lattice_equivalence}).
Notably, the effective NHH of the lattice could be characterized by a lower decay rate of the observables, resulting in better visibility of the EP.

\subsubsection{Moments-based non-Hermitian Hamiltonian of generic quadratic Liouvillians}

In a general case, the correspondence between $\boldsymbol{M}_A$ and the matrix form of the NHH $\Hef$, drawn in Eq.~(\ref{MH}), no longer holds. Instead, one has:
\begin{equation}\label{MHgen} 
 \dline H_{\rm eff} \equiv i\dline\eta_1\dline M_A+i\dline\eta_2\dline M_A^{\dagger}\dline\eta_3,
\end{equation}
where 
\begin{equation}\label{F}  
\dline\eta_1=\overset{\sim}{\bigoplus\limits_n}\begin{pmatrix}
1 &  \\
 & 0
\end{pmatrix},\hspace{1.5mm}\dline\eta_2=\overset{\sim}{\bigoplus\limits_n}\begin{pmatrix}
0 &  \\
 & 1
\end{pmatrix},\hspace{1.5mm}\dline\eta_3=\overset{\sim}{\bigoplus\limits_n}\begin{pmatrix}
-1 &  \\
 & 1
\end{pmatrix}.
\end{equation}
In Eq.~(\ref{F}), the symbol $\overset{\sim}{\bigoplus}$ means a direct sum (not the Kronecker one).
In Eq.~(\ref{MHgen}), without loss of generality, we have also dropped a constant term related to the field frequencies. 

 Equation~(\ref{MHgen}) reads as follows. The first term on its right-hand side takes into account the dynamics of only the annihilation operators $\hat a_k$, i.e., the odd elements of the vector $\hat A$ in Eq.~(\ref{Agen}). This term is equivalent to the right-hand side of Eq.~(\ref{MH}), when the system is $U(1)$ symmetric and linear.
The second term in Eq.~(\ref{MHgen}) accounts for the dynamics of  the creation operators of the fields, i.e., the even elements of the vector $\hat A$.

Contrary to the  case of linear systems, Eq.~(\ref{MHgen}) implies that the evolution matrix $\dline M_A$ and the NHH $\Hef$ are, in general, not simply related. 
Even though the critical points of the NHH and Eq.~(\ref{MHgen}) coincide, the spectral properties of these matrices might differ. 
And it is the spectral degeneracies of the evolution matrices that determine the properties of the coherence and spectral functions, which are experimentally accessible via photocount or homodyne measurements. 

Despite the fact that now there is no one-to-one correspondence between the evolution matrix for the first-order field moments and effective NHH in the case of generic nonlinear quadratic Liouvillians, according to Eq.~(\ref{MHgen}), one still can map the evolution matrix for higher-order moments to a new moments-based NHH, i.e.,
\begin{equation}\label{Hbgen}
    \dline M_\gamma \to \dline H_\gamma^{\rm mb}. 
\end{equation}
The mapping in Eq.~(\ref{Hbgen}), in general, might require additional operations over the matrix $\dline M_\gamma$, compared to Eqs.~(\ref{Hb}) and (\ref{MHgen}), e.g., row and column permutations, which depends on how a corresponding vector of operators $\vec{\gamma}$ is constructed from the initial vector $\vec{A}$.

As shown in Ref.~\cite{Miri2016,Wang2019}, even for a class of {\it Hermitian} quadratic Hamiltonians, i.e., without dissipation, and which describes optical nonlinear processes,  the corresponding evolution matrix $\dline M_A$ can reveal an EP.   Moreover, one can easily map the matrix $\dline M_A$ to a different  NHH as in Eq.~(\ref{MH})~\cite{Miri2016}.  Based on the latter, one then can straightforwardly apply  Eq.~(\ref{Hb}) for the construction of different moments-based NHHs exhibiting higher-order EPs.

Within this description, we can properly define the moments-based NHH of composite systems of higher-order moments. The advantages of this procedure are manifold: (i) The moments-based NHH now correctly takes into account the effects of quantum jumps. (ii) These NHHs can now be correctly quantized, and the commutation rules emerging from the physical moments-based NHH can correspond to those of the original system. In other words, the computation of quantum-relevant variables is unaffected by the algebraic construction. (iii) The procedure can be easily implemented numerically, even for large systems.

We also stress that such constructed NHHs and their properties are meaningful only in the synthetic space of field moments, encoded in the very form of the moments-based NHH. Stemming from the evolution matrices of {\it nonlinear} quadratic systems, the NHH does not necessarily imply that the corresponding spatial lattice of coupled resonators exhibits, e.g., nonclassicality.
Indeed, such nonclassicality, which can be revealed by the negative-definite field moments matrices, can only be universally revealed by considering {\it normally-ordered} moments~~\cite{AgarwalBook}, which is not the case under consideration 
[c.f.  Eq.~(\ref{Mgamma})].


\section{Engineering Higher-order exceptional points from quadratic Liouvillians} \label{IV}
In the previous discussion, we proved that it is possible to define a physical NHH using the full-Liouvillian description if one focuses on the dynamics of moments. Despite the lack of the correspondence between the eigenvectors describing the state and operators evolution (namely, the right- and left-hand-side eigenstates of the Liouvillian) there is a correspondence between their eigenvalues and their degeneracies. Indeed, if the states have an EP, so do the moments.
Here, by considering the spectra of evolution matrices for higher-order field moments, and exploiting the above-described mapping, we show how to engineer moments-based NHHs with higher-order EPs in any quadratic Liouvillian systems.

\subsection{High-order exceptional points}
Due to the structure of the equations of motion of the moments, the spectral degeneracies of an evolution matrix $\dline M_\gamma$  can be directly obtained from those of $\dline M_A$.
Given the Kronecker sum in Eq.~(\ref{Mgamma}), one has the following relation between eigenvalues of the matrices~\cite{Lototsky2015}:
\begin{equation}\label{lambdagamma} 
    \lambda_{ij}(\dline M_\gamma)=\lambda_i(\dline M_\alpha)+\lambda_j(\dline M_\beta), 
\end{equation}
with $i=1,\dots,N_\alpha$ and $j=1,\dots,N_\beta$.
Therefore, with each new term $\dline M_A$ in Eq.~(\ref{Mgamma}), the order of an EP of $\dline M_\gamma$ increases by one. 
Note that, although the \textit{algebraic} degeneracy of the eigenvalues in Eq.~(\ref{lambdagamma}) increases proportionally to $N_\gamma$, its \textit{geometric} multiplicity does not.
Indeed, if $\dline M_A$ has an EP of order 2, the evolution matrix $\dline M_\gamma=\bigoplus\limits_{i=1}^{m}\dline M_A$  has an EP of order $(m+1)$.

This result is an analogous proof to that in Ref.~\cite{Arkhipov2020b}, where it was demonstrated that once the evolution matrix $\dline M_A$ has an EP of second order, it immediately implies the presence of an EP of any higher-order $n\geq 2$ in the Liouvillian eigenspace. This eigenspace determines the evolution matrices $\dline M_\gamma$ of field moments. 
Consequently, the quantized moments and the corresponding NHHs given in Eqs.~(\ref{Hb}) and (\ref{Hbgen}) are characterized by the same degeneracy as that of $\dline M_\gamma$.

The matrix $\dline M_\gamma$ determines also the dynamics of high-order correlation functions, according to the quantum regression theorem~\cite{CarmichaelBook}. 
Wick's theorem for Gaussian states (i.e., quadratic systems) indicates that correlation functions of any order can be expressed as a sum of products of correlation functions of lower orders. 
This implies that if the matrix $\dline M_A$, which determines the first-order coherence function, exhibits an EP,  higher-order coherence functions reveal a higher order EP~\cite{Arkhipov2020b}.
The same conclusion can alternatively be drawn using the properties of the matrix exponential of a Kronecker sum~\cite{Neudecker1969}. 

Nevertheless, when constructing the moments-based NHH, the reduced matrix $\dline M_{\gamma}^{\rm red}$ is used instead, according to Eqs.~(\ref{Hb}) and (\ref{Hbgen}).  As a result,  the order of an EP of the reduced matrix would correspond to its dimension, e.g., for the $U(1)$ case, $N({\dline M_{\gamma}^{\rm red}})=m+1$, which is the same as the order of the EP. As a result, the  corresponding moments-based NHH $\hat H_\gamma^{\rm mb}$, describing $(m+1)$ modes (resonators) would have an EP of the order $(m+1)$.

Moreover,  from Eqs.~(\ref{Mgamma}) and (\ref{lambdagamma}) it is evident that the construction of the moments-based NHH $\hat{H}_{\gamma}^{\rm mb}$ for higher-order EPs can be realized with arbitrary matrices $\dline M_\alpha$ and $\dline M_\beta$.
We also note that although our method implicitly assumes that the evolution matrix $\dline M_A$ for first-order field moments  already has an EP, 
there are methods which allow to construct a new matrix having an EP from a combination of generic complex matrices with no initial degeneracies~\cite{Muhic2014}.


\section{Example of a $U(1)$ anti-$\cal PT$-symmetric cavity} \label{V}
In this section, we implement our scheme for the example of a $U(1)$-symmetric two-mode cavity with incoherent mode coupling, which additionally possesses the anti-$\cal PT$-symmetry. 
Namely, we show how a Liouvillian eigenspace of such a two-mode system, expressed via field moments and their evolution matrices,  can be mapped to a new moments-based NHH representing a multimode system.

In other words, first we  reveal EPs of any order arising from the evolution matrices of the bimodal system under consideration. Then, for any evolution matrix which exhibits an EP of order $n>2$, we assign a new moments-based NHH, which can, thus, correspond to a new coupled $n$-mode system with higher-order EPs.

\begin{figure}[t!] 
\includegraphics[width=0.33\textwidth]{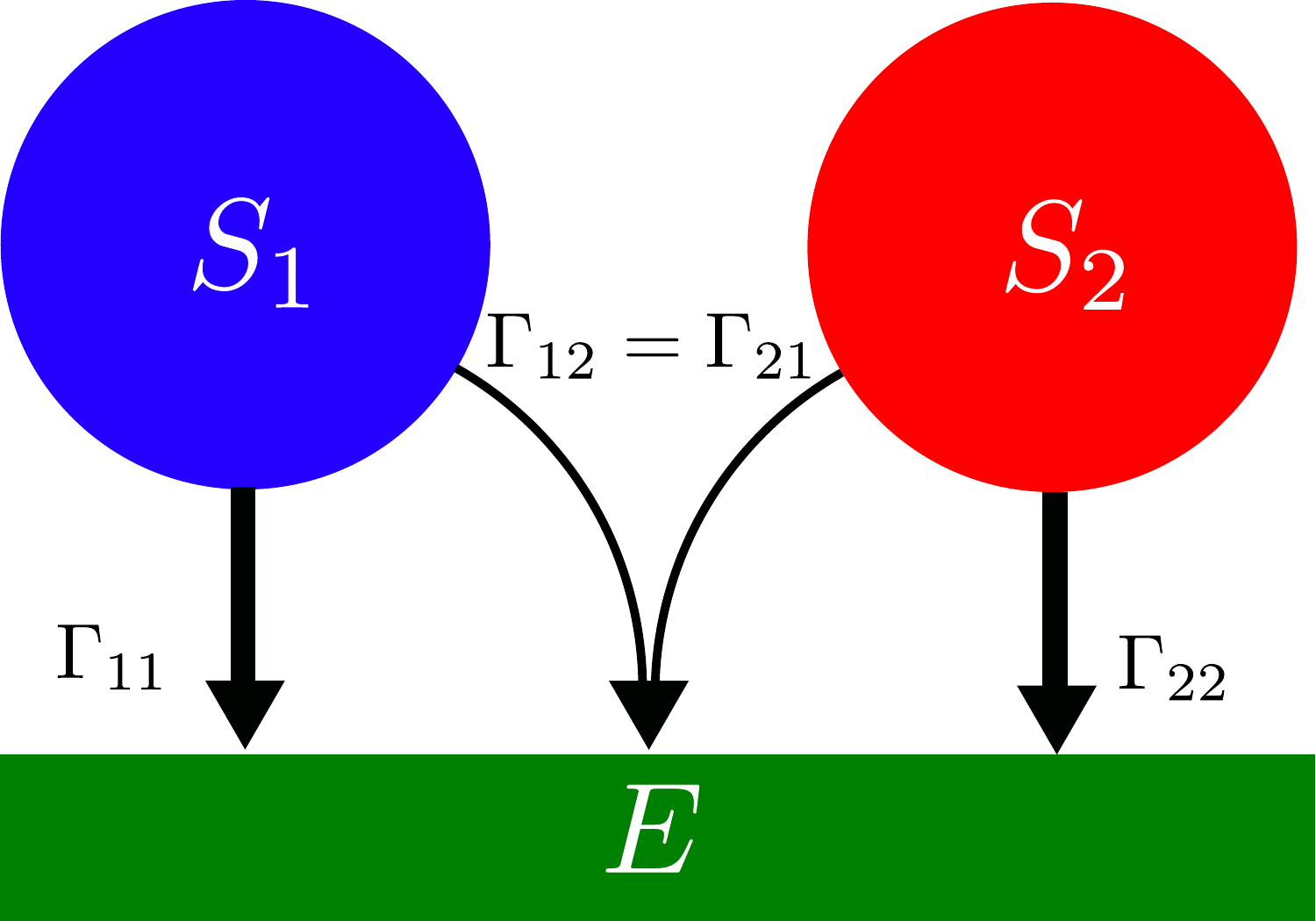}
\caption{Schematic representation of the model described by the Liouvillian superoperator $\cal L$ in Eq.~(\ref{eq4}). $S_{1}$ and $S_2$ are two system modes, which dissipate to the environment $E$ with rates $\Gamma_{11}$ and $\Gamma_{22}$, respectively. The modes are also dissipatively coupled to the environment with the incoherent coupling strength $\Gamma_{12}=\Gamma_{21}$. }\label{fig1}
\end{figure}

\subsection{Model of an anti-$\cal PT$-symmetric bimodal cavity}
The model under consideration is the same as in Ref.~\cite{Arkhipov2020b}.
Namely, we consider the Lindblad master equation in Eq.~(\ref{ME})
with the following Liouvillian superoperator~\cite{Haake2003,Franke19,Franke2020}:
\begin{equation}\label{eq4} 
{\cal L}\hat\rho=-i\left(\hat H_{\rm eff}\hat\rho-\hat\rho\hat H^{\dagger}_{\rm eff}\right)+\sum\limits_{j,k=1,2}\Gamma_{jk}\hat a_{j}\hat\rho\hat a_{k}^{\dagger},
\end{equation}
with the effective NHH:
\begin{equation}\label{Heff} 
\Hef=\sum\limits_{j=1,2}\omega_j\hat a^{\dagger}_j\hat a_j-{i}\sum\limits_{j,k=1,2}\Gamma_{jk}\hat
a_j^{\dagger}\hat a_k.
\end{equation}
The Liouvillian in Eq.~(\ref{eq4}) describes a dissipative linear system of two incoherently coupled modes. A schematic diagram of the model under study is shown in Fig.~\ref{fig1}.
In Eq.~(\ref{eq4}), $\hat a_j$ ($\hat a_j^{\dagger}$) is the
annihilation (creation) operator of  mode $j$ with a bare frequency $\omega_j$; the diagonal
damping coefficient $\Gamma_{kk}$ denotes the inner $k$th mode
{decay rate}, while the off-diagonal coefficient $\Gamma_{jk}=\Gamma_{kj}$ (for $j\neq k$)
accounts for the {\it incoherent} coupling strength between modes $j$ and $k$,
due to the interaction of both modes with the
environment~\cite{Haake2003}. That is, without loss of generality, we focus on a symmetric form of the decoherence matrix in Eq.~(\ref{eq4}), although, in general, $\Gamma_{jk}\neq\Gamma_{kj}$. The latter case can also result into the chiral character of the interaction between modes.

\subsection{Second-order EP}
In our previous study~\cite{Arkhipov2020b}, we analyzed EPs, up to their third order, of such an anti-$\cal PT$-symmetric bimodal cavity.
The evolution matrix for the first-order field moments $\langle\hat A\rangle=[\langle\hat a_1\rangle,\langle\hat a_2\rangle]^T$ in Eq.~(\ref{RES}) takes the form~\cite{Arkhipov2020b}:
\begin{equation} \label{MAex} 
\dline M_A=\begin{pmatrix}
-i\Delta-{\Gamma} & -\Gamma_{12} \\
-\Gamma_{12} & i\Delta-\Gamma
\end{pmatrix},
\end{equation}
where $\Delta =\omega_2-\omega_1$ is the frequency difference between the two modes, $\Gamma=\Gamma_{11}=\Gamma_{22}$ is an inner loss rate of each mode, and $\Gamma_{12}=\Gamma_{21}$ is an incoherent mode coupling strength (for details, see  Ref.~\cite{Arkhipov2020b}).

According to Eq.~(\ref{MAex}), the corresponding NHH is anti-$\cal PT$-symmetric, since it anticommutes with the parity-time  $\cal PT$ operator, i.e., 
\begin{equation}\label{APT} 
{\cal PT}\hat H_{\rm eff}{\cal PT}=-\hat H_{\rm eff},   
\end{equation}
which implies the $\cal PT$-symmetry of the evolution matrix $\dline M_A$. Moreover, by appropriately rotating the anti-$\cal PT$-symmetric NHH $\hat H_{\rm eff}$, one can switch it to a passive $\cal PT$-symmetric system~\cite{Arkhipov2020b}.

The eigenvalues of the matrix $\dline M_A$ in Eq.~(\ref{MAex}) read
\begin{equation}\label{eigA} 
\lambda_{1,2}=-\Gamma\pm\sqrt{\Gamma_{12}^2-\Delta^2}.
\end{equation}
Thus, the  EP of the system (which is of second order for the evolution matrix $\dline M_A$) is observed at the point 
\begin{equation}\label{EP} 
    \Gamma_{12}^{\rm EP}=|\Delta|.
\end{equation}
As a consequence, at the EP, the matrix $\dline M_A$, and thus $\Hef$, acquire a Jordan form, i.e., they become non-diagonalizable.

\subsection{Third-order EP}
According to Eqs.~(\ref{B}) and (\ref{MB}), the matrix $\dline M_B=\dline M_A\oplus\dline M_A$ written for the non-Hermitian second-order moments $\langle\vec B\rangle$ [in the form given in Eq.~(\ref{B})] reads as: 
\begin{equation}\label{MBex} 
   \dline M_B=
    \left( \begin {array}{cccc} -2i\Delta-2\Gamma&-\Gamma_{12}&-\Gamma_{12}
&0\\ \noalign{\medskip}-\Gamma_{12}&-2\Gamma&0&-\Gamma_{12}
\\ \noalign{\medskip}-\Gamma_{12}&0&-2\Gamma&-\Gamma_{12}
\\ \noalign{\medskip}0&-\Gamma_{12}&-\Gamma_{12}&2i\Delta-2\Gamma
\end {array} \right),
\end{equation}
would attain the same EP but of its third order. Indeed, 
the eigenvalues of this matrix are:
\begin{equation}\label{eig} 
\lambda_{1,2}=-2\Gamma\pm2\sqrt{\Gamma_{12}^2-\Delta^2}, \quad \lambda_{3,4}=-2\Gamma,
\end{equation}
which at the EP in Eq.~(\ref{EP}) become identical. 
The plots for these eigenvalues were presented in Ref.~\cite{Arkhipov2020b}.
Although the \textit{algebraic} multiplicity of the eigenvalues in Eq.~(\ref{eig}) at the EP equals four, the \textit{geometric} multiplicity equals three; thus, indicating the coalescence of three modes. The latter fact points that the EP is, indeed, of  third order.  
 Moreover, for the matrix $\dline M_B$ in Eq.~(\ref{MBex}), as mentioned above, one can effectively reduce its dimension to three, keeping the same order of the EP. That is,  by reducing the vector of moments
\begin{equation} 
    \begin{pmatrix}
    \langle\hat a_1^2\rangle \\
     \langle\hat a_2\hat a_1\rangle \\
      \langle\hat a_1\hat a_2\rangle \\
       \langle\hat a_2^2\rangle
    \end{pmatrix}\to
     \begin{pmatrix}
    \langle\hat a_1^2\rangle \\
     \langle\hat a_1\hat a_2\rangle \\
       \langle\hat a_2^2\rangle
    \end{pmatrix},
\end{equation}
one obtains the following effective matrix 
\begin{equation}\label{Meffex} 
    \dline M_B\to\dline M_B^{\rm red}=
    \left( \begin {array}{ccc} -2i\Delta-2\Gamma&-2\Gamma_{12}&0
\\ \noalign{\medskip}-\Gamma_{12}&-2\Gamma&-\Gamma_{12}
\\ \noalign{\medskip}0&-2\Gamma_{12}&2i\Delta-2\Gamma\end {array}
 \right). 
\end{equation}
By comparing Eqs.~(\ref{MBex}) and (\ref{Meffex}), one can see that the reduced matrix $\dline M_B^{\rm red}$ has lost the symmetry of the initial matrix $\dline M_B$, i.e.,  $\dline M^{\rm red}_B\neq(\dline M^{\rm red}_B)^T$. 

One can directly obtain a NHH from $\dline M_B^{\rm red}$, according to Eq.~(\ref{Hb}).
Namely, by additionally discarding the inner decoherence terms $2\Gamma$ in Eq.~(\ref{Meffex}), the corresponding dissipative system for this model is a lattice of three bosonic modes, interacting via the effective NHH
\begin{eqnarray}
    \hat{H}_{\rm eff} = &&2 \Delta \left(\hat{b}_1^\dagger\hat{b}_1 -\hat{b}_3^\dagger\hat{b}_3\right)\nonumber \\
    &&-i \Gamma_{12} \left(\hat{b}_1 \hat{b}_2^\dagger + 2\hat{b}_1^\dagger \hat{b}_2 +\hat{b}_3 \hat{b}_2^\dagger + 2\hat{b}_3^\dagger \hat{b}_2  \right),
\end{eqnarray}
and whose explicit Liouvillian dynamics reads
\begin{equation}\label{L3mode}
\mathcal{L}\hat{\rho}(t) = -i (\hat{H}_{\rm eff} \hat{\rho}(t)  -\hat{\rho}(t) \hat{H}_{\rm eff} )
     +2\sum_{i, \, j} \gamma_{ij} \left(\hat{b}_i \hat{\rho} \hat{b}_j^\dagger\right),
\end{equation}
where the decoherence matrix in Eq.~(\ref{L3mode}) has only nonzero off-diagonal elements $\gamma_{ji}=2\gamma_{ij}=2\Gamma_{12}$ with $i=1,3$, $j=2$.

\begin{figure} 
\includegraphics[width=0.48\textwidth]{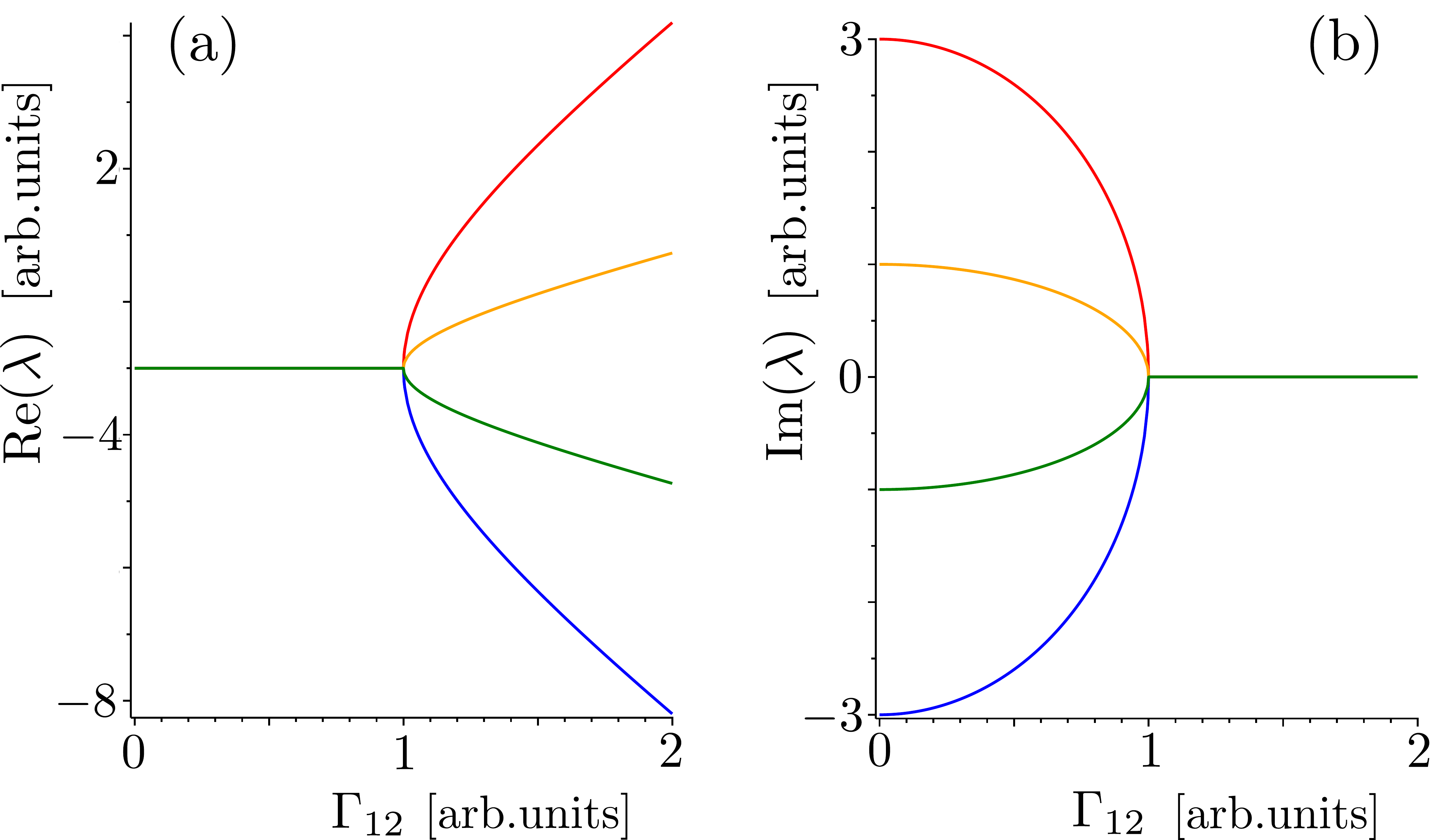}
\caption{(a) Real  and (b) imaginary parts of the eigenvalues
$\lambda$, according to Eq.~(\ref{nu}), of the effective evolution matrix $\hat M^{\rm eff}_C$, given in Eq.~(\ref{MC}), for the third-order field moments. Its four eigenvalues coalesce at the EP in Eq.~(\ref{EP}), thus indicating that the EP is of the fourth order. System parameters are set as $\Gamma=1$ [arb. units] and $\Delta=1$ [arb. units].}\label{fig2}
\end{figure}

\subsection{Fourth-order EP}
In a similar way, one can generate an NHH with an EP of the fourth order, whose matrix form reads as follows
\begin{equation} 
 \dline H^{\rm mb}_{\gamma}=i\dline M_C=i\bigoplus\limits_{n=1}^{3}\dline M_A,    
\end{equation}
 where $\vec{\gamma}=\vec{A}\otimes \vec{A} \otimes \vec{A}$,  and $\dline M_A$ is given in Eqs.~(\ref{MBex}) and (\ref{MAex}), respectively.
 Note that the resulting matrix $\hat M_C$ has dimension $2^3\times 2^3$. In other words, the number of required modes increases to eight. Nonetheless, again, one can contract the resulting matrix to dimension four, but at the expense of losing the mode-coupling symmetry. 
Namely, the evolution matrix $\dline M_C$ can be defined by eight moments of the form $\langle \hat a_i\hat a_j\hat a_k\rangle$, where $i,j,k=1,2$. As such, there are two sets of three equivalent moments $\langle\hat a_i^2\hat a_j\rangle=\langle\hat a_i\hat a_j\hat a_i\rangle=\langle\hat a_j\hat a_i^2\rangle$ for $i,j=1,2$ and $i\neq j$. As a result, only \textit{four} non-degenerate moments remain out of \textit{eight}, which, thus, defines the $4\times4$ reduced matrix $\dline M_C^{\rm red}$, and which attains the following form
\begin{eqnarray}\label{MC} 
    &&\dline M_C^{\rm red} = \nonumber \\
 &&   \left( \begin {array}{cccc} -3i\Delta-3\Gamma&-3\Gamma_{12}&0&0
\\ \noalign{\medskip}-\Gamma_{12}&-i\Delta-3\Gamma&-2\Gamma_{12}&0
\\ \noalign{\medskip}0&-2\Gamma_{12}&i\Delta-3\Gamma&-\Gamma_{12}
\\ \noalign{\medskip}0&0&-3\Gamma_{12}&3i\Delta-3\Gamma
\end {array} \right), \nonumber \\
\end{eqnarray}
and its eigenvalues read
\begin{eqnarray}\label{nu} 
\lambda_{1,2}&=&-3\Gamma\pm\sqrt{\Gamma_{12}^2-\Delta^2}, \nonumber \\
\lambda_{3,4}&=&-3\Gamma\pm3\sqrt{\Gamma_{12}^2-\Delta^2}.
\end{eqnarray}
We plot these eigenvalues in Fig.~\ref{fig2}.

\subsection{$(N+1)$th-order EP}
Clearly, one can obtain the effective evolution matrices for any dimension $(N+1)$ as 
\begin{eqnarray}\label{MN} 
   \dline M_{N+1}^{\rm eff} = \left( \begin {array}{ccccc} \Delta_0&-N\Gamma_{12}& & & 
\\ \noalign{\medskip}\ldots&\ldots&\ldots& \ldots & \ldots
\\ \noalign{\medskip}&-n\Gamma_{12}&\Delta_n&-(N-n)\Gamma_{12} &
\\ \noalign{\medskip}\ldots&\ldots&\ldots& \ldots & \ldots
\\ \noalign{\medskip}&& &-N\Gamma_{12}&\Delta_N  
\end {array} \right), \nonumber \\
\end{eqnarray}
where $\Delta_n=i(-N+2n)\Delta-N\Gamma$ for $n=0,\ldots,N$.
Combining now Eqs.~(\ref{lambdagamma}) and (\ref{eigA}), the eigenvalues of this matrix can be written  as
\begin{equation} 
    \lambda_{n}=-N\Gamma\pm (N-2n)\sqrt{\Gamma_{12}^2-\Delta^2}, \quad n=0,\ldots,N.
\end{equation}
Hence, a dissipative system described by the following effective NHH (expressed via matrix form):  
\begin{equation}\label{eq31} 
     \dline H_{\gamma}^{\rm mb} =i\dline M_{N+1}^{\rm red},
\end{equation}
which is defined for $(N+1)$ modes $\vec{\gamma}'$,  exhibits the EP of  order $(N+1)$.

Therefore, by recursively repeating the same procedure, one can create new effective NHHs with higher-order EPs out of various evolution matrices, generated by the initial matrix $\boldsymbol{M}_A$ corresponding to a bimodal anti-$\cal PT$-symmetric cavity. Importantly, the same conclusions can be drawn for any two-mode linear $U(1)$-symmetric open systems exhibiting an EP, including those with coherent mode coupling.

We note, when constructing moments-based NHH, according to Eq.~(\ref{eq31}), one can always  rescale the inner decoherence rate $\Gamma$ to ensure that the achieved resolution of a given high-order EPs via coherence function and spectra is maximal. 




\section{ Discussion and Conclusions}\label{VI}
In this article, 
we have proposed a method to properly define a new class of NHHs (so-called moments-based NHH) for quadratic Liouvillian systems, and which can exhibit higher-order quantum EPs.
These moments-based NHHs are derived via a quantization of moments of system operators obtained via a given quadratic Liouvillian. This approach correctly captures the dynamics of high-order system operators without semiclassical approximations, i.e., including the effect of quantum jumps~\cite{Minganti2019,Arkhipov2020}. In other words, we have proposed a simple and effective method for engineering higher-order quantum EPs based on the moments-based NHH.

The dynamics of the moments-based NHHs can be associated with networks of dissipative coupled resonators. 
That is, by starting from a quadratic Liouvillian, describing, e.g., a two-mode system,
the resulting evolution matrices, governing the higher-order field moments, can be cast
to the new NHHs, which can, thus, describe networks of coupled cavities instead. 
To put it another way, one can assign to moments-based NHHs a clear physical meaning by mapping the synthetic space of field moments to spatial lattice of coupled resonators, which exhibits a similar Liouvillian dynamics.
This can be prove also useful to design lattice of $U(1)$ resonators with higher-order EPs, which could have applications for transport properties \cite{Khandelwal2021}.

 Compared to other existing methods for constructing NHHs with high-order EPs, based on spatial lattices~\cite{Teimourpour2014,Zhang2020c,Sahin2020}, the main advantage of our approach lies in its \textit{simplicity and preservation of the commensurate character of mode coupling strengths.} Also, compared to other approaches based on synthetic dimensions realized, e.g., in Fock space~\cite{Quiroz19,Tschernig20}, the synthetic {\it field moments space} can precisely simulate a spatial lattice of cavities in the pure quantum regime,
where quantum jumps can play the dominant role, and the postselection procedure becomes experimentally demanding~\cite{Quiroz19,Tschernig20}. Moreover, when simulating such phenomenological or postselected NHHs in the quantum regime, one might require the implementation of a specific metric operator, in order to correctly determine the physical states and the averages of operators within the quantum NHH formalism~\cite{MOSTAFAZADEH_2010}.

On the other hand, the Liouvillian eigenspace genuinely incorporates the effect of quantum jumps, 
the {\it averaged} effect of which is  captured in the dynamics of the field moments expressed via the corresponding evolution matrices.
As such, the mapping between the evolution matrices, governing the moments of the system operators, and moments-based NHHs ensures that the latter has a clear physical meaning also in the quantum limit.
Note that the presence of quantum jumps can profoundly affect the system {\it dynamical} spectra, e.g., 
 in some finite dimensional systems, the jumps can even lead to a shift of an EP in the parameter space~\cite{Minganti2020}.

As an example, we have analyzed a $U(1)$-symmetric cavity with incoherent mode coupling, which can also possess  anti-$\cal PT$-symmetry. 
The system studied in Sec.~\ref{V} can serve as a promising platform for implementing structures with high-order EPs.
Indeed, such systems with incoherent mode coupling can combine features related to both $\cal PT$ and anti-$\cal PT$-symmetries, as has been demonstrated in Ref.~\cite{Arkhipov2020b}.
The construction of the physical moments-based NHH allows to obtain such EPs without building complex networks of coupled cavities, compared to previous works~\cite{Teimourpour2014,Nada2017,Zhang2020c,Sahin2020}. Instead, one can just excite additional modes in a multimode cavity, ensuring that the system modes interact incoherently. Moreover, according to the previous section, the incoherent mode coupling strengths in such systems are commensurate and do not require fine tuning as, e.g., in Ref.~\cite{Teimourpour2014}, where coupling constants have incommensurate irrational prefactors, due to the expansion of the Hilbert space of an effective NHH. This conclusion is also valid when the intermode interaction is coherent.

The  method proposed here allows to construct moments-based NHHs which can be both symmetric and asymmetric.

Such asymmetrical mode coupling can be engineered via backscattering processes, as has been shown in Refs.~\cite{Vivieskas2004,Kullig2018,Kullig2019}. Moreover, exploiting such asymmetric incoherent mode interactions, one can also implement a recently proposed scheme~\cite{Sahin2020}, which is based on the chiral nature of coherently coupled cavities,  but rather using a much simpler \textit{single multimode} cavity. Using our approach, one can also engineer block triangular moments-based NHHs highlighting the chirality in the mode coupling~\cite{Sahin2020}. The latter can be obtained by combining Eq.~(\ref{Mgamma}) and Schur's triangularization theorem~\cite{HornBook}.

Apart from the interest in exploiting higher-order quantum Liouvillian EPs for constructing new NHHs in synthetic space of field moments, 
it also prompts the question of a further utility of such EPs, which can be revealed by coherence and spectral functions for a given open quantum system~\cite{Arkhipov2020b}. After all, it is exactly the Liouvillian eigenspace which correctly captures the dynamics of the fields, not phenomenological NHHs, based on the semiclassical approximation or postselection procedure.

The proposed method  is  universal in its nature. According to Eqs.~(\ref{Mgamma}) and (\ref{MHgen}), it can be applied to any physically realizable matrices $\dline M_\alpha$ and $\dline M_\beta$, each exhibiting an EP of some order, and which can be related to arbitrary quadratic Liouvillians.
For instance, matrices $\dline M_\alpha$ and $\dline M_\beta$ can represent a $Z_2$-symmetric quadratic Liouvillian which describes a nonlinear parametric dissipative process.

It is also noteworthy that other moments-based methods have found a wide range of applications (e.g.,~\cite{Lin1996,Lin2002,Zueco2018,Purkayastha2020}).
Moreover, various sufficient and necessary criteria of nonclassical
correlations can be derived from such matrices of moments. These include criteria of:
(i) quantum entanglement (by applying the partial transposition~\cite{Shchukin2005,Miranowicz2006}
or other positive maps ~\cite{Miranowicz2009} to the matrices of moments),
(ii) quantum steering~\cite{Kogias2015},
(iii) Bell nonlocality~\cite{Navascus2007}, as well as
(iv) spatial~\cite{Richter2002} and spatio-temporal nonclassicality~\cite{Vogel2008,Miranowicz2010}.

Additionally, with the help of the technique described here, one can also analyze spectral degeneracies in systems with a weak nonlinearity, e.g., with Kerr-like interactions between modes.
In this case, one can still invoke the Gaussian approximation for the fields fluctuations near the steady state, where the fields intensities are assumed to be fixed. 
As such, one can analytically derive and experimentally probe the system dynamical critical points by means of higher-order coherence and spectral functions of the fields fluctuations even in the presence of weak nonlinearities.

\begin{acknowledgments}
I.A.
thanks the Grant Agency of the Czech Republic (Project
No.~18-08874S), and Project No.
CZ.02.1.01\/0.0\/0.0\/16\_019\/0000754 of the Ministry of
Education, Youth and Sports of the Czech Republic. A.M. was
supported by the Polish National Science Centre (NCN) under the
Maestro Grant No. DEC-2019/34/A/ST2/00081. F.N. is supported in part by:
Nippon Telegraph and Telephone Corporation (NTT) Research,
the Japan Science and Technology Agency (JST) [via
the Quantum Leap Flagship Program (Q-LEAP),
the Moonshot R\&D Grant Number JPMJMS2061, and
the Centers of Research Excellence in Science and Technology (CREST) Grant No. JPMJCR1676],
the Japan Society for the Promotion of Science (JSPS)
[via the Grants-in-Aid for Scientific Research (KAKENHI) Grant No. JP20H00134 and the
JSPS–RFBR Grant No. JPJSBP120194828],
the Army Research Office (ARO) (Grant No. W911NF-18-1-0358),
the Asian Office of Aerospace Research and Development (AOARD) (via Grant No. FA2386-20-1-4069), and
the Foundational Questions Institute Fund (FQXi) via Grant No. FQXi-IAF19-06.\end{acknowledgments}
\bibliography{references}

\begin{thebibliography}{114}%
\makeatletter
\providecommand \@ifxundefined [1]{%
 \@ifx{#1\undefined}
}%
\providecommand \@ifnum [1]{%
 \ifnum #1\expandafter \@firstoftwo
 \else \expandafter \@secondoftwo
 \fi
}%
\providecommand \@ifx [1]{%
 \ifx #1\expandafter \@firstoftwo
 \else \expandafter \@secondoftwo
 \fi
}%
\providecommand \natexlab [1]{#1}%
\providecommand \enquote  [1]{``#1''}%
\providecommand \bibnamefont  [1]{#1}%
\providecommand \bibfnamefont [1]{#1}%
\providecommand \citenamefont [1]{#1}%
\providecommand \href@noop [0]{\@secondoftwo}%
\providecommand \href [0]{\begingroup \@sanitize@url \@href}%
\providecommand \@href[1]{\@@startlink{#1}\@@href}%
\providecommand \@@href[1]{\endgroup#1\@@endlink}%
\providecommand \@sanitize@url [0]{\catcode `\\12\catcode `\$12\catcode
  `\&12\catcode `\#12\catcode `\^12\catcode `\_12\catcode `\%12\relax}%
\providecommand \@@startlink[1]{}%
\providecommand \@@endlink[0]{}%
\providecommand \url  [0]{\begingroup\@sanitize@url \@url }%
\providecommand \@url [1]{\endgroup\@href {#1}{\urlprefix }}%
\providecommand \urlprefix  [0]{URL }%
\providecommand \Eprint [0]{\href }%
\providecommand \doibase [0]{http://dx.doi.org/}%
\providecommand \selectlanguage [0]{\@gobble}%
\providecommand \bibinfo  [0]{\@secondoftwo}%
\providecommand \bibfield  [0]{\@secondoftwo}%
\providecommand \translation [1]{[#1]}%
\providecommand \BibitemOpen [0]{}%
\providecommand \bibitemStop [0]{}%
\providecommand \bibitemNoStop [0]{.\EOS\space}%
\providecommand \EOS [0]{\spacefactor3000\relax}%
\providecommand \BibitemShut  [1]{\csname bibitem#1\endcsname}%
\let\auto@bib@innerbib\@empty
\bibitem [{\citenamefont {\c{S}. K.~{\"O}zdemir}\ \emph
  {et~al.}(2019)\citenamefont {\c{S}. K.~{\"O}zdemir}, \citenamefont {Rotter},
  \citenamefont {Nori},\ and\ \citenamefont {Yang}}]{Ozdemir2019}%
  \BibitemOpen
  \bibfield  {author} {\bibinfo {author} {\bibnamefont {\c{S}.
  K.~{\"O}zdemir}}, \bibinfo {author} {\bibfnamefont {S.}~\bibnamefont
  {Rotter}}, \bibinfo {author} {\bibfnamefont {F.}~\bibnamefont {Nori}}, \ and\
  \bibinfo {author} {\bibfnamefont {L.}~\bibnamefont {Yang}},\ }\bibfield
  {title} {\enquote {\bibinfo {title} {Parity-time symmetry and exceptional
  points in photonics},}\ }\href {\doibase 10.1038/s41563-019-0304-9}
  {\bibfield  {journal} {\bibinfo  {journal} {Nat. Mater.}\ }\textbf {\bibinfo
  {volume} {18}},\ \bibinfo {pages} {783} (\bibinfo {year} {2019})}\BibitemShut
  {NoStop}%
\bibitem [{\citenamefont {Miri}\ and\ \citenamefont {Al\`u}(2019)}]{Miri2019}%
  \BibitemOpen
  \bibfield  {author} {\bibinfo {author} {\bibfnamefont {M.}~\bibnamefont
  {Miri}}\ and\ \bibinfo {author} {\bibfnamefont {A.}~\bibnamefont {Al\`u}},\
  }\bibfield  {title} {\enquote {\bibinfo {title} {Exceptional points in optics
  and photonics},}\ }\href {https://doi.org/10.1126/science.aar7709} {\bibfield
   {journal} {\bibinfo  {journal} {Science}\ }\textbf {\bibinfo {volume}
  {363}},\ \bibinfo {pages} {7709} (\bibinfo {year} {2019})}\BibitemShut
  {NoStop}%
\bibitem [{\citenamefont {Ashida}\ \emph {et~al.}(2020)\citenamefont {Ashida},
  \citenamefont {Gong},\ and\ \citenamefont {Ueda}}]{Ashida2020}%
  \BibitemOpen
  \bibfield  {author} {\bibinfo {author} {\bibfnamefont {Y.}~\bibnamefont
  {Ashida}}, \bibinfo {author} {\bibfnamefont {Z.}~\bibnamefont {Gong}}, \ and\
  \bibinfo {author} {\bibfnamefont {M.}~\bibnamefont {Ueda}},\ }\bibfield
  {title} {\enquote {\bibinfo {title} {Non-{H}ermitian physics},}\ }\href
  {\doibase 10.1080/00018732.2021.1876991} {\bibfield  {journal} {\bibinfo
  {journal} {Adv. Phys.}\ }\textbf {\bibinfo {volume} {69}},\ \bibinfo {pages}
  {249–435} (\bibinfo {year} {2020})}\BibitemShut {NoStop}%
\bibitem [{\citenamefont {Bender}\ and\ \citenamefont
  {Boettcher}(1998)}]{Bender1998}%
  \BibitemOpen
  \bibfield  {author} {\bibinfo {author} {\bibfnamefont {C.~M.}\ \bibnamefont
  {Bender}}\ and\ \bibinfo {author} {\bibfnamefont {S.}~\bibnamefont
  {Boettcher}},\ }\bibfield  {title} {\enquote {\bibinfo {title} {Real spectra
  in non-{H}ermitian {H}amiltonians having $\mathcal{PT}$ symmetry},}\ }\href
  {\doibase 10.1103/PhysRevLett.80.5243} {\bibfield  {journal} {\bibinfo
  {journal} {Phys. Rev. Lett.}\ }\textbf {\bibinfo {volume} {80}},\ \bibinfo
  {pages} {5243} (\bibinfo {year} {1998})}\BibitemShut {NoStop}%
\bibitem [{\citenamefont {Ju}\ \emph {et~al.}(2019)\citenamefont {Ju},
  \citenamefont {Miranowicz}, \citenamefont {Chen},\ and\ \citenamefont
  {Nori}}]{Ju2019}%
  \BibitemOpen
  \bibfield  {author} {\bibinfo {author} {\bibfnamefont {C.-Y.}\ \bibnamefont
  {Ju}}, \bibinfo {author} {\bibfnamefont {A.}~\bibnamefont {Miranowicz}},
  \bibinfo {author} {\bibfnamefont {G.-Y.}\ \bibnamefont {Chen}}, \ and\
  \bibinfo {author} {\bibfnamefont {F.}~\bibnamefont {Nori}},\ }\bibfield
  {title} {\enquote {\bibinfo {title} {Non-{H}ermitian {H}amiltonians and no-go
  theorems in quantum information},}\ }\href {\doibase
  10.1103/PhysRevA.100.062118} {\bibfield  {journal} {\bibinfo  {journal}
  {Phys. Rev. A}\ }\textbf {\bibinfo {volume} {100}},\ \bibinfo {pages}
  {062118} (\bibinfo {year} {2019})}\BibitemShut {NoStop}%
\bibitem [{\citenamefont {Mostafazadeh}(2002)}]{Mostafazadeh2002}%
  \BibitemOpen
  \bibfield  {author} {\bibinfo {author} {\bibfnamefont {A.}~\bibnamefont
  {Mostafazadeh}},\ }\bibfield  {title} {\enquote {\bibinfo {title}
  {Pseudo-{H}ermiticity versus $\mathcal{PT}$-symmetry: {T}he necessary
  condition for the reality of the spectrum of a non-{H}ermitian
  {H}amiltonian},}\ }\href {\doibase 10.1063/1.1418246} {\bibfield  {journal}
  {\bibinfo  {journal} {J. of Math. Phys.}\ }\textbf {\bibinfo {volume} {43}},\
  \bibinfo {pages} {205} (\bibinfo {year} {2002})}\BibitemShut {NoStop}%
\bibitem [{\citenamefont {Roy}\ \emph {et~al.}(2021{\natexlab{a}})\citenamefont
  {Roy}, \citenamefont {Jahani}, \citenamefont {Guo}, \citenamefont {Dutt},
  \citenamefont {Fan}, \citenamefont {Miri},\ and\ \citenamefont
  {Marandi}}]{Roy2020}%
  \BibitemOpen
  \bibfield  {author} {\bibinfo {author} {\bibfnamefont {A.}~\bibnamefont
  {Roy}}, \bibinfo {author} {\bibfnamefont {S.}~\bibnamefont {Jahani}},
  \bibinfo {author} {\bibfnamefont {Q.}~\bibnamefont {Guo}}, \bibinfo {author}
  {\bibfnamefont {A.}~\bibnamefont {Dutt}}, \bibinfo {author} {\bibfnamefont
  {S.}~\bibnamefont {Fan}}, \bibinfo {author} {\bibfnamefont {M.-A.}\
  \bibnamefont {Miri}}, \ and\ \bibinfo {author} {\bibfnamefont
  {A.}~\bibnamefont {Marandi}},\ }\bibfield  {title} {\enquote {\bibinfo
  {title} {Nondissipative non-{H}ermitian dynamics and exceptional points in
  coupled optical parametric oscillators},}\ }\href {\doibase
  10.1364/optica.415569} {\bibfield  {journal} {\bibinfo  {journal} {Optica}\
  }\textbf {\bibinfo {volume} {8}},\ \bibinfo {pages} {415} (\bibinfo {year}
  {2021}{\natexlab{a}})}\BibitemShut {NoStop}%
\bibitem [{\citenamefont {Roy}\ \emph {et~al.}(2021{\natexlab{b}})\citenamefont
  {Roy}, \citenamefont {Jahani}, \citenamefont {Langrock}, \citenamefont
  {Fejer},\ and\ \citenamefont {Marandi}}]{Roy2021}%
  \BibitemOpen
  \bibfield  {author} {\bibinfo {author} {\bibfnamefont {A.}~\bibnamefont
  {Roy}}, \bibinfo {author} {\bibfnamefont {S.}~\bibnamefont {Jahani}},
  \bibinfo {author} {\bibfnamefont {C.}~\bibnamefont {Langrock}}, \bibinfo
  {author} {\bibfnamefont {M.}~\bibnamefont {Fejer}}, \ and\ \bibinfo {author}
  {\bibfnamefont {A.}~\bibnamefont {Marandi}},\ }\bibfield  {title} {\enquote
  {\bibinfo {title} {Spectral phase transitions in optical parametric
  oscillators},}\ }\href {http://dx.doi.org/10.1038/s41467-021-21048-z}
  {\bibfield  {journal} {\bibinfo  {journal} {Nat. Commun.}\ }\textbf {\bibinfo
  {volume} {12}} (\bibinfo {year} {2021}{\natexlab{b}})}\BibitemShut {NoStop}%
\bibitem [{\citenamefont {Jahani}\ \emph {et~al.}(2021)\citenamefont {Jahani},
  \citenamefont {Roy},\ and\ \citenamefont {Marandi}}]{Jahani2021}%
  \BibitemOpen
  \bibfield  {author} {\bibinfo {author} {\bibfnamefont {S.}~\bibnamefont
  {Jahani}}, \bibinfo {author} {\bibfnamefont {A.}~\bibnamefont {Roy}}, \ and\
  \bibinfo {author} {\bibfnamefont {A.}~\bibnamefont {Marandi}},\ }\bibfield
  {title} {\enquote {\bibinfo {title} {Wavelength-scale optical parametric
  oscillators},}\ }\href {\doibase 10.1364/optica.411708} {\bibfield  {journal}
  {\bibinfo  {journal} {Optica}\ }\textbf {\bibinfo {volume} {8}},\ \bibinfo
  {pages} {262} (\bibinfo {year} {2021})}\BibitemShut {NoStop}%
\bibitem [{\citenamefont {Schindler}\ \emph {et~al.}(2011)\citenamefont
  {Schindler}, \citenamefont {Li}, \citenamefont {Zheng}, \citenamefont
  {Ellis},\ and\ \citenamefont {Kottos}}]{Schindler2011}%
  \BibitemOpen
  \bibfield  {author} {\bibinfo {author} {\bibfnamefont {J.}~\bibnamefont
  {Schindler}}, \bibinfo {author} {\bibfnamefont {A.}~\bibnamefont {Li}},
  \bibinfo {author} {\bibfnamefont {M.C.}\ \bibnamefont {Zheng}}, \bibinfo
  {author} {\bibfnamefont {F.~M.}\ \bibnamefont {Ellis}}, \ and\ \bibinfo
  {author} {\bibfnamefont {T.}~\bibnamefont {Kottos}},\ }\bibfield  {title}
  {\enquote {\bibinfo {title} {Experimental study of active {LRC} circuits with
  $\mathcal{PT}$ symmetries},}\ }\href
  {https://doi.org/10.1103/PhysRevA.84.040101} {\bibfield  {journal} {\bibinfo
  {journal} {Phys. Rev. A}\ }\textbf {\bibinfo {volume} {84}},\ \bibinfo
  {pages} {040101(R)} (\bibinfo {year} {2011})}\BibitemShut {NoStop}%
\bibitem [{\citenamefont {Jing}\ \emph {et~al.}(2014)\citenamefont {Jing},
  \citenamefont {\c{S}. K.~{\"O}zdemir}, \citenamefont {L{\"u}}, \citenamefont
  {Zhang}, \citenamefont {Yang},\ and\ \citenamefont {Nori}}]{Jing2014}%
  \BibitemOpen
  \bibfield  {author} {\bibinfo {author} {\bibfnamefont {H.}~\bibnamefont
  {Jing}}, \bibinfo {author} {\bibnamefont {\c{S}. K.~{\"O}zdemir}}, \bibinfo
  {author} {\bibfnamefont {X.-Y.}\ \bibnamefont {L{\"u}}}, \bibinfo {author}
  {\bibfnamefont {J.}~\bibnamefont {Zhang}}, \bibinfo {author} {\bibfnamefont
  {L.}~\bibnamefont {Yang}}, \ and\ \bibinfo {author} {\bibfnamefont
  {F.}~\bibnamefont {Nori}},\ }\bibfield  {title} {\enquote {\bibinfo {title}
  {$\mathcal{PT}$-symmetric phonon laser},}\ }\href
  {https://doi.org/10.1103/PhysRevLett.113.053604} {\bibfield  {journal}
  {\bibinfo  {journal} {Phys. Rev. Lett.}\ }\textbf {\bibinfo {volume} {113}},\
  \bibinfo {pages} {053604} (\bibinfo {year} {2014})}\BibitemShut {NoStop}%
\bibitem [{\citenamefont {Jing}\ \emph {et~al.}(2015)\citenamefont {Jing},
  \citenamefont {\c{S}. K.~{\"O}zdemir}, \citenamefont {Geng}, \citenamefont
  {Zhang}, \citenamefont {L{\"u}}, \citenamefont {Peng}, \citenamefont {Yang},\
  and\ \citenamefont {Nori}}]{Jing2015}%
  \BibitemOpen
  \bibfield  {author} {\bibinfo {author} {\bibfnamefont {H.}~\bibnamefont
  {Jing}}, \bibinfo {author} {\bibnamefont {\c{S}. K.~{\"O}zdemir}}, \bibinfo
  {author} {\bibfnamefont {Z.}~\bibnamefont {Geng}}, \bibinfo {author}
  {\bibfnamefont {J.}~\bibnamefont {Zhang}}, \bibinfo {author} {\bibfnamefont
  {X.-Y.}\ \bibnamefont {L{\"u}}}, \bibinfo {author} {\bibfnamefont
  {B.}~\bibnamefont {Peng}}, \bibinfo {author} {\bibfnamefont {L.}~\bibnamefont
  {Yang}}, \ and\ \bibinfo {author} {\bibfnamefont {F.}~\bibnamefont {Nori}},\
  }\bibfield  {title} {\enquote {\bibinfo {title} {Optomechanically-induced
  transparency in parity-time-symmetric microresonators},}\ }\href
  {https://doi.org/10.1038/srep09663} {\bibfield  {journal} {\bibinfo
  {journal} {Sci. Rep.}\ }\textbf {\bibinfo {volume} {5}},\ \bibinfo {pages}
  {9663} (\bibinfo {year} {2015})}\BibitemShut {NoStop}%
\bibitem [{\citenamefont {Xu}\ \emph {et~al.}(2016)\citenamefont {Xu},
  \citenamefont {Mason}, \citenamefont {Jiang},\ and\ \citenamefont
  {Harris}}]{Harris2016}%
  \BibitemOpen
  \bibfield  {author} {\bibinfo {author} {\bibfnamefont {H.}~\bibnamefont
  {Xu}}, \bibinfo {author} {\bibfnamefont {D.}~\bibnamefont {Mason}}, \bibinfo
  {author} {\bibfnamefont {L.}~\bibnamefont {Jiang}}, \ and\ \bibinfo {author}
  {\bibfnamefont {J.~G.~E.}\ \bibnamefont {Harris}},\ }\bibfield  {title}
  {\enquote {\bibinfo {title} {Topological energy transfer in an optomechanical
  system with exceptional points},}\ }\href
  {https://doi.org/10.1038/nature18604} {\bibfield  {journal} {\bibinfo
  {journal} {Nature (London)}\ }\textbf {\bibinfo {volume} {537}},\ \bibinfo
  {pages} {80} (\bibinfo {year} {2016})}\BibitemShut {NoStop}%
\bibitem [{\citenamefont {Jing}\ \emph {et~al.}(2017)\citenamefont {Jing},
  \citenamefont {\c{S}. K.~{\"O}zdemir}, \citenamefont {L{\"u}},\ and\
  \citenamefont {Nori}}]{Jing2017}%
  \BibitemOpen
  \bibfield  {author} {\bibinfo {author} {\bibfnamefont {H.}~\bibnamefont
  {Jing}}, \bibinfo {author} {\bibnamefont {\c{S}. K.~{\"O}zdemir}}, \bibinfo
  {author} {\bibfnamefont {H.}~\bibnamefont {L{\"u}}}, \ and\ \bibinfo {author}
  {\bibfnamefont {F.}~\bibnamefont {Nori}},\ }\bibfield  {title} {\enquote
  {\bibinfo {title} {High-order exceptional points in optomechanics},}\ }\href
  {https://doi.org/10.1038/s41598-017-03546-7} {\bibfield  {journal} {\bibinfo
  {journal} {Sci. Rep.}\ }\textbf {\bibinfo {volume} {7}},\ \bibinfo {pages}
  {3386} (\bibinfo {year} {2017})}\BibitemShut {NoStop}%
\bibitem [{\citenamefont {Zhu}\ \emph {et~al.}(2014)\citenamefont {Zhu},
  \citenamefont {Ramezani}, \citenamefont {Shi}, \citenamefont {Zhu},\ and\
  \citenamefont {Zhang}}]{Zhu2014}%
  \BibitemOpen
  \bibfield  {author} {\bibinfo {author} {\bibfnamefont {X.}~\bibnamefont
  {Zhu}}, \bibinfo {author} {\bibfnamefont {H.}~\bibnamefont {Ramezani}},
  \bibinfo {author} {\bibfnamefont {C.}~\bibnamefont {Shi}}, \bibinfo {author}
  {\bibfnamefont {J.}~\bibnamefont {Zhu}}, \ and\ \bibinfo {author}
  {\bibfnamefont {X.}~\bibnamefont {Zhang}},\ }\bibfield  {title} {\enquote
  {\bibinfo {title} {$\mathcal{PT}$-symmetric acoustics},}\ }\href
  {https://doi.org/10.1103/PhysRevX.4.031042} {\bibfield  {journal} {\bibinfo
  {journal} {Phys. Rev. X}\ }\textbf {\bibinfo {volume} {4}},\ \bibinfo {pages}
  {031042} (\bibinfo {year} {2014})}\BibitemShut {NoStop}%
\bibitem [{\citenamefont {Fleury}\ \emph {et~al.}(2015)\citenamefont {Fleury},
  \citenamefont {Sounas},\ and\ \citenamefont {Al\`u}}]{Alu2015}%
  \BibitemOpen
  \bibfield  {author} {\bibinfo {author} {\bibfnamefont {R.}~\bibnamefont
  {Fleury}}, \bibinfo {author} {\bibfnamefont {D.}~\bibnamefont {Sounas}}, \
  and\ \bibinfo {author} {\bibfnamefont {A.}~\bibnamefont {Al\`u}},\ }\bibfield
   {title} {\enquote {\bibinfo {title} {An invisible acoustic sensor based on
  parity-time symmetry},}\ }\href {https://doi.org/10.1038/ncomms6905}
  {\bibfield  {journal} {\bibinfo  {journal} {Nat. Commun.}\ }\textbf {\bibinfo
  {volume} {6}},\ \bibinfo {pages} {5905} (\bibinfo {year} {2015})}\BibitemShut
  {NoStop}%
\bibitem [{\citenamefont {Benisty}\ \emph {et~al.}(2011)\citenamefont
  {Benisty}, \citenamefont {Degiron}, \citenamefont {Lupu}, \citenamefont
  {Lustrac}, \citenamefont {Chenais}, \citenamefont {Forget}, \citenamefont
  {Besbes}, \citenamefont {Barbillon}, \citenamefont {Bruyant}, \citenamefont
  {Blaize},\ and\ \citenamefont {Lerondel}}]{Benisty2011}%
  \BibitemOpen
  \bibfield  {author} {\bibinfo {author} {\bibfnamefont {H.}~\bibnamefont
  {Benisty}}, \bibinfo {author} {\bibfnamefont {A.}~\bibnamefont {Degiron}},
  \bibinfo {author} {\bibfnamefont {A.}~\bibnamefont {Lupu}}, \bibinfo {author}
  {\bibfnamefont {A.~De}\ \bibnamefont {Lustrac}}, \bibinfo {author}
  {\bibfnamefont {S.}~\bibnamefont {Chenais}}, \bibinfo {author} {\bibfnamefont
  {S.}~\bibnamefont {Forget}}, \bibinfo {author} {\bibfnamefont
  {M.}~\bibnamefont {Besbes}}, \bibinfo {author} {\bibfnamefont
  {G.}~\bibnamefont {Barbillon}}, \bibinfo {author} {\bibfnamefont
  {A.}~\bibnamefont {Bruyant}}, \bibinfo {author} {\bibfnamefont
  {S.}~\bibnamefont {Blaize}}, \ and\ \bibinfo {author} {\bibfnamefont
  {G.}~\bibnamefont {Lerondel}},\ }\bibfield  {title} {\enquote {\bibinfo
  {title} {Implementation of $\mathcal{PT}$ symmetric devices using plasmonics:
  principle and applications},}\ }\href {https://doi.org/10.1364/OE.19.018004}
  {\bibfield  {journal} {\bibinfo  {journal} {Opt. Express}\ }\textbf {\bibinfo
  {volume} {19}},\ \bibinfo {pages} {18004} (\bibinfo {year}
  {2011})}\BibitemShut {NoStop}%
\bibitem [{\citenamefont {Kang}\ \emph {et~al.}(2013)\citenamefont {Kang},
  \citenamefont {Liu},\ and\ \citenamefont {Li}}]{Kang2013}%
  \BibitemOpen
  \bibfield  {author} {\bibinfo {author} {\bibfnamefont {M.}~\bibnamefont
  {Kang}}, \bibinfo {author} {\bibfnamefont {F.}~\bibnamefont {Liu}}, \ and\
  \bibinfo {author} {\bibfnamefont {J.}~\bibnamefont {Li}},\ }\bibfield
  {title} {\enquote {\bibinfo {title} {Effective spontaneous
  $\mathcal{PT}$-symmetry breaking in hybridized metamaterials},}\ }\href
  {https://doi.org/10.1103/PhysRevA.87.053824} {\bibfield  {journal} {\bibinfo
  {journal} {Phys. Rev. A}\ }\textbf {\bibinfo {volume} {87}},\ \bibinfo
  {pages} {053824} (\bibinfo {year} {2013})}\BibitemShut {NoStop}%
\bibitem [{\citenamefont {Ding}\ \emph {et~al.}(2021)\citenamefont {Ding},
  \citenamefont {Shi}, \citenamefont {Zhang}, \citenamefont {Shen},
  \citenamefont {Zhang},\ and\ \citenamefont {Zhang}}]{Ding2021}%
  \BibitemOpen
  \bibfield  {author} {\bibinfo {author} {\bibfnamefont {L.}~\bibnamefont
  {Ding}}, \bibinfo {author} {\bibfnamefont {K.}~\bibnamefont {Shi}}, \bibinfo
  {author} {\bibfnamefont {Q.}~\bibnamefont {Zhang}}, \bibinfo {author}
  {\bibfnamefont {D.}~\bibnamefont {Shen}}, \bibinfo {author} {\bibfnamefont
  {X.}~\bibnamefont {Zhang}}, \ and\ \bibinfo {author} {\bibfnamefont
  {W.}~\bibnamefont {Zhang}},\ }\bibfield  {title} {\enquote {\bibinfo {title}
  {Experimental determination of $\mathcal{P}\mathcal{T}$-symmetric exceptional
  points in a single trapped ion},}\ }\href {\doibase
  10.1103/PhysRevLett.126.083604} {\bibfield  {journal} {\bibinfo  {journal}
  {Phys. Rev. Lett.}\ }\textbf {\bibinfo {volume} {126}},\ \bibinfo {pages}
  {083604} (\bibinfo {year} {2021})}\BibitemShut {NoStop}%
\bibitem [{\citenamefont {Heiss}\ and\ \citenamefont
  {Harney}(2001)}]{Heiss2001}%
  \BibitemOpen
  \bibfield  {author} {\bibinfo {author} {\bibfnamefont {W.D.}\ \bibnamefont
  {Heiss}}\ and\ \bibinfo {author} {\bibfnamefont {H.L.}\ \bibnamefont
  {Harney}},\ }\bibfield  {title} {\enquote {\bibinfo {title} {The chirality of
  exceptional points},}\ }\href {\doibase 10.1007/s100530170017} {\bibfield
  {journal} {\bibinfo  {journal} {Eur. Phys. J. D}\ }\textbf {\bibinfo {volume}
  {17}},\ \bibinfo {pages} {149} (\bibinfo {year} {2001})}\BibitemShut
  {NoStop}%
\bibitem [{\citenamefont {Dembowski}\ \emph {et~al.}(2003)\citenamefont
  {Dembowski}, \citenamefont {Dietz}, \citenamefont {Gr\"af}, \citenamefont
  {Harney}, \citenamefont {Heine}, \citenamefont {Heiss},\ and\ \citenamefont
  {Richter}}]{Dembowski2003}%
  \BibitemOpen
  \bibfield  {author} {\bibinfo {author} {\bibfnamefont {C.}~\bibnamefont
  {Dembowski}}, \bibinfo {author} {\bibfnamefont {B.}~\bibnamefont {Dietz}},
  \bibinfo {author} {\bibfnamefont {H.-D.}\ \bibnamefont {Gr\"af}}, \bibinfo
  {author} {\bibfnamefont {H.~L.}\ \bibnamefont {Harney}}, \bibinfo {author}
  {\bibfnamefont {A.}~\bibnamefont {Heine}}, \bibinfo {author} {\bibfnamefont
  {W.~D.}\ \bibnamefont {Heiss}}, \ and\ \bibinfo {author} {\bibfnamefont
  {A.}~\bibnamefont {Richter}},\ }\bibfield  {title} {\enquote {\bibinfo
  {title} {Observation of a chiral state in a microwave cavity},}\ }\href
  {\doibase 10.1103/PhysRevLett.90.034101} {\bibfield  {journal} {\bibinfo
  {journal} {Phys. Rev. Lett.}\ }\textbf {\bibinfo {volume} {90}},\ \bibinfo
  {pages} {034101} (\bibinfo {year} {2003})}\BibitemShut {NoStop}%
\bibitem [{\citenamefont {Lin}\ \emph {et~al.}(2011)\citenamefont {Lin},
  \citenamefont {Ramezani}, \citenamefont {Eichelkraut}, \citenamefont
  {Kottos}, \citenamefont {Cao},\ and\ \citenamefont
  {Christodoulides}}]{Lin2011}%
  \BibitemOpen
  \bibfield  {author} {\bibinfo {author} {\bibfnamefont {Z.}~\bibnamefont
  {Lin}}, \bibinfo {author} {\bibfnamefont {H.}~\bibnamefont {Ramezani}},
  \bibinfo {author} {\bibfnamefont {T.}~\bibnamefont {Eichelkraut}}, \bibinfo
  {author} {\bibfnamefont {T.}~\bibnamefont {Kottos}}, \bibinfo {author}
  {\bibfnamefont {H.}~\bibnamefont {Cao}}, \ and\ \bibinfo {author}
  {\bibfnamefont {D.~N.}\ \bibnamefont {Christodoulides}},\ }\bibfield  {title}
  {\enquote {\bibinfo {title} {Unidirectional invisibility induced by
  $\mathcal{PT}$-symmetric periodic structures},}\ }\href
  {https://doi.org/10.1103/PhysRevLett.106.213901} {\bibfield  {journal}
  {\bibinfo  {journal} {Phys. Rev. Lett.}\ }\textbf {\bibinfo {volume} {106}},\
  \bibinfo {pages} {213901} (\bibinfo {year} {2011})}\BibitemShut {NoStop}%
\bibitem [{\citenamefont {Regensburger}\ \emph {et~al.}(2012)\citenamefont
  {Regensburger}, \citenamefont {Bersch}, \citenamefont {Miri}, \citenamefont
  {Onishchukov}, \citenamefont {Christodoulides},\ and\ \citenamefont
  {Peschel}}]{Regensburger2012}%
  \BibitemOpen
  \bibfield  {author} {\bibinfo {author} {\bibfnamefont {A.}~\bibnamefont
  {Regensburger}}, \bibinfo {author} {\bibfnamefont {C.}~\bibnamefont
  {Bersch}}, \bibinfo {author} {\bibfnamefont {M.-A.}\ \bibnamefont {Miri}},
  \bibinfo {author} {\bibfnamefont {G.}~\bibnamefont {Onishchukov}}, \bibinfo
  {author} {\bibfnamefont {D.~N.}\ \bibnamefont {Christodoulides}}, \ and\
  \bibinfo {author} {\bibfnamefont {U.}~\bibnamefont {Peschel}},\ }\bibfield
  {title} {\enquote {\bibinfo {title} {Parity–time synthetic photonic
  lattices},}\ }\href {https://doi.org/10.1038/nature1129} {\bibfield
  {journal} {\bibinfo  {journal} {Nature}\ }\textbf {\bibinfo {volume} {488}},\
  \bibinfo {pages} {167} (\bibinfo {year} {2012})}\BibitemShut {NoStop}%
\bibitem [{\citenamefont {Feng}\ \emph {et~al.}(2014)\citenamefont {Feng},
  \citenamefont {Wong}, \citenamefont {Ma}, \citenamefont {Wang},\ and\
  \citenamefont {Zhang}}]{Feng2014}%
  \BibitemOpen
  \bibfield  {author} {\bibinfo {author} {\bibfnamefont {L.}~\bibnamefont
  {Feng}}, \bibinfo {author} {\bibfnamefont {Z.~J.}\ \bibnamefont {Wong}},
  \bibinfo {author} {\bibfnamefont {R.-M.}\ \bibnamefont {Ma}}, \bibinfo
  {author} {\bibfnamefont {Y.}~\bibnamefont {Wang}}, \ and\ \bibinfo {author}
  {\bibfnamefont {X.}~\bibnamefont {Zhang}},\ }\bibfield  {title} {\enquote
  {\bibinfo {title} {Single-mode laser by parity-time symmetry breaking},}\
  }\href {https://doi.org/10.1126/science.1258479} {\bibfield  {journal}
  {\bibinfo  {journal} {Science}\ }\textbf {\bibinfo {volume} {346}},\ \bibinfo
  {pages} {972} (\bibinfo {year} {2014})}\BibitemShut {NoStop}%
\bibitem [{\citenamefont {Hodaei}\ \emph {et~al.}(2014)\citenamefont {Hodaei},
  \citenamefont {Miri}, \citenamefont {Heinrich}, \citenamefont
  {Christodoulides},\ and\ \citenamefont {Khajavikhan}}]{Hodaei2014}%
  \BibitemOpen
  \bibfield  {author} {\bibinfo {author} {\bibfnamefont {H.}~\bibnamefont
  {Hodaei}}, \bibinfo {author} {\bibfnamefont {M.-A.}\ \bibnamefont {Miri}},
  \bibinfo {author} {\bibfnamefont {M.}~\bibnamefont {Heinrich}}, \bibinfo
  {author} {\bibfnamefont {D.~N.}\ \bibnamefont {Christodoulides}}, \ and\
  \bibinfo {author} {\bibfnamefont {M.}~\bibnamefont {Khajavikhan}},\
  }\bibfield  {title} {\enquote {\bibinfo {title} {Parity-time-symmetric
  microring lasers},}\ }\href {https://doi.org/10.1126/science.1258480}
  {\bibfield  {journal} {\bibinfo  {journal} {Science}\ }\textbf {\bibinfo
  {volume} {346}},\ \bibinfo {pages} {975} (\bibinfo {year}
  {2014})}\BibitemShut {NoStop}%
\bibitem [{\citenamefont {Peng}\ \emph
  {et~al.}(2014{\natexlab{a}})\citenamefont {Peng}, \citenamefont {\c{S}.
  K.~{\"O}zdemir}, \citenamefont {Lei}, \citenamefont {Monifi}, \citenamefont
  {Gianfreda}, \citenamefont {Long}, \citenamefont {Fan}, \citenamefont {Nori},
  \citenamefont {Bender},\ and\ \citenamefont {Yang}}]{Peng2014}%
  \BibitemOpen
  \bibfield  {author} {\bibinfo {author} {\bibfnamefont {B.}~\bibnamefont
  {Peng}}, \bibinfo {author} {\bibnamefont {\c{S}. K.~{\"O}zdemir}}, \bibinfo
  {author} {\bibfnamefont {F.}~\bibnamefont {Lei}}, \bibinfo {author}
  {\bibfnamefont {F.}~\bibnamefont {Monifi}}, \bibinfo {author} {\bibfnamefont
  {M.}~\bibnamefont {Gianfreda}}, \bibinfo {author} {\bibfnamefont {G.~L.}\
  \bibnamefont {Long}}, \bibinfo {author} {\bibfnamefont {S.}~\bibnamefont
  {Fan}}, \bibinfo {author} {\bibfnamefont {F.}~\bibnamefont {Nori}}, \bibinfo
  {author} {\bibfnamefont {C.}~\bibnamefont {Bender}}, \ and\ \bibinfo {author}
  {\bibfnamefont {L.}~\bibnamefont {Yang}},\ }\bibfield  {title} {\enquote
  {\bibinfo {title} {Parity-time-symmetric whispering-gallery microcavities},}\
  }\href {https://doi.org/10.1038/nphys2927} {\bibfield  {journal} {\bibinfo
  {journal} {Nat. Phys.}\ }\textbf {\bibinfo {volume} {10}},\ \bibinfo {pages}
  {394} (\bibinfo {year} {2014}{\natexlab{a}})}\BibitemShut {NoStop}%
\bibitem [{\citenamefont {Chang}\ \emph {et~al.}(2014)\citenamefont {Chang},
  \citenamefont {Jiang}, \citenamefont {Hua}, \citenamefont {Yang},
  \citenamefont {Wen}, \citenamefont {Jiang}, \citenamefont {Li}, \citenamefont
  {Wang},\ and\ \citenamefont {Xiao}}]{Chang2014}%
  \BibitemOpen
  \bibfield  {author} {\bibinfo {author} {\bibfnamefont {L.}~\bibnamefont
  {Chang}}, \bibinfo {author} {\bibfnamefont {X.}~\bibnamefont {Jiang}},
  \bibinfo {author} {\bibfnamefont {S.}~\bibnamefont {Hua}}, \bibinfo {author}
  {\bibfnamefont {C.}~\bibnamefont {Yang}}, \bibinfo {author} {\bibfnamefont
  {J.}~\bibnamefont {Wen}}, \bibinfo {author} {\bibfnamefont {L.}~\bibnamefont
  {Jiang}}, \bibinfo {author} {\bibfnamefont {G.}~\bibnamefont {Li}}, \bibinfo
  {author} {\bibfnamefont {G.}~\bibnamefont {Wang}}, \ and\ \bibinfo {author}
  {\bibfnamefont {M.}~\bibnamefont {Xiao}},\ }\bibfield  {title} {\enquote
  {\bibinfo {title} {Parity-time symmetry and variable optical isolation in
  active-passive-coupled microresonators},}\ }\href
  {https://doi.org/10.1038/nphoton.2014.133} {\bibfield  {journal} {\bibinfo
  {journal} {Nat. Photon.}\ }\textbf {\bibinfo {volume} {8}},\ \bibinfo {pages}
  {524} (\bibinfo {year} {2014})}\BibitemShut {NoStop}%
\bibitem [{\citenamefont {Arkhipov}\ \emph {et~al.}(2019)\citenamefont
  {Arkhipov}, \citenamefont {Miranowicz}, \citenamefont {Di~Stefano},
  \citenamefont {Stassi}, \citenamefont {Savasta}, \citenamefont {Nori},\ and\
  \citenamefont {\"Ozdemir}}]{Arkhipov2019b}%
  \BibitemOpen
  \bibfield  {author} {\bibinfo {author} {\bibfnamefont {I.~I.}\ \bibnamefont
  {Arkhipov}}, \bibinfo {author} {\bibfnamefont {A.}~\bibnamefont
  {Miranowicz}}, \bibinfo {author} {\bibfnamefont {O.}~\bibnamefont
  {Di~Stefano}}, \bibinfo {author} {\bibfnamefont {R.}~\bibnamefont {Stassi}},
  \bibinfo {author} {\bibfnamefont {S.}~\bibnamefont {Savasta}}, \bibinfo
  {author} {\bibfnamefont {F.}~\bibnamefont {Nori}}, \ and\ \bibinfo {author}
  {\bibfnamefont {\c{S}.~K.}\ \bibnamefont {\"Ozdemir}},\ }\bibfield  {title}
  {\enquote {\bibinfo {title} {Scully-{L}amb quantum laser model for
  parity-time-symmetric whispering-gallery microcavities: {G}ain saturation
  effects and nonreciprocity},}\ }\href {\doibase 10.1103/PhysRevA.99.053806}
  {\bibfield  {journal} {\bibinfo  {journal} {Phys. Rev. A}\ }\textbf {\bibinfo
  {volume} {99}},\ \bibinfo {pages} {053806} (\bibinfo {year}
  {2019})}\BibitemShut {NoStop}%
\bibitem [{\citenamefont {Huang}\ \emph {et~al.}(2020)\citenamefont {Huang},
  \citenamefont {\c{S}. K.~{\"O}zdemir}, \citenamefont {Liao}, \citenamefont
  {Minganti}, \citenamefont {Kuang}, \citenamefont {Nori},\ and\ \citenamefont
  {Jing}}]{Huang2020}%
  \BibitemOpen
  \bibfield  {author} {\bibinfo {author} {\bibfnamefont {R.}~\bibnamefont
  {Huang}}, \bibinfo {author} {\bibnamefont {\c{S}. K.~{\"O}zdemir}}, \bibinfo
  {author} {\bibfnamefont {J.~Q.}\ \bibnamefont {Liao}}, \bibinfo {author}
  {\bibfnamefont {F.}~\bibnamefont {Minganti}}, \bibinfo {author}
  {\bibfnamefont {L.~M.}\ \bibnamefont {Kuang}}, \bibinfo {author}
  {\bibfnamefont {F.}~\bibnamefont {Nori}}, \ and\ \bibinfo {author}
  {\bibfnamefont {H.}~\bibnamefont {Jing}},\ }\href@noop {} {\enquote {\bibinfo
  {title} {Exceptional photon blockade},}\ } (\bibinfo {year} {2020}),\ \Eprint
  {http://arxiv.org/abs/2001.09492} {arXiv:2001.09492} \BibitemShut {NoStop}%
\bibitem [{\citenamefont {Brandstetter}\ \emph {et~al.}(2014)\citenamefont
  {Brandstetter}, \citenamefont {Liertzer}, \citenamefont {Deutsch},
  \citenamefont {Klang}, \citenamefont {Schoberl}, \citenamefont {Tureci},
  \citenamefont {Strasser}, \citenamefont {Unterrainer},\ and\ \citenamefont
  {Rotter}}]{Brands2014a}%
  \BibitemOpen
  \bibfield  {author} {\bibinfo {author} {\bibfnamefont {M.}~\bibnamefont
  {Brandstetter}}, \bibinfo {author} {\bibfnamefont {M.}~\bibnamefont
  {Liertzer}}, \bibinfo {author} {\bibfnamefont {C.}~\bibnamefont {Deutsch}},
  \bibinfo {author} {\bibfnamefont {P.}~\bibnamefont {Klang}}, \bibinfo
  {author} {\bibfnamefont {J.}~\bibnamefont {Schoberl}}, \bibinfo {author}
  {\bibfnamefont {H.~E.}\ \bibnamefont {Tureci}}, \bibinfo {author}
  {\bibfnamefont {G.}~\bibnamefont {Strasser}}, \bibinfo {author}
  {\bibfnamefont {K.}~\bibnamefont {Unterrainer}}, \ and\ \bibinfo {author}
  {\bibfnamefont {S.}~\bibnamefont {Rotter}},\ }\bibfield  {title} {\enquote
  {\bibinfo {title} {Reversing the pump dependence of a laser at an exceptional
  point},}\ }\href {https://doi.org/10.1038/ncomms5034} {\bibfield  {journal}
  {\bibinfo  {journal} {Nat. Commun.}\ }\textbf {\bibinfo {volume} {5}},\
  \bibinfo {pages} {4034} (\bibinfo {year} {2014})}\BibitemShut {NoStop}%
\bibitem [{\citenamefont {Peng}\ \emph
  {et~al.}(2014{\natexlab{b}})\citenamefont {Peng}, \citenamefont {\c{S}.
  K.~{\"O}zdemir}, \citenamefont {Rotter}, \citenamefont {Yilmaz},
  \citenamefont {Liertzer}, \citenamefont {Monifi}, \citenamefont {Bender},
  \citenamefont {Nori},\ and\ \citenamefont {Yang}}]{Peng2014a}%
  \BibitemOpen
  \bibfield  {author} {\bibinfo {author} {\bibfnamefont {B.}~\bibnamefont
  {Peng}}, \bibinfo {author} {\bibnamefont {\c{S}. K.~{\"O}zdemir}}, \bibinfo
  {author} {\bibfnamefont {S.}~\bibnamefont {Rotter}}, \bibinfo {author}
  {\bibfnamefont {H.}~\bibnamefont {Yilmaz}}, \bibinfo {author} {\bibfnamefont
  {M.}~\bibnamefont {Liertzer}}, \bibinfo {author} {\bibfnamefont
  {F.}~\bibnamefont {Monifi}}, \bibinfo {author} {\bibfnamefont {C.~M.}\
  \bibnamefont {Bender}}, \bibinfo {author} {\bibfnamefont {F.}~\bibnamefont
  {Nori}}, \ and\ \bibinfo {author} {\bibfnamefont {L.}~\bibnamefont {Yang}},\
  }\bibfield  {title} {\enquote {\bibinfo {title} {Loss-induced suppression and
  revival of lasing},}\ }\href {https://doi.org/10.1126/science.1258004}
  {\bibfield  {journal} {\bibinfo  {journal} {Science}\ }\textbf {\bibinfo
  {volume} {346}},\ \bibinfo {pages} {328} (\bibinfo {year}
  {2014}{\natexlab{b}})}\BibitemShut {NoStop}%
\bibitem [{\citenamefont {Pe\ifmmode~\check{r}\else \v{r}\fi{}ina}\ \emph
  {et~al.}(2019)\citenamefont {Pe\ifmmode~\check{r}\else \v{r}\fi{}ina},
  \citenamefont {Luk\ifmmode~\check{s}\else \v{s}\fi{}}, \citenamefont
  {Kalaga}, \citenamefont {Leo\ifmmode~\acute{n}\else \'{n}\fi{}ski},\ and\
  \citenamefont {Miranowicz}}]{Perina2019b}%
  \BibitemOpen
  \bibfield  {author} {\bibinfo {author} {\bibfnamefont {J.}~\bibnamefont
  {Pe\ifmmode~\check{r}\else \v{r}\fi{}ina}}, \bibinfo {author} {\bibfnamefont
  {A.}~\bibnamefont {Luk\ifmmode~\check{s}\else \v{s}\fi{}}}, \bibinfo {author}
  {\bibfnamefont {J.~K.}\ \bibnamefont {Kalaga}}, \bibinfo {author}
  {\bibfnamefont {W.}~\bibnamefont {Leo\ifmmode~\acute{n}\else \'{n}\fi{}ski}},
  \ and\ \bibinfo {author} {\bibfnamefont {A.}~\bibnamefont {Miranowicz}},\
  }\bibfield  {title} {\enquote {\bibinfo {title} {Nonclassical light at
  exceptional points of a quantum $\mathcal{PT}$-symmetric two-mode system},}\
  }\href {\doibase 10.1103/PhysRevA.100.053820} {\bibfield  {journal} {\bibinfo
   {journal} {Phys. Rev. A}\ }\textbf {\bibinfo {volume} {100}},\ \bibinfo
  {pages} {053820} (\bibinfo {year} {2019})}\BibitemShut {NoStop}%
\bibitem [{\citenamefont {Lange}\ \emph {et~al.}(2020)\citenamefont {Lange},
  \citenamefont {Chimczak}, \citenamefont {Kowalewska-Kudlaszyk},\ and\
  \citenamefont {Bartkiewicz}}]{Lange2020}%
  \BibitemOpen
  \bibfield  {author} {\bibinfo {author} {\bibfnamefont {E.}~\bibnamefont
  {Lange}}, \bibinfo {author} {\bibfnamefont {G.}~\bibnamefont {Chimczak}},
  \bibinfo {author} {\bibfnamefont {A.}~\bibnamefont {Kowalewska-Kudlaszyk}}, \
  and\ \bibinfo {author} {\bibfnamefont {K.}~\bibnamefont {Bartkiewicz}},\
  }\bibfield  {title} {\enquote {\bibinfo {title} {Rotation-time symmetry in
  bosonic systems and the existence of exceptional points in the absence of
  $\mathcal{PT}$ symmetry},}\ }\href
  {https://doi.org/10.1038/s41598-020-76787-8} {\bibfield  {journal} {\bibinfo
  {journal} {Sci. Rep.}\ }\textbf {\bibinfo {volume} {10}},\ \bibinfo {pages}
  {19906} (\bibinfo {year} {2020})}\BibitemShut {NoStop}%
\bibitem [{\citenamefont {Kuo}\ \emph {et~al.}(2020)\citenamefont {Kuo},
  \citenamefont {Lambert}, \citenamefont {Miranowicz}, \citenamefont {Chen},
  \citenamefont {Chen}, \citenamefont {Chen},\ and\ \citenamefont
  {Nori}}]{Kuo2020}%
  \BibitemOpen
  \bibfield  {author} {\bibinfo {author} {\bibfnamefont {P.-C.}\ \bibnamefont
  {Kuo}}, \bibinfo {author} {\bibfnamefont {N.}~\bibnamefont {Lambert}},
  \bibinfo {author} {\bibfnamefont {A.}~\bibnamefont {Miranowicz}}, \bibinfo
  {author} {\bibfnamefont {H.-B.}\ \bibnamefont {Chen}}, \bibinfo {author}
  {\bibfnamefont {G.-Y.}\ \bibnamefont {Chen}}, \bibinfo {author}
  {\bibfnamefont {Y.-N.}\ \bibnamefont {Chen}}, \ and\ \bibinfo {author}
  {\bibfnamefont {F.}~\bibnamefont {Nori}},\ }\bibfield  {title} {\enquote
  {\bibinfo {title} {Collectively induced exceptional points of quantum
  emitters coupled to nanoparticle surface plasmons},}\ }\href {\doibase
  10.1103/PhysRevA.101.013814} {\bibfield  {journal} {\bibinfo  {journal}
  {Phys. Rev. A}\ }\textbf {\bibinfo {volume} {101}},\ \bibinfo {pages}
  {013814} (\bibinfo {year} {2020})}\BibitemShut {NoStop}%
\bibitem [{\citenamefont {Wiersig}(2014)}]{Wiersig2014}%
  \BibitemOpen
  \bibfield  {author} {\bibinfo {author} {\bibfnamefont {J.}~\bibnamefont
  {Wiersig}},\ }\bibfield  {title} {\enquote {\bibinfo {title} {Enhancing the
  sensitivity of frequency and energy splitting detection by using exceptional
  points: Application to microcavity sensors for single-particle detection},}\
  }\href {https://link.aps.org/doi/10.1103/PhysRevLett.112.203901} {\bibfield
  {journal} {\bibinfo  {journal} {Phys. Rev. Lett.}\ }\textbf {\bibinfo
  {volume} {112}},\ \bibinfo {pages} {203901} (\bibinfo {year}
  {2014})}\BibitemShut {NoStop}%
\bibitem [{\citenamefont {Zhang}\ \emph {et~al.}(2015)\citenamefont {Zhang},
  \citenamefont {Liu}, \citenamefont {Wang}, \citenamefont {Gu}, \citenamefont
  {Li}, \citenamefont {Yi}, \citenamefont {Xiao},\ and\ \citenamefont
  {Song}}]{Zhang2015}%
  \BibitemOpen
  \bibfield  {author} {\bibinfo {author} {\bibfnamefont {N.}~\bibnamefont
  {Zhang}}, \bibinfo {author} {\bibfnamefont {S.}~\bibnamefont {Liu}}, \bibinfo
  {author} {\bibfnamefont {K.}~\bibnamefont {Wang}}, \bibinfo {author}
  {\bibfnamefont {Z.}~\bibnamefont {Gu}}, \bibinfo {author} {\bibfnamefont
  {M.}~\bibnamefont {Li}}, \bibinfo {author} {\bibfnamefont {N.}~\bibnamefont
  {Yi}}, \bibinfo {author} {\bibfnamefont {S.}~\bibnamefont {Xiao}}, \ and\
  \bibinfo {author} {\bibfnamefont {Q.}~\bibnamefont {Song}},\ }\bibfield
  {title} {\enquote {\bibinfo {title} {Single nanoparticle detection using
  far-field emission of photonic molecule around the exceptional point},}\
  }\href {https://doi.org/10.1038/srep11912} {\bibfield  {journal} {\bibinfo
  {journal} {Sci. Rep.}\ }\textbf {\bibinfo {volume} {5}},\ \bibinfo {pages}
  {11912} (\bibinfo {year} {2015})}\BibitemShut {NoStop}%
\bibitem [{\citenamefont {Wiersig}(2016)}]{Wiersig2016}%
  \BibitemOpen
  \bibfield  {author} {\bibinfo {author} {\bibfnamefont {J.}~\bibnamefont
  {Wiersig}},\ }\bibfield  {title} {\enquote {\bibinfo {title} {Sensors
  operating at exceptional points: General theory},}\ }\href
  {https://link.aps.org/doi/10.1103/PhysRevA.93.033809} {\bibfield  {journal}
  {\bibinfo  {journal} {Phys. Rev. A}\ }\textbf {\bibinfo {volume} {93}},\
  \bibinfo {pages} {033809} (\bibinfo {year} {2016})}\BibitemShut {NoStop}%
\bibitem [{\citenamefont {Ren}\ \emph {et~al.}(2017)\citenamefont {Ren},
  \citenamefont {Hodaei}, \citenamefont {Harari}, \citenamefont {Hassan},
  \citenamefont {Chow}, \citenamefont {Soltani}, \citenamefont
  {Christodoulides},\ and\ \citenamefont {Khajavikhan}}]{Ren2017}%
  \BibitemOpen
  \bibfield  {author} {\bibinfo {author} {\bibfnamefont {J.}~\bibnamefont
  {Ren}}, \bibinfo {author} {\bibfnamefont {H.}~\bibnamefont {Hodaei}},
  \bibinfo {author} {\bibfnamefont {G.}~\bibnamefont {Harari}}, \bibinfo
  {author} {\bibfnamefont {A.~U.}\ \bibnamefont {Hassan}}, \bibinfo {author}
  {\bibfnamefont {W.}~\bibnamefont {Chow}}, \bibinfo {author} {\bibfnamefont
  {M.}~\bibnamefont {Soltani}}, \bibinfo {author} {\bibfnamefont
  {D.}~\bibnamefont {Christodoulides}}, \ and\ \bibinfo {author} {\bibfnamefont
  {M.}~\bibnamefont {Khajavikhan}},\ }\bibfield  {title} {\enquote {\bibinfo
  {title} {Ultrasensitive micro-scale parity-time-symmetric ring laser
  gyroscope},}\ }\href {https://doi.org/10.1364/ol.42.001556} {\bibfield
  {journal} {\bibinfo  {journal} {Opt. Lett.}\ }\textbf {\bibinfo {volume}
  {42}},\ \bibinfo {pages} {1556} (\bibinfo {year} {2017})}\BibitemShut
  {NoStop}%
\bibitem [{\citenamefont {Chen}\ \emph {et~al.}(2017)\citenamefont {Chen},
  \citenamefont {\"{O}zdemir}, \citenamefont {Zhao}, \citenamefont {Wiersig},\
  and\ \citenamefont {Yang}}]{Chen2017}%
  \BibitemOpen
  \bibfield  {author} {\bibinfo {author} {\bibfnamefont {W.}~\bibnamefont
  {Chen}}, \bibinfo {author} {\bibfnamefont {{\c{S}}.~Kaya}\ \bibnamefont
  {\"{O}zdemir}}, \bibinfo {author} {\bibfnamefont {G.}~\bibnamefont {Zhao}},
  \bibinfo {author} {\bibfnamefont {J.}~\bibnamefont {Wiersig}}, \ and\
  \bibinfo {author} {\bibfnamefont {L.}~\bibnamefont {Yang}},\ }\bibfield
  {title} {\enquote {\bibinfo {title} {Exceptional points enhance sensing in an
  optical microcavity},}\ }\href {https://doi.org/10.1038/nature23281}
  {\bibfield  {journal} {\bibinfo  {journal} {Nature (London)}\ }\textbf
  {\bibinfo {volume} {548}},\ \bibinfo {pages} {192} (\bibinfo {year}
  {2017})}\BibitemShut {NoStop}%
\bibitem [{\citenamefont {Hodaei}\ \emph {et~al.}(2017)\citenamefont {Hodaei},
  \citenamefont {Hassan}, \citenamefont {Wittek}, \citenamefont
  {Garcia-Gracia}, \citenamefont {El-Ganainy}, \citenamefont
  {Christodoulides},\ and\ \citenamefont {Khajavikhan}}]{Hodaei2017}%
  \BibitemOpen
  \bibfield  {author} {\bibinfo {author} {\bibfnamefont {H.}~\bibnamefont
  {Hodaei}}, \bibinfo {author} {\bibfnamefont {A.~U.}\ \bibnamefont {Hassan}},
  \bibinfo {author} {\bibfnamefont {S.}~\bibnamefont {Wittek}}, \bibinfo
  {author} {\bibfnamefont {H.}~\bibnamefont {Garcia-Gracia}}, \bibinfo {author}
  {\bibfnamefont {R.}~\bibnamefont {El-Ganainy}}, \bibinfo {author}
  {\bibfnamefont {D.~N.}\ \bibnamefont {Christodoulides}}, \ and\ \bibinfo
  {author} {\bibfnamefont {M.}~\bibnamefont {Khajavikhan}},\ }\bibfield
  {title} {\enquote {\bibinfo {title} {Enhanced sensitivity at higher-order
  exceptional points},}\ }\href {https://doi.org/10.1038/nature23280}
  {\bibfield  {journal} {\bibinfo  {journal} {Nature (London)}\ }\textbf
  {\bibinfo {volume} {548}},\ \bibinfo {pages} {187} (\bibinfo {year}
  {2017})}\BibitemShut {NoStop}%
\bibitem [{\citenamefont {Chen}\ \emph {et~al.}(2018)\citenamefont {Chen},
  \citenamefont {Sakhdari}, \citenamefont {Hajizadegan}, \citenamefont {Cui},
  \citenamefont {Cheng}, \citenamefont {El-Ganainy},\ and\ \citenamefont
  {Al{\`{u}}}}]{ChenNat2018}%
  \BibitemOpen
  \bibfield  {author} {\bibinfo {author} {\bibfnamefont {P.-Y.}\ \bibnamefont
  {Chen}}, \bibinfo {author} {\bibfnamefont {M.}~\bibnamefont {Sakhdari}},
  \bibinfo {author} {\bibfnamefont {M.}~\bibnamefont {Hajizadegan}}, \bibinfo
  {author} {\bibfnamefont {Q.}~\bibnamefont {Cui}}, \bibinfo {author}
  {\bibfnamefont {M.~M.-C.}\ \bibnamefont {Cheng}}, \bibinfo {author}
  {\bibfnamefont {R.}~\bibnamefont {El-Ganainy}}, \ and\ \bibinfo {author}
  {\bibfnamefont {A.}~\bibnamefont {Al{\`{u}}}},\ }\bibfield  {title} {\enquote
  {\bibinfo {title} {Generalized parity{\textendash}time symmetry condition for
  enhanced sensor telemetry},}\ }\href
  {https://doi.org/10.1038/s41928-018-0072-6} {\bibfield  {journal} {\bibinfo
  {journal} {Nat. Electron.}\ }\textbf {\bibinfo {volume} {1}},\ \bibinfo
  {pages} {297} (\bibinfo {year} {2018})}\BibitemShut {NoStop}%
\bibitem [{\citenamefont {Liu}\ \emph {et~al.}(2016)\citenamefont {Liu},
  \citenamefont {Zhang}, \citenamefont {\c{S}. K.~{\"O}zdemir}, \citenamefont
  {Peng}, \citenamefont {Jing}, \citenamefont {L{\"u}}, \citenamefont {Li},
  \citenamefont {Yang}, \citenamefont {Nori},\ and\ \citenamefont
  {Liu}}]{Liu2016}%
  \BibitemOpen
  \bibfield  {author} {\bibinfo {author} {\bibfnamefont {Z.-P.}\ \bibnamefont
  {Liu}}, \bibinfo {author} {\bibfnamefont {J.}~\bibnamefont {Zhang}}, \bibinfo
  {author} {\bibnamefont {\c{S}. K.~{\"O}zdemir}}, \bibinfo {author}
  {\bibfnamefont {B.}~\bibnamefont {Peng}}, \bibinfo {author} {\bibfnamefont
  {H.}~\bibnamefont {Jing}}, \bibinfo {author} {\bibfnamefont {X.-Y.}\
  \bibnamefont {L{\"u}}}, \bibinfo {author} {\bibfnamefont {C.-W.}\
  \bibnamefont {Li}}, \bibinfo {author} {\bibfnamefont {L.}~\bibnamefont
  {Yang}}, \bibinfo {author} {\bibfnamefont {F.}~\bibnamefont {Nori}}, \ and\
  \bibinfo {author} {\bibfnamefont {Y.-X.}\ \bibnamefont {Liu}},\ }\bibfield
  {title} {\enquote {\bibinfo {title} {Metrology with $\mathcal{PT}$-symmetric
  cavities: {E}nhanced sensitivity near the $\mathcal{PT}$-phase transition},}\
  }\href {https://doi.org/10.1103/PhysRevLett.117.110802} {\bibfield  {journal}
  {\bibinfo  {journal} {Phys. Rev. Lett.}\ }\textbf {\bibinfo {volume} {117}},\
  \bibinfo {pages} {110802} (\bibinfo {year} {2016})}\BibitemShut {NoStop}%
\bibitem [{\citenamefont {Mortensen}\ \emph {et~al.}(2018)\citenamefont
  {Mortensen}, \citenamefont {Gon{\c{c}}alves}, \citenamefont {Khajavikhan},
  \citenamefont {Christodoulides}, \citenamefont {Tserkezis},\ and\
  \citenamefont {Wolff}}]{Mortensen2018}%
  \BibitemOpen
  \bibfield  {author} {\bibinfo {author} {\bibfnamefont {N.~A.}\ \bibnamefont
  {Mortensen}}, \bibinfo {author} {\bibfnamefont {P.~A.~D.}\ \bibnamefont
  {Gon{\c{c}}alves}}, \bibinfo {author} {\bibfnamefont {M.}~\bibnamefont
  {Khajavikhan}}, \bibinfo {author} {\bibfnamefont {D.~N.}\ \bibnamefont
  {Christodoulides}}, \bibinfo {author} {\bibfnamefont {C.}~\bibnamefont
  {Tserkezis}}, \ and\ \bibinfo {author} {\bibfnamefont {C.}~\bibnamefont
  {Wolff}},\ }\bibfield  {title} {\enquote {\bibinfo {title} {Fluctuations and
  noise-limited sensing near the exceptional point of parity-time-symmetric
  resonator systems},}\ }\href {https://doi.org/10.1364/optica.5.001342}
  {\bibfield  {journal} {\bibinfo  {journal} {Optica}\ }\textbf {\bibinfo
  {volume} {5}},\ \bibinfo {pages} {1342} (\bibinfo {year} {2018})}\BibitemShut
  {NoStop}%
\bibitem [{\citenamefont {Wiersig}(2020{\natexlab{a}})}]{Wiersig2020}%
  \BibitemOpen
  \bibfield  {author} {\bibinfo {author} {\bibfnamefont {J.}~\bibnamefont
  {Wiersig}},\ }\bibfield  {title} {\enquote {\bibinfo {title} {Robustness of
  exceptional point-based sensors against parametric noise: {T}he role of
  {H}amiltonian and {L}iouvillian degeneracies},}\ }\href
  {https://link.aps.org/doi/10.1103/PhysRevA.101.053846} {\bibfield  {journal}
  {\bibinfo  {journal} {Phys. Rev. A}\ }\textbf {\bibinfo {volume} {101}},\
  \bibinfo {pages} {053846} (\bibinfo {year} {2020}{\natexlab{a}})}\BibitemShut
  {NoStop}%
\bibitem [{\citenamefont {Wiersig}(2020{\natexlab{b}})}]{Wiersig2020b}%
  \BibitemOpen
  \bibfield  {author} {\bibinfo {author} {\bibfnamefont {J.}~\bibnamefont
  {Wiersig}},\ }\bibfield  {title} {\enquote {\bibinfo {title} {Prospects and
  fundamental limits in exceptional point-based sensing},}\ }\href
  {https://doi.org/10.1038/s41467-020-16373-8} {\bibfield  {journal} {\bibinfo
  {journal} {Nat. Commun.}\ }\textbf {\bibinfo {volume} {11}},\ \bibinfo
  {pages} {2454} (\bibinfo {year} {2020}{\natexlab{b}})}\BibitemShut {NoStop}%
\bibitem [{\citenamefont {Wang}\ \emph {et~al.}(2020)\citenamefont {Wang},
  \citenamefont {Lai}, \citenamefont {Yuan}, \citenamefont {Suh},\ and\
  \citenamefont {Vahala}}]{Wang2020}%
  \BibitemOpen
  \bibfield  {author} {\bibinfo {author} {\bibfnamefont {H.}~\bibnamefont
  {Wang}}, \bibinfo {author} {\bibfnamefont {Y.-H.}\ \bibnamefont {Lai}},
  \bibinfo {author} {\bibfnamefont {Z.}~\bibnamefont {Yuan}}, \bibinfo {author}
  {\bibfnamefont {M.-G.}\ \bibnamefont {Suh}}, \ and\ \bibinfo {author}
  {\bibfnamefont {K.}~\bibnamefont {Vahala}},\ }\bibfield  {title} {\enquote
  {\bibinfo {title} {Petermann-factor sensitivity limit near an exceptional
  point in a {B}rillouin ring laser gyroscope},}\ }\href
  {https://doi.org/10.1038/s41467-020-15341-6} {\bibfield  {journal} {\bibinfo
  {journal} {Nat. Commun.}\ }\textbf {\bibinfo {volume} {11}},\ \bibinfo
  {pages} {1610} (\bibinfo {year} {2020})}\BibitemShut {NoStop}%
\bibitem [{\citenamefont {Chen}\ \emph {et~al.}(2019)\citenamefont {Chen},
  \citenamefont {Jin},\ and\ \citenamefont {Liu}}]{Chen2019}%
  \BibitemOpen
  \bibfield  {author} {\bibinfo {author} {\bibfnamefont {C.}~\bibnamefont
  {Chen}}, \bibinfo {author} {\bibfnamefont {L.}~\bibnamefont {Jin}}, \ and\
  \bibinfo {author} {\bibfnamefont {R.-B.}\ \bibnamefont {Liu}},\ }\bibfield
  {title} {\enquote {\bibinfo {title} {Sensitivity of parameter estimation near
  the exceptional point of a non-{H}ermitian system},}\ }\href
  {https://doi.org/10.1088%2F1367-2630%2Fab32ab} {\bibfield  {journal}
  {\bibinfo  {journal} {New J. Phys.}\ }\textbf {\bibinfo {volume} {21}},\
  \bibinfo {pages} {083002} (\bibinfo {year} {2019})}\BibitemShut {NoStop}%
\bibitem [{\citenamefont {Lau}\ and\ \citenamefont {Clerk}(2018)}]{Lau2018}%
  \BibitemOpen
  \bibfield  {author} {\bibinfo {author} {\bibfnamefont {H.-K.}\ \bibnamefont
  {Lau}}\ and\ \bibinfo {author} {\bibfnamefont {A.~A.}\ \bibnamefont
  {Clerk}},\ }\bibfield  {title} {\enquote {\bibinfo {title} {Fundamental
  limits and non-reciprocal approaches in non-{H}ermitian quantum sensing},}\
  }\href {https://doi.org/10.1038/s41467-018-06477-7} {\bibfield  {journal}
  {\bibinfo  {journal} {Nat. Commun.}\ }\textbf {\bibinfo {volume} {9}},\
  \bibinfo {pages} {4320} (\bibinfo {year} {2018})}\BibitemShut {NoStop}%
\bibitem [{\citenamefont {Zhang}\ \emph {et~al.}(2019)\citenamefont {Zhang},
  \citenamefont {Sweeney}, \citenamefont {Hsu}, \citenamefont {Yang},
  \citenamefont {Stone},\ and\ \citenamefont {Jiang}}]{Zhang2019}%
  \BibitemOpen
  \bibfield  {author} {\bibinfo {author} {\bibfnamefont {M.}~\bibnamefont
  {Zhang}}, \bibinfo {author} {\bibfnamefont {W.}~\bibnamefont {Sweeney}},
  \bibinfo {author} {\bibfnamefont {C.~W.}\ \bibnamefont {Hsu}}, \bibinfo
  {author} {\bibfnamefont {L.}~\bibnamefont {Yang}}, \bibinfo {author}
  {\bibfnamefont {A.~D.}\ \bibnamefont {Stone}}, \ and\ \bibinfo {author}
  {\bibfnamefont {L.}~\bibnamefont {Jiang}},\ }\bibfield  {title} {\enquote
  {\bibinfo {title} {Quantum noise theory of exceptional point amplifying
  sensors},}\ }\href {\doibase 10.1103/PhysRevLett.123.180501} {\bibfield
  {journal} {\bibinfo  {journal} {Phys. Rev. Lett.}\ }\textbf {\bibinfo
  {volume} {123}},\ \bibinfo {pages} {180501} (\bibinfo {year}
  {2019})}\BibitemShut {NoStop}%
\bibitem [{\citenamefont {Langbein}(2018)}]{Langbein2018}%
  \BibitemOpen
  \bibfield  {author} {\bibinfo {author} {\bibfnamefont {W.}~\bibnamefont
  {Langbein}},\ }\bibfield  {title} {\enquote {\bibinfo {title} {No exceptional
  precision of exceptional-point sensors},}\ }\href
  {https://link.aps.org/doi/10.1103/PhysRevA.98.023805} {\bibfield  {journal}
  {\bibinfo  {journal} {Phys. Rev. A}\ }\textbf {\bibinfo {volume} {98}},\
  \bibinfo {pages} {023805} (\bibinfo {year} {2018})}\BibitemShut {NoStop}%
\bibitem [{\citenamefont {Yu}\ \emph {et~al.}(2020)\citenamefont {Yu},
  \citenamefont {Meng}, \citenamefont {Tang}, \citenamefont {Xu}, \citenamefont
  {Wang}, \citenamefont {Yin}, \citenamefont {Ke}, \citenamefont {Liu},
  \citenamefont {Li}, \citenamefont {Yang}, \citenamefont {Chen}, \citenamefont
  {Han}, \citenamefont {Li},\ and\ \citenamefont {Guo}}]{Yu2020}%
  \BibitemOpen
  \bibfield  {author} {\bibinfo {author} {\bibfnamefont {S.}~\bibnamefont
  {Yu}}, \bibinfo {author} {\bibfnamefont {Y.}~\bibnamefont {Meng}}, \bibinfo
  {author} {\bibfnamefont {J.-S.}\ \bibnamefont {Tang}}, \bibinfo {author}
  {\bibfnamefont {X.-Y.}\ \bibnamefont {Xu}}, \bibinfo {author} {\bibfnamefont
  {Y.-T.}\ \bibnamefont {Wang}}, \bibinfo {author} {\bibfnamefont
  {P.}~\bibnamefont {Yin}}, \bibinfo {author} {\bibfnamefont {Z.-J.}\
  \bibnamefont {Ke}}, \bibinfo {author} {\bibfnamefont {W.}~\bibnamefont
  {Liu}}, \bibinfo {author} {\bibfnamefont {Z.-P.}\ \bibnamefont {Li}},
  \bibinfo {author} {\bibfnamefont {Y.-Z.}\ \bibnamefont {Yang}}, \bibinfo
  {author} {\bibfnamefont {G.}~\bibnamefont {Chen}}, \bibinfo {author}
  {\bibfnamefont {Y.-J.}\ \bibnamefont {Han}}, \bibinfo {author} {\bibfnamefont
  {C.-F.}\ \bibnamefont {Li}}, \ and\ \bibinfo {author} {\bibfnamefont {G.-C.}\
  \bibnamefont {Guo}},\ }\bibfield  {title} {\enquote {\bibinfo {title}
  {Experimental investigation of quantum $\mathcal{P}\mathcal{T}$-enhanced
  sensor},}\ }\href {\doibase 10.1103/PhysRevLett.125.240506} {\bibfield
  {journal} {\bibinfo  {journal} {Phys. Rev. Lett.}\ }\textbf {\bibinfo
  {volume} {125}},\ \bibinfo {pages} {240506} (\bibinfo {year}
  {2020})}\BibitemShut {NoStop}%
\bibitem [{\citenamefont {Teimourpour}\ \emph {et~al.}(2014)\citenamefont
  {Teimourpour}, \citenamefont {El-Ganainy}, \citenamefont {Eisfeld},
  \citenamefont {Szameit},\ and\ \citenamefont
  {Christodoulides}}]{Teimourpour2014}%
  \BibitemOpen
  \bibfield  {author} {\bibinfo {author} {\bibfnamefont {M.~H.}\ \bibnamefont
  {Teimourpour}}, \bibinfo {author} {\bibfnamefont {R.}~\bibnamefont
  {El-Ganainy}}, \bibinfo {author} {\bibfnamefont {A.}~\bibnamefont {Eisfeld}},
  \bibinfo {author} {\bibfnamefont {A.}~\bibnamefont {Szameit}}, \ and\
  \bibinfo {author} {\bibfnamefont {D.~N.}\ \bibnamefont {Christodoulides}},\
  }\bibfield  {title} {\enquote {\bibinfo {title} {Light transport in
  $\mathcal{PT}$-invariant photonic structures with hidden symmetries},}\
  }\href {\doibase 10.1103/PhysRevA.90.053817} {\bibfield  {journal} {\bibinfo
  {journal} {Phys. Rev. A}\ }\textbf {\bibinfo {volume} {90}},\ \bibinfo
  {pages} {053817} (\bibinfo {year} {2014})}\BibitemShut {NoStop}%
\bibitem [{\citenamefont {Nada}\ \emph {et~al.}(2017)\citenamefont {Nada},
  \citenamefont {Othman},\ and\ \citenamefont {Capolino}}]{Nada2017}%
  \BibitemOpen
  \bibfield  {author} {\bibinfo {author} {\bibfnamefont {M.~Y.}\ \bibnamefont
  {Nada}}, \bibinfo {author} {\bibfnamefont {M.~A.~K.}\ \bibnamefont {Othman}},
  \ and\ \bibinfo {author} {\bibfnamefont {F.}~\bibnamefont {Capolino}},\
  }\bibfield  {title} {\enquote {\bibinfo {title} {Theory of coupled resonator
  optical waveguides exhibiting high-order exceptional points of degeneracy},}\
  }\href {\doibase 10.1103/PhysRevB.96.184304} {\bibfield  {journal} {\bibinfo
  {journal} {Phys. Rev. B}\ }\textbf {\bibinfo {volume} {96}},\ \bibinfo
  {pages} {184304} (\bibinfo {year} {2017})}\BibitemShut {NoStop}%
\bibitem [{\citenamefont {Wu}\ \emph {et~al.}(2018)\citenamefont {Wu},
  \citenamefont {Zheng}, \citenamefont {Chen},\ and\ \citenamefont
  {Liu}}]{Wu2018}%
  \BibitemOpen
  \bibfield  {author} {\bibinfo {author} {\bibfnamefont {R.-B.}\ \bibnamefont
  {Wu}}, \bibinfo {author} {\bibfnamefont {Y.}~\bibnamefont {Zheng}}, \bibinfo
  {author} {\bibfnamefont {Q.-M.}\ \bibnamefont {Chen}}, \ and\ \bibinfo
  {author} {\bibfnamefont {Y.-x.}\ \bibnamefont {Liu}},\ }\bibfield  {title}
  {\enquote {\bibinfo {title} {Synthesizing exceptional points with three
  resonators},}\ }\href {\doibase 10.1103/PhysRevA.98.033817} {\bibfield
  {journal} {\bibinfo  {journal} {Phys. Rev. A}\ }\textbf {\bibinfo {volume}
  {98}},\ \bibinfo {pages} {033817} (\bibinfo {year} {2018})}\BibitemShut
  {NoStop}%
\bibitem [{\citenamefont {Zhang}\ \emph
  {et~al.}(2020{\natexlab{a}})\citenamefont {Zhang}, \citenamefont {Zhang},
  \citenamefont {Jin},\ and\ \citenamefont {Song}}]{Zhang2020c}%
  \BibitemOpen
  \bibfield  {author} {\bibinfo {author} {\bibfnamefont {S.~M.}\ \bibnamefont
  {Zhang}}, \bibinfo {author} {\bibfnamefont {X.~Z.}\ \bibnamefont {Zhang}},
  \bibinfo {author} {\bibfnamefont {L.}~\bibnamefont {Jin}}, \ and\ \bibinfo
  {author} {\bibfnamefont {Z.}~\bibnamefont {Song}},\ }\bibfield  {title}
  {\enquote {\bibinfo {title} {High-order exceptional points in supersymmetric
  arrays},}\ }\href {\doibase 10.1103/PhysRevA.101.033820} {\bibfield
  {journal} {\bibinfo  {journal} {Phys. Rev. A}\ }\textbf {\bibinfo {volume}
  {101}},\ \bibinfo {pages} {033820} (\bibinfo {year}
  {2020}{\natexlab{a}})}\BibitemShut {NoStop}%
\bibitem [{\citenamefont {Zhong}\ \emph {et~al.}(2020)\citenamefont {Zhong},
  \citenamefont {Kou}, \citenamefont {\"Ozdemir},\ and\ \citenamefont
  {El-Ganainy}}]{Sahin2020}%
  \BibitemOpen
  \bibfield  {author} {\bibinfo {author} {\bibfnamefont {Q.}~\bibnamefont
  {Zhong}}, \bibinfo {author} {\bibfnamefont {J.}~\bibnamefont {Kou}}, \bibinfo
  {author} {\bibfnamefont {\ifmmode \mbox{\c{S}}\else \c{S}\fi{}.~K.}\
  \bibnamefont {\"Ozdemir}}, \ and\ \bibinfo {author} {\bibfnamefont
  {R.}~\bibnamefont {El-Ganainy}},\ }\bibfield  {title} {\enquote {\bibinfo
  {title} {Hierarchical construction of higher-order exceptional points},}\
  }\href {\doibase 10.1103/PhysRevLett.125.203602} {\bibfield  {journal}
  {\bibinfo  {journal} {Phys. Rev. Lett.}\ }\textbf {\bibinfo {volume} {125}},\
  \bibinfo {pages} {203602} (\bibinfo {year} {2020})}\BibitemShut {NoStop}%
\bibitem [{\citenamefont {Tschernig}\ \emph {et~al.}(2018)\citenamefont
  {Tschernig}, \citenamefont {Le\'on-Montiel}, \citenamefont
  {Maga{\~{n}}a-Loaiza}, \citenamefont {Szameit}, \citenamefont {Busch},\ and\
  \citenamefont {P\'erez-Leija}}]{Tschernig18}%
  \BibitemOpen
  \bibfield  {author} {\bibinfo {author} {\bibfnamefont {K.}~\bibnamefont
  {Tschernig}}, \bibinfo {author} {\bibfnamefont {R.~J.}\ \bibnamefont
  {Le\'on-Montiel}}, \bibinfo {author} {\bibfnamefont {O.~S.}\ \bibnamefont
  {Maga{\~{n}}a-Loaiza}}, \bibinfo {author} {\bibfnamefont {A.}~\bibnamefont
  {Szameit}}, \bibinfo {author} {\bibfnamefont {K.}~\bibnamefont {Busch}}, \
  and\ \bibinfo {author} {\bibfnamefont {A.}~\bibnamefont {P\'erez-Leija}},\
  }\bibfield  {title} {\enquote {\bibinfo {title} {Multiphoton discrete
  fractional {F}ourier dynamics in waveguide beam splitters},}\ }\href
  {\doibase 10.1364/JOSAB.35.001985} {\bibfield  {journal} {\bibinfo  {journal}
  {J. Opt. Soc. Am. B}\ }\textbf {\bibinfo {volume} {35}},\ \bibinfo {pages}
  {1985} (\bibinfo {year} {2018})}\BibitemShut {NoStop}%
\bibitem [{\citenamefont {Tschernig}\ \emph {et~al.}(2020)\citenamefont
  {Tschernig}, \citenamefont {Le\'{o}n-Montiel}, \citenamefont
  {P\'{e}rez-Leija},\ and\ \citenamefont {Busch}}]{Tschernig20}%
  \BibitemOpen
  \bibfield  {author} {\bibinfo {author} {\bibfnamefont {K.}~\bibnamefont
  {Tschernig}}, \bibinfo {author} {\bibfnamefont {R.~J.}\ \bibnamefont
  {Le\'{o}n-Montiel}}, \bibinfo {author} {\bibfnamefont {A.}~\bibnamefont
  {P\'{e}rez-Leija}}, \ and\ \bibinfo {author} {\bibfnamefont {K.}~\bibnamefont
  {Busch}},\ }\bibfield  {title} {\enquote {\bibinfo {title} {Multiphoton
  synthetic lattices in multiport waveguide arrays: synthetic atoms and {F}ock
  graphs},}\ }\href {\doibase 10.1364/PRJ.382831} {\bibfield  {journal}
  {\bibinfo  {journal} {Photon. Res.}\ }\textbf {\bibinfo {volume} {8}},\
  \bibinfo {pages} {1161} (\bibinfo {year} {2020})}\BibitemShut {NoStop}%
\bibitem [{\citenamefont {Quiroz-Ju\'{a}rez}\ \emph {et~al.}(2019)\citenamefont
  {Quiroz-Ju\'{a}rez}, \citenamefont {Perez-Leija}, \citenamefont {Tschernig},
  \citenamefont {Rodr\'{i}guez-Lara}, \citenamefont {Maga{\~{n}}a-Loaiza},
  \citenamefont {Busch}, \citenamefont {Joglekar},\ and\ \citenamefont
  {Le\'{o}n-Montiel}}]{Quiroz19}%
  \BibitemOpen
  \bibfield  {author} {\bibinfo {author} {\bibfnamefont {M.~A.}\ \bibnamefont
  {Quiroz-Ju\'{a}rez}}, \bibinfo {author} {\bibfnamefont {A.}~\bibnamefont
  {Perez-Leija}}, \bibinfo {author} {\bibfnamefont {K.}~\bibnamefont
  {Tschernig}}, \bibinfo {author} {\bibfnamefont {B.~M.}\ \bibnamefont
  {Rodr\'{i}guez-Lara}}, \bibinfo {author} {\bibfnamefont {O.~S.}\ \bibnamefont
  {Maga{\~{n}}a-Loaiza}}, \bibinfo {author} {\bibfnamefont {K.}~\bibnamefont
  {Busch}}, \bibinfo {author} {\bibfnamefont {Y.~N.}\ \bibnamefont {Joglekar}},
  \ and\ \bibinfo {author} {\bibfnamefont {R.~J.}\ \bibnamefont
  {Le\'{o}n-Montiel}},\ }\bibfield  {title} {\enquote {\bibinfo {title}
  {Exceptional points of any order in a single, lossy waveguide beam splitter
  by photon-number-resolved detection},}\ }\href {\doibase
  10.1364/PRJ.7.000862} {\bibfield  {journal} {\bibinfo  {journal} {Photon.
  Res.}\ }\textbf {\bibinfo {volume} {7}},\ \bibinfo {pages} {862} (\bibinfo
  {year} {2019})}\BibitemShut {NoStop}%
\bibitem [{\citenamefont {Mostafazadeh}(2010)}]{MOSTAFAZADEH_2010}%
  \BibitemOpen
  \bibfield  {author} {\bibinfo {author} {\bibfnamefont {A.}~\bibnamefont
  {Mostafazadeh}},\ }\bibfield  {title} {\enquote {\bibinfo {title}
  {Pseudo-{H}ermitian representation of quantum mechanics},}\ }\href {\doibase
  10.1142/s0219887810004816} {\bibfield  {journal} {\bibinfo  {journal} {Int.
  J. Geom. Meth. Mod. Phys.}\ }\textbf {\bibinfo {volume} {07}},\ \bibinfo
  {pages} {1191–1306} (\bibinfo {year} {2010})}\BibitemShut {NoStop}%
\bibitem [{\citenamefont {Minganti}\ \emph {et~al.}(2019)\citenamefont
  {Minganti}, \citenamefont {Miranowicz}, \citenamefont {Chhajlany},\ and\
  \citenamefont {Nori}}]{Minganti2019}%
  \BibitemOpen
  \bibfield  {author} {\bibinfo {author} {\bibfnamefont {F.}~\bibnamefont
  {Minganti}}, \bibinfo {author} {\bibfnamefont {A.}~\bibnamefont
  {Miranowicz}}, \bibinfo {author} {\bibfnamefont {R.~W.}\ \bibnamefont
  {Chhajlany}}, \ and\ \bibinfo {author} {\bibfnamefont {F.}~\bibnamefont
  {Nori}},\ }\bibfield  {title} {\enquote {\bibinfo {title} {Quantum
  exceptional points of non-{H}ermitian {H}amiltonians and {L}iouvillians:
  {T}he effects of quantum jumps},}\ }\href {\doibase
  10.1103/PhysRevA.100.062131} {\bibfield  {journal} {\bibinfo  {journal}
  {Phys. Rev. A}\ }\textbf {\bibinfo {volume} {100}},\ \bibinfo {pages}
  {062131} (\bibinfo {year} {2019})}\BibitemShut {NoStop}%
\bibitem [{\citenamefont {Prosen}(2012)}]{Prosen2012}%
  \BibitemOpen
  \bibfield  {author} {\bibinfo {author} {\bibfnamefont {T.}~\bibnamefont
  {Prosen}},\ }\bibfield  {title} {\enquote {\bibinfo {title}
  {$\mathcal{PT}$-symmetric quantum {L}iouvillean dynamics},}\ }\href {\doibase
  10.1103/PhysRevLett.109.090404} {\bibfield  {journal} {\bibinfo  {journal}
  {Phys. Rev. Lett.}\ }\textbf {\bibinfo {volume} {109}},\ \bibinfo {pages}
  {090404} (\bibinfo {year} {2012})}\BibitemShut {NoStop}%
\bibitem [{\citenamefont {Arkhipov}\ \emph
  {et~al.}(2020{\natexlab{a}})\citenamefont {Arkhipov}, \citenamefont
  {Miranowicz}, \citenamefont {Minganti},\ and\ \citenamefont
  {Nori}}]{Arkhipov2020}%
  \BibitemOpen
  \bibfield  {author} {\bibinfo {author} {\bibfnamefont {I.~I.}\ \bibnamefont
  {Arkhipov}}, \bibinfo {author} {\bibfnamefont {A.}~\bibnamefont
  {Miranowicz}}, \bibinfo {author} {\bibfnamefont {F.}~\bibnamefont
  {Minganti}}, \ and\ \bibinfo {author} {\bibfnamefont {F.}~\bibnamefont
  {Nori}},\ }\bibfield  {title} {\enquote {\bibinfo {title} {Quantum and
  semiclassical exceptional points of a linear system of coupled cavities with
  losses and gain within the {S}cully-{L}amb laser theory},}\ }\href {\doibase
  10.1103/PhysRevA.101.013812} {\bibfield  {journal} {\bibinfo  {journal}
  {Phys. Rev. A}\ }\textbf {\bibinfo {volume} {101}},\ \bibinfo {pages}
  {013812} (\bibinfo {year} {2020}{\natexlab{a}})}\BibitemShut {NoStop}%
\bibitem [{\citenamefont {Minganti}\ \emph {et~al.}(2020)\citenamefont
  {Minganti}, \citenamefont {Miranowicz}, \citenamefont {Chhajlany},
  \citenamefont {Arkhipov},\ and\ \citenamefont {Nori}}]{Minganti2020}%
  \BibitemOpen
  \bibfield  {author} {\bibinfo {author} {\bibfnamefont {F.}~\bibnamefont
  {Minganti}}, \bibinfo {author} {\bibfnamefont {A.}~\bibnamefont
  {Miranowicz}}, \bibinfo {author} {\bibfnamefont {R.~W.}\ \bibnamefont
  {Chhajlany}}, \bibinfo {author} {\bibfnamefont {I.~I.}\ \bibnamefont
  {Arkhipov}}, \ and\ \bibinfo {author} {\bibfnamefont {F.}~\bibnamefont
  {Nori}},\ }\bibfield  {title} {\enquote {\bibinfo {title}
  {Hybrid-{L}iouvillian formalism connecting exceptional points of
  non-{H}ermitian {H}amiltonians and {L}iouvillians via postselection of
  quantum trajectories},}\ }\href {\doibase 10.1103/PhysRevA.101.062112}
  {\bibfield  {journal} {\bibinfo  {journal} {Phys. Rev. A}\ }\textbf {\bibinfo
  {volume} {101}},\ \bibinfo {pages} {062112} (\bibinfo {year}
  {2020})}\BibitemShut {NoStop}%
\bibitem [{\citenamefont {Jaramillo~\'Avila}\ \emph {et~al.}(2020)\citenamefont
  {Jaramillo~\'Avila}, \citenamefont {Ventura-Vel\'azquez}, \citenamefont
  {Le\'on-Montiel}, \citenamefont {Joglekar},\ and\ \citenamefont
  {Rodr\'iguez-Lara}}]{Jaramillo2020}%
  \BibitemOpen
  \bibfield  {author} {\bibinfo {author} {\bibfnamefont {B.}~\bibnamefont
  {Jaramillo~\'Avila}}, \bibinfo {author} {\bibfnamefont {C.}~\bibnamefont
  {Ventura-Vel\'azquez}}, \bibinfo {author} {\bibfnamefont {R.~de~J.}\
  \bibnamefont {Le\'on-Montiel}}, \bibinfo {author} {\bibfnamefont {Y.~N.}\
  \bibnamefont {Joglekar}}, \ and\ \bibinfo {author} {\bibfnamefont {B.~M.}\
  \bibnamefont {Rodr\'iguez-Lara}},\ }\bibfield  {title} {\enquote {\bibinfo
  {title} {$\mathcal{PT}$ -symmetry from {L}indblad dynamics in a linearized
  optomechanical system},}\ }\href {\doibase 10.1038/s41598-020-58582-7}
  {\bibfield  {journal} {\bibinfo  {journal} {Sci. Rep}\ }\textbf {\bibinfo
  {volume} {10}},\ \bibinfo {pages} {1761} (\bibinfo {year}
  {2020})}\BibitemShut {NoStop}%
\bibitem [{\citenamefont {Huber}\ \emph {et~al.}(2020)\citenamefont {Huber},
  \citenamefont {Kirton}, \citenamefont {Rotter},\ and\ \citenamefont
  {Rabl}}]{Huber2020}%
  \BibitemOpen
  \bibfield  {author} {\bibinfo {author} {\bibfnamefont {J.}~\bibnamefont
  {Huber}}, \bibinfo {author} {\bibfnamefont {P.}~\bibnamefont {Kirton}},
  \bibinfo {author} {\bibfnamefont {S.}~\bibnamefont {Rotter}}, \ and\ \bibinfo
  {author} {\bibfnamefont {P.}~\bibnamefont {Rabl}},\ }\bibfield  {title}
  {\enquote {\bibinfo {title} {Emergence of $\mathcal{PT}$-symmetry breaking in
  open quantum systems},}\ }\href {\doibase 10.21468/SciPostPhys.9.4.052}
  {\bibfield  {journal} {\bibinfo  {journal} {SciPost Phys.}\ }\textbf
  {\bibinfo {volume} {9}},\ \bibinfo {pages} {52} (\bibinfo {year}
  {2020})}\BibitemShut {NoStop}%
\bibitem [{\citenamefont {Nakanishi}\ and\ \citenamefont
  {Sasamoto}(2021)}]{Nakanishi2021}%
  \BibitemOpen
  \bibfield  {author} {\bibinfo {author} {\bibfnamefont {Y.}~\bibnamefont
  {Nakanishi}}\ and\ \bibinfo {author} {\bibfnamefont {T.}~\bibnamefont
  {Sasamoto}},\ }\href@noop {} {\enquote {\bibinfo {title} {$\cal{PT}$ phase
  transition in open quantum systems with {L}indblad dynamics},}\ } (\bibinfo
  {year} {2021}),\ \Eprint {http://arxiv.org/abs/2104.07349} {arXiv:2104.07349}
  \BibitemShut {NoStop}%
\bibitem [{\citenamefont {Arkhipov}\ \emph
  {et~al.}(2020{\natexlab{b}})\citenamefont {Arkhipov}, \citenamefont
  {Miranowicz}, \citenamefont {Minganti},\ and\ \citenamefont
  {Nori}}]{Arkhipov2020b}%
  \BibitemOpen
  \bibfield  {author} {\bibinfo {author} {\bibfnamefont {I.~I.}\ \bibnamefont
  {Arkhipov}}, \bibinfo {author} {\bibfnamefont {A.}~\bibnamefont
  {Miranowicz}}, \bibinfo {author} {\bibfnamefont {F.}~\bibnamefont
  {Minganti}}, \ and\ \bibinfo {author} {\bibfnamefont {F.}~\bibnamefont
  {Nori}},\ }\bibfield  {title} {\enquote {\bibinfo {title} {Liouvillian
  exceptional points of any order in dissipative linear bosonic systems:
  {C}oherence functions and switching between $\mathcal{PT}$ and
  anti-$\mathcal{PT}$ symmetries},}\ }\href {\doibase
  10.1103/PhysRevA.102.033715} {\bibfield  {journal} {\bibinfo  {journal}
  {Phys. Rev. A}\ }\textbf {\bibinfo {volume} {102}},\ \bibinfo {pages}
  {033715} (\bibinfo {year} {2020}{\natexlab{b}})}\BibitemShut {NoStop}%
\bibitem [{\citenamefont {W\"unsche}(1996)}]{Wunsche1996}%
  \BibitemOpen
  \bibfield  {author} {\bibinfo {author} {\bibfnamefont {A.}~\bibnamefont
  {W\"unsche}},\ }\bibfield  {title} {\enquote {\bibinfo {title} {Tomographic
  reconstruction of the density operator from its normally ordered moments},}\
  }\href {\doibase 10.1103/PhysRevA.54.5291} {\bibfield  {journal} {\bibinfo
  {journal} {Phys. Rev. A}\ }\textbf {\bibinfo {volume} {54}},\ \bibinfo
  {pages} {5291--5294} (\bibinfo {year} {1996})}\BibitemShut {NoStop}%
\bibitem [{\citenamefont {Zhang}\ \emph
  {et~al.}(2020{\natexlab{b}})\citenamefont {Zhang}, \citenamefont {Feng},
  \citenamefont {Chen}, \citenamefont {Ge},\ and\ \citenamefont
  {Wan}}]{Zhang2020}%
  \BibitemOpen
  \bibfield  {author} {\bibinfo {author} {\bibfnamefont {F.}~\bibnamefont
  {Zhang}}, \bibinfo {author} {\bibfnamefont {Y.}~\bibnamefont {Feng}},
  \bibinfo {author} {\bibfnamefont {X.}~\bibnamefont {Chen}}, \bibinfo {author}
  {\bibfnamefont {L.}~\bibnamefont {Ge}}, \ and\ \bibinfo {author}
  {\bibfnamefont {W.}~\bibnamefont {Wan}},\ }\bibfield  {title} {\enquote
  {\bibinfo {title} {Synthetic anti-$\mathcal{PT}$ symmetry in a single
  microcavity},}\ }\href {\doibase 10.1103/PhysRevLett.124.053901} {\bibfield
  {journal} {\bibinfo  {journal} {Phys. Rev. Lett.}\ }\textbf {\bibinfo
  {volume} {124}},\ \bibinfo {pages} {053901} (\bibinfo {year}
  {2020}{\natexlab{b}})}\BibitemShut {NoStop}%
\bibitem [{\citenamefont {Zhang}\ \emph
  {et~al.}(2020{\natexlab{c}})\citenamefont {Zhang}, \citenamefont {Huang},
  \citenamefont {Zhang}, \citenamefont {Li}, \citenamefont {Qiu}, \citenamefont
  {Nori},\ and\ \citenamefont {Jing}}]{Zhang_2020}%
  \BibitemOpen
  \bibfield  {author} {\bibinfo {author} {\bibfnamefont {H.}~\bibnamefont
  {Zhang}}, \bibinfo {author} {\bibfnamefont {R.}~\bibnamefont {Huang}},
  \bibinfo {author} {\bibfnamefont {S.-D.}\ \bibnamefont {Zhang}}, \bibinfo
  {author} {\bibfnamefont {Y.}~\bibnamefont {Li}}, \bibinfo {author}
  {\bibfnamefont {C.-W.}\ \bibnamefont {Qiu}}, \bibinfo {author} {\bibfnamefont
  {F.}~\bibnamefont {Nori}}, \ and\ \bibinfo {author} {\bibfnamefont
  {H.}~\bibnamefont {Jing}},\ }\bibfield  {title} {\enquote {\bibinfo {title}
  {Breaking anti-$\cal{PT}$ symmetry by spinning a resonator},}\ }\href
  {\doibase 10.1021/acs.nanolett.0c03119} {\bibfield  {journal} {\bibinfo
  {journal} {Nano Lett.}\ }\textbf {\bibinfo {volume} {20}},\ \bibinfo {pages}
  {7594–7599} (\bibinfo {year} {2020}{\natexlab{c}})}\BibitemShut {NoStop}%
\bibitem [{\citenamefont {Fan}\ \emph {et~al.}(2020)\citenamefont {Fan},
  \citenamefont {Chen}, \citenamefont {Zhao}, \citenamefont {Wen},\ and\
  \citenamefont {Huang}}]{Fan2020}%
  \BibitemOpen
  \bibfield  {author} {\bibinfo {author} {\bibfnamefont {H.}~\bibnamefont
  {Fan}}, \bibinfo {author} {\bibfnamefont {J.}~\bibnamefont {Chen}}, \bibinfo
  {author} {\bibfnamefont {Z.}~\bibnamefont {Zhao}}, \bibinfo {author}
  {\bibfnamefont {J.}~\bibnamefont {Wen}}, \ and\ \bibinfo {author}
  {\bibfnamefont {Y.}~\bibnamefont {Huang}},\ }\bibfield  {title} {\enquote
  {\bibinfo {title} {Anti-parity-time symmetry in passive nanophotonics},}\
  }\href {https://doi.org/10.1021/acsphotonics.0c01053} {\bibfield  {journal}
  {\bibinfo  {journal} {{ACS} {P}hotonics}\ }\textbf {\bibinfo {volume} {7}},\
  \bibinfo {pages} {3035} (\bibinfo {year} {2020})}\BibitemShut {NoStop}%
\bibitem [{\citenamefont {Choi}\ \emph {et~al.}(2018)\citenamefont {Choi},
  \citenamefont {Hahn}, \citenamefont {Yoon},\ and\ \citenamefont
  {Song}}]{Choi2018}%
  \BibitemOpen
  \bibfield  {author} {\bibinfo {author} {\bibfnamefont {Y.}~\bibnamefont
  {Choi}}, \bibinfo {author} {\bibfnamefont {C.}~\bibnamefont {Hahn}}, \bibinfo
  {author} {\bibfnamefont {J.~W.}\ \bibnamefont {Yoon}}, \ and\ \bibinfo
  {author} {\bibfnamefont {S.~H.}\ \bibnamefont {Song}},\ }\bibfield  {title}
  {\enquote {\bibinfo {title} {Observation of an anti-$\cal{PT}$-symmetric
  exceptional point and energy-difference conserving dynamics in electrical
  circuit resonators},}\ }\href {\doibase 10.1038/s41467-018-04690-y}
  {\bibfield  {journal} {\bibinfo  {journal} {Nat. Commun.}\ }\textbf {\bibinfo
  {volume} {9}},\ \bibinfo {pages} {2182} (\bibinfo {year} {2018})}\BibitemShut
  {NoStop}%
\bibitem [{\citenamefont {Nair}\ \emph {et~al.}(2021)\citenamefont {Nair},
  \citenamefont {Mukhopadhyay},\ and\ \citenamefont {Agarwal}}]{Nair2020}%
  \BibitemOpen
  \bibfield  {author} {\bibinfo {author} {\bibfnamefont {J.~M.~P.}\
  \bibnamefont {Nair}}, \bibinfo {author} {\bibfnamefont {D.}~\bibnamefont
  {Mukhopadhyay}}, \ and\ \bibinfo {author} {\bibfnamefont {G.~S.}\
  \bibnamefont {Agarwal}},\ }\bibfield  {title} {\enquote {\bibinfo {title}
  {Enhanced sensing of weak anharmonicities through coherences in dissipatively
  coupled anti-{PT} symmetric systems},}\ }\href {\doibase
  10.1103/PhysRevLett.126.180401} {\bibfield  {journal} {\bibinfo  {journal}
  {Phys. Rev. Lett.}\ }\textbf {\bibinfo {volume} {126}},\ \bibinfo {pages}
  {180401} (\bibinfo {year} {2021})}\BibitemShut {NoStop}%
\bibitem [{\citenamefont {Peng}\ \emph {et~al.}(2016)\citenamefont {Peng},
  \citenamefont {Cao}, \citenamefont {Shen}, \citenamefont {Qu}, \citenamefont
  {Wen}, \citenamefont {Jiang},\ and\ \citenamefont {Xiao}}]{Peng2016}%
  \BibitemOpen
  \bibfield  {author} {\bibinfo {author} {\bibfnamefont {P.}~\bibnamefont
  {Peng}}, \bibinfo {author} {\bibfnamefont {W.}~\bibnamefont {Cao}}, \bibinfo
  {author} {\bibfnamefont {C.}~\bibnamefont {Shen}}, \bibinfo {author}
  {\bibfnamefont {W.}~\bibnamefont {Qu}}, \bibinfo {author} {\bibfnamefont
  {J.}~\bibnamefont {Wen}}, \bibinfo {author} {\bibfnamefont {L.}~\bibnamefont
  {Jiang}}, \ and\ \bibinfo {author} {\bibfnamefont {Y.}~\bibnamefont {Xiao}},\
  }\bibfield  {title} {\enquote {\bibinfo {title} {Anti-parity–time symmetry
  with flying atoms},}\ }\href {\doibase 10.1038/nphys3842} {\bibfield
  {journal} {\bibinfo  {journal} {Nat. Phys.}\ }\textbf {\bibinfo {volume}
  {12}},\ \bibinfo {pages} {1139} (\bibinfo {year} {2016})}\BibitemShut
  {NoStop}%
\bibitem [{\citenamefont {Mukherjee}\ \emph {et~al.}(2017)\citenamefont
  {Mukherjee}, \citenamefont {Mogilevtsev}, \citenamefont {Slepyan},
  \citenamefont {Doherty}, \citenamefont {Thomson},\ and\ \citenamefont
  {Korolkova}}]{Mukherjee2017}%
  \BibitemOpen
  \bibfield  {author} {\bibinfo {author} {\bibfnamefont {S.}~\bibnamefont
  {Mukherjee}}, \bibinfo {author} {\bibfnamefont {D.}~\bibnamefont
  {Mogilevtsev}}, \bibinfo {author} {\bibfnamefont {G.~Y.}\ \bibnamefont
  {Slepyan}}, \bibinfo {author} {\bibfnamefont {T.~H.}\ \bibnamefont
  {Doherty}}, \bibinfo {author} {\bibfnamefont {R.~R.}\ \bibnamefont
  {Thomson}}, \ and\ \bibinfo {author} {\bibfnamefont {N.}~\bibnamefont
  {Korolkova}},\ }\bibfield  {title} {\enquote {\bibinfo {title} {Dissipatively
  coupled waveguide networks for coherent diffusive photonics},}\ }\href
  {http://dx.doi.org/10.1038/s41467-017-02048-4} {\bibfield  {journal}
  {\bibinfo  {journal} {Nat. Commun.}\ }\textbf {\bibinfo {volume} {8}}
  (\bibinfo {year} {2017})}\BibitemShut {NoStop}%
\bibitem [{\citenamefont {Metelmann}\ and\ \citenamefont
  {Clerk}(2015)}]{Metelmann2015}%
  \BibitemOpen
  \bibfield  {author} {\bibinfo {author} {\bibfnamefont {A.}~\bibnamefont
  {Metelmann}}\ and\ \bibinfo {author} {\bibfnamefont {A.~A.}\ \bibnamefont
  {Clerk}},\ }\bibfield  {title} {\enquote {\bibinfo {title} {Nonreciprocal
  photon transmission and amplification via reservoir engineering},}\ }\href
  {\doibase 10.1103/PhysRevX.5.021025} {\bibfield  {journal} {\bibinfo
  {journal} {Phys. Rev. X}\ }\textbf {\bibinfo {volume} {5}},\ \bibinfo {pages}
  {021025} (\bibinfo {year} {2015})}\BibitemShut {NoStop}%
\bibitem [{\citenamefont {Kullig}\ \emph {et~al.}(2018)\citenamefont {Kullig},
  \citenamefont {Yi},\ and\ \citenamefont {Wiersig}}]{Kullig2018}%
  \BibitemOpen
  \bibfield  {author} {\bibinfo {author} {\bibfnamefont {J.}~\bibnamefont
  {Kullig}}, \bibinfo {author} {\bibfnamefont {C.-H.}\ \bibnamefont {Yi}}, \
  and\ \bibinfo {author} {\bibfnamefont {J.}~\bibnamefont {Wiersig}},\
  }\bibfield  {title} {\enquote {\bibinfo {title} {Exceptional points by
  coupling of modes with different angular momenta in deformed microdisks: A
  perturbative analysis},}\ }\href {\doibase 10.1103/PhysRevA.98.023851}
  {\bibfield  {journal} {\bibinfo  {journal} {Phys. Rev. A}\ }\textbf {\bibinfo
  {volume} {98}},\ \bibinfo {pages} {023851} (\bibinfo {year}
  {2018})}\BibitemShut {NoStop}%
\bibitem [{\citenamefont {Kullig}\ and\ \citenamefont
  {Wiersig}(2019)}]{Kullig2019}%
  \BibitemOpen
  \bibfield  {author} {\bibinfo {author} {\bibfnamefont {J.}~\bibnamefont
  {Kullig}}\ and\ \bibinfo {author} {\bibfnamefont {J.}~\bibnamefont
  {Wiersig}},\ }\bibfield  {title} {\enquote {\bibinfo {title} {High-order
  exceptional points of counterpropagating waves in weakly deformed microdisk
  cavities},}\ }\href {\doibase 10.1103/PhysRevA.100.043837} {\bibfield
  {journal} {\bibinfo  {journal} {Phys. Rev. A}\ }\textbf {\bibinfo {volume}
  {100}},\ \bibinfo {pages} {043837} (\bibinfo {year} {2019})}\BibitemShut
  {NoStop}%
\bibitem [{\citenamefont {Ding}\ \emph {et~al.}(2019)\citenamefont {Ding},
  \citenamefont {Belykh}, \citenamefont {Marandi},\ and\ \citenamefont
  {Miri}}]{Ding2019}%
  \BibitemOpen
  \bibfield  {author} {\bibinfo {author} {\bibfnamefont {J.}~\bibnamefont
  {Ding}}, \bibinfo {author} {\bibfnamefont {I.}~\bibnamefont {Belykh}},
  \bibinfo {author} {\bibfnamefont {A.}~\bibnamefont {Marandi}}, \ and\
  \bibinfo {author} {\bibfnamefont {M.-A.}\ \bibnamefont {Miri}},\ }\bibfield
  {title} {\enquote {\bibinfo {title} {Dispersive versus dissipative coupling
  for frequency synchronization in lasers},}\ }\href {\doibase
  10.1103/PhysRevApplied.12.054039} {\bibfield  {journal} {\bibinfo  {journal}
  {Phys. Rev. Applied}\ }\textbf {\bibinfo {volume} {12}},\ \bibinfo {pages}
  {054039} (\bibinfo {year} {2019})}\BibitemShut {NoStop}%
\bibitem [{\citenamefont {Qin}\ \emph {et~al.}(2021)\citenamefont {Qin},
  \citenamefont {Yin},\ and\ \citenamefont {Ding}}]{Qin2021}%
  \BibitemOpen
  \bibfield  {author} {\bibinfo {author} {\bibfnamefont {H.}~\bibnamefont
  {Qin}}, \bibinfo {author} {\bibfnamefont {Y.}~\bibnamefont {Yin}}, \ and\
  \bibinfo {author} {\bibfnamefont {M.}~\bibnamefont {Ding}},\ }\bibfield
  {title} {\enquote {\bibinfo {title} {Sensing and induced transparency with a
  synthetic anti-{PT} symmetric optical resonator},}\ }\href {\doibase
  10.1021/acsomega.0c05673} {\bibfield  {journal} {\bibinfo  {journal} {ACS
  Omega}\ }\textbf {\bibinfo {volume} {6}},\ \bibinfo {pages} {5463--5470}
  (\bibinfo {year} {2021})}\BibitemShut {NoStop}%
\bibitem [{\citenamefont {Marandi}\ \emph {et~al.}(2014)\citenamefont
  {Marandi}, \citenamefont {Wang}, \citenamefont {Takata}, \citenamefont
  {Byer},\ and\ \citenamefont {Yamamoto}}]{Marandi2014}%
  \BibitemOpen
  \bibfield  {author} {\bibinfo {author} {\bibfnamefont {A.}~\bibnamefont
  {Marandi}}, \bibinfo {author} {\bibfnamefont {Z.}~\bibnamefont {Wang}},
  \bibinfo {author} {\bibfnamefont {K.}~\bibnamefont {Takata}}, \bibinfo
  {author} {\bibfnamefont {R.~L.}\ \bibnamefont {Byer}}, \ and\ \bibinfo
  {author} {\bibfnamefont {Y.}~\bibnamefont {Yamamoto}},\ }\bibfield  {title}
  {\enquote {\bibinfo {title} {Network of time-multiplexed optical parametric
  oscillators as a coherent {I}sing machine},}\ }\href {\doibase
  10.1038/nphoton.2014.249} {\bibfield  {journal} {\bibinfo  {journal} {Nat.
  Photon.}\ }\textbf {\bibinfo {volume} {8}},\ \bibinfo {pages} {937–942}
  (\bibinfo {year} {2014})}\BibitemShut {NoStop}%
\bibitem [{\citenamefont {Inagaki}\ \emph {et~al.}(2016)\citenamefont
  {Inagaki}, \citenamefont {Haribara}, \citenamefont {Igarashi}, \citenamefont
  {Sonobe}, \citenamefont {Tamate}, \citenamefont {Honjo}, \citenamefont
  {Marandi}, \citenamefont {McMahon}, \citenamefont {Umeki}, \citenamefont
  {Enbutsu}, \citenamefont {Tadanaga}, \citenamefont {Takenouchi},
  \citenamefont {Aihara}, \citenamefont {Kawarabayashi}, \citenamefont {Inoue},
  \citenamefont {Utsunomiya},\ and\ \citenamefont {Takesue}}]{Inagaki2016}%
  \BibitemOpen
  \bibfield  {author} {\bibinfo {author} {\bibfnamefont {T.}~\bibnamefont
  {Inagaki}}, \bibinfo {author} {\bibfnamefont {Y.}~\bibnamefont {Haribara}},
  \bibinfo {author} {\bibfnamefont {K.}~\bibnamefont {Igarashi}}, \bibinfo
  {author} {\bibfnamefont {T.}~\bibnamefont {Sonobe}}, \bibinfo {author}
  {\bibfnamefont {S.}~\bibnamefont {Tamate}}, \bibinfo {author} {\bibfnamefont
  {T.}~\bibnamefont {Honjo}}, \bibinfo {author} {\bibfnamefont
  {A.}~\bibnamefont {Marandi}}, \bibinfo {author} {\bibfnamefont {P.~L.}\
  \bibnamefont {McMahon}}, \bibinfo {author} {\bibfnamefont {T.}~\bibnamefont
  {Umeki}}, \bibinfo {author} {\bibfnamefont {K.}~\bibnamefont {Enbutsu}},
  \bibinfo {author} {\bibfnamefont {O.}~\bibnamefont {Tadanaga}}, \bibinfo
  {author} {\bibfnamefont {H.}~\bibnamefont {Takenouchi}}, \bibinfo {author}
  {\bibfnamefont {K.}~\bibnamefont {Aihara}}, \bibinfo {author} {\bibfnamefont
  {K.-i.}\ \bibnamefont {Kawarabayashi}}, \bibinfo {author} {\bibfnamefont
  {K.}~\bibnamefont {Inoue}}, \bibinfo {author} {\bibfnamefont
  {S.}~\bibnamefont {Utsunomiya}}, \ and\ \bibinfo {author} {\bibfnamefont
  {H.}~\bibnamefont {Takesue}},\ }\bibfield  {title} {\enquote {\bibinfo
  {title} {A coherent {I}sing machine for 2000-node optimization problems},}\
  }\href {\doibase 10.1126/science.aah4243} {\bibfield  {journal} {\bibinfo
  {journal} {Science}\ }\textbf {\bibinfo {volume} {354}},\ \bibinfo {pages}
  {603--606} (\bibinfo {year} {2016})}\BibitemShut {NoStop}%
\bibitem [{\citenamefont {Pe\v{r}ina}(1991)}]{Perina1991Book}%
  \BibitemOpen
  \bibfield  {author} {\bibinfo {author} {\bibfnamefont {J.}~\bibnamefont
  {Pe\v{r}ina}},\ }\href@noop {} {\emph {\bibinfo {title} {Quantum Statistics
  of Linear and Nonlinear Optical Phenomena}}}\ (\bibinfo  {publisher} {Kluwer,
  Dordrecht},\ \bibinfo {year} {1991})\BibitemShut {NoStop}%
\bibitem [{\citenamefont {Dalibard}\ \emph {et~al.}(1992)\citenamefont
  {Dalibard}, \citenamefont {Castin},\ and\ \citenamefont
  {M\o{}lmer}}]{Dalibard92}%
  \BibitemOpen
  \bibfield  {author} {\bibinfo {author} {\bibfnamefont {J.}~\bibnamefont
  {Dalibard}}, \bibinfo {author} {\bibfnamefont {Y.}~\bibnamefont {Castin}}, \
  and\ \bibinfo {author} {\bibfnamefont {K.}~\bibnamefont {M\o{}lmer}},\
  }\bibfield  {title} {\enquote {\bibinfo {title} {Wave-function approach to
  dissipative processes in quantum optics},}\ }\href
  {https://link.aps.org/doi/10.1103/PhysRevLett.68.580} {\bibfield  {journal}
  {\bibinfo  {journal} {Phys. Rev. Lett.}\ }\textbf {\bibinfo {volume} {68}},\
  \bibinfo {pages} {580--583} (\bibinfo {year} {1992})}\BibitemShut {NoStop}%
\bibitem [{\citenamefont {M{\o}lmer}\ \emph {et~al.}(1993)\citenamefont
  {M{\o}lmer}, \citenamefont {Castin},\ and\ \citenamefont
  {Dalibard}}]{Molmer93}%
  \BibitemOpen
  \bibfield  {author} {\bibinfo {author} {\bibfnamefont {K.}~\bibnamefont
  {M{\o}lmer}}, \bibinfo {author} {\bibfnamefont {Y.}~\bibnamefont {Castin}}, \
  and\ \bibinfo {author} {\bibfnamefont {J.}~\bibnamefont {Dalibard}},\
  }\bibfield  {title} {\enquote {\bibinfo {title} {Monte {C}arlo wave-function
  method in quantum optics},}\ }\href
  {http://josab.osa.org/abstract.cfm?URI=josab-10-3-524} {\bibfield  {journal}
  {\bibinfo  {journal} {J. Opt. Soc. Am. B}\ }\textbf {\bibinfo {volume}
  {10}},\ \bibinfo {pages} {524--538} (\bibinfo {year} {1993})}\BibitemShut
  {NoStop}%
\bibitem [{\citenamefont {Haroche}\ and\ \citenamefont
  {Raimond}(2006)}]{HarocheBook}%
  \BibitemOpen
  \bibfield  {author} {\bibinfo {author} {\bibfnamefont {S.}~\bibnamefont
  {Haroche}}\ and\ \bibinfo {author} {\bibfnamefont {J.~M.}\ \bibnamefont
  {Raimond}},\ }\href@noop {} {\emph {\bibinfo {title} {Exploring the
  {Q}uantum: {A}toms, {C}avities, and {P}hotons}}}\ (\bibinfo  {publisher}
  {Oxford University Press, Oxford},\ \bibinfo {year} {2006})\BibitemShut
  {NoStop}%
\bibitem [{\citenamefont {Naghiloo}\ \emph {et~al.}(2019)\citenamefont
  {Naghiloo}, \citenamefont {Abbasi}, \citenamefont {Joglekar},\ and\
  \citenamefont {Murch}}]{Naghiloo19}%
  \BibitemOpen
  \bibfield  {author} {\bibinfo {author} {\bibfnamefont {M.}~\bibnamefont
  {Naghiloo}}, \bibinfo {author} {\bibfnamefont {M.}~\bibnamefont {Abbasi}},
  \bibinfo {author} {\bibfnamefont {Yogesh~N.}\ \bibnamefont {Joglekar}}, \
  and\ \bibinfo {author} {\bibfnamefont {K.~W.}\ \bibnamefont {Murch}},\
  }\bibfield  {title} {\enquote {\bibinfo {title} {Quantum state tomography
  across the exceptional point in a single dissipative qubit},}\ }\href
  {\doibase 10.1038/s41567-019-0652-z} {\bibfield  {journal} {\bibinfo
  {journal} {Nature Physics}\ }\textbf {\bibinfo {volume} {15}},\ \bibinfo
  {pages} {1232} (\bibinfo {year} {2019})}\BibitemShut {NoStop}%
\bibitem [{\citenamefont {Agarwal}(2013)}]{AgarwalBook}%
  \BibitemOpen
  \bibfield  {author} {\bibinfo {author} {\bibfnamefont {G.}~\bibnamefont
  {Agarwal}},\ }\href@noop {} {\emph {\bibinfo {title} {Quantum Optics}}}\
  (\bibinfo  {publisher} {Cambridge University Press, Cambridge, UK},\ \bibinfo
  {year} {2013})\BibitemShut {NoStop}%
\bibitem [{\citenamefont {Carmichael}(2010)}]{CarmichaelBook}%
  \BibitemOpen
  \bibfield  {author} {\bibinfo {author} {\bibfnamefont {H.~J.}\ \bibnamefont
  {Carmichael}},\ }\href@noop {} {\emph {\bibinfo {title} {Statistical Methods
  in Quantum Optics 1}}}\ (\bibinfo  {publisher} {Springer, Berlin},\ \bibinfo
  {year} {2010})\BibitemShut {NoStop}%
\bibitem [{\citenamefont {Lototsky}(2015)}]{Lototsky2015}%
  \BibitemOpen
  \bibfield  {author} {\bibinfo {author} {\bibfnamefont {S.V.}\ \bibnamefont
  {Lototsky}},\ }\bibfield  {title} {\enquote {\bibinfo {title} {Simple
  spectral bounds for sums of certain {K}ronecker products},}\ }\href {\doibase
  https://doi.org/10.1016/j.laa.2014.11.026} {\bibfield  {journal} {\bibinfo
  {journal} {Linear Algebra and its Applications}\ }\textbf {\bibinfo {volume}
  {469}},\ \bibinfo {pages} {114 -- 129} (\bibinfo {year} {2015})}\BibitemShut
  {NoStop}%
\bibitem [{\citenamefont {Quijandr\'{\i}a}\ \emph {et~al.}(2018)\citenamefont
  {Quijandr\'{\i}a}, \citenamefont {Naether}, \citenamefont {\"Ozdemir},
  \citenamefont {Nori},\ and\ \citenamefont {Zueco}}]{Zueco2018}%
  \BibitemOpen
  \bibfield  {author} {\bibinfo {author} {\bibfnamefont {F.}~\bibnamefont
  {Quijandr\'{\i}a}}, \bibinfo {author} {\bibfnamefont {U.}~\bibnamefont
  {Naether}}, \bibinfo {author} {\bibfnamefont {S.~K.}\ \bibnamefont
  {\"Ozdemir}}, \bibinfo {author} {\bibfnamefont {F.}~\bibnamefont {Nori}}, \
  and\ \bibinfo {author} {\bibfnamefont {D.}~\bibnamefont {Zueco}},\ }\bibfield
   {title} {\enquote {\bibinfo {title} {$\mathcal{PT}$-symmetric circuit
  {QED}},}\ }\href {\doibase 10.1103/PhysRevA.97.053846} {\bibfield  {journal}
  {\bibinfo  {journal} {Phys. Rev. A}\ }\textbf {\bibinfo {volume} {97}},\
  \bibinfo {pages} {053846} (\bibinfo {year} {2018})}\BibitemShut {NoStop}%
\bibitem [{\citenamefont {Downing}\ \emph {et~al.}(2020)\citenamefont
  {Downing}, \citenamefont {Zueco},\ and\ \citenamefont
  {Mart\'in-Moreno}}]{Downing2020}%
  \BibitemOpen
  \bibfield  {author} {\bibinfo {author} {\bibfnamefont {C.~A.}\ \bibnamefont
  {Downing}}, \bibinfo {author} {\bibfnamefont {D.}~\bibnamefont {Zueco}}, \
  and\ \bibinfo {author} {\bibfnamefont {L.}~\bibnamefont {Mart\'in-Moreno}},\
  }\bibfield  {title} {\enquote {\bibinfo {title} {Chiral current circulation
  and $\mathcal{PT}$ symmetry in a trimer of oscillators},}\ }\href
  {https://doi.org/10.1021/acsphotonics.0c01208} {\bibfield  {journal}
  {\bibinfo  {journal} {ACS Photonics}\ }\textbf {\bibinfo {volume} {7}},\
  \bibinfo {pages} {3401} (\bibinfo {year} {2020})}\BibitemShut {NoStop}%
\bibitem [{\citenamefont {Purkayastha}\ \emph {et~al.}(2020)\citenamefont
  {Purkayastha}, \citenamefont {Kulkarni},\ and\ \citenamefont
  {Joglekar}}]{Purkayastha2020}%
  \BibitemOpen
  \bibfield  {author} {\bibinfo {author} {\bibfnamefont {A.}~\bibnamefont
  {Purkayastha}}, \bibinfo {author} {\bibfnamefont {M.}~\bibnamefont
  {Kulkarni}}, \ and\ \bibinfo {author} {\bibfnamefont {Y.~N.}\ \bibnamefont
  {Joglekar}},\ }\bibfield  {title} {\enquote {\bibinfo {title} {Emergent
  $\mathcal{PT}$ symmetry in a double-quantum-dot circuit {QED} setup},}\
  }\href {\doibase 10.1103/PhysRevResearch.2.043075} {\bibfield  {journal}
  {\bibinfo  {journal} {Phys. Rev. Research}\ }\textbf {\bibinfo {volume}
  {2}},\ \bibinfo {pages} {043075} (\bibinfo {year} {2020})}\BibitemShut
  {NoStop}%
\bibitem [{\citenamefont {Miri}\ and\ \citenamefont {Al\`u}(2016)}]{Miri2016}%
  \BibitemOpen
  \bibfield  {author} {\bibinfo {author} {\bibfnamefont {M.-A.}\ \bibnamefont
  {Miri}}\ and\ \bibinfo {author} {\bibfnamefont {A.}~\bibnamefont {Al\`u}},\
  }\bibfield  {title} {\enquote {\bibinfo {title} {Nonlinearity-induced
  $\mathcal{PT}$-symmetry without material gain},}\ }\href
  {https://doi.org/10.1088/1367-2630/18/6/065001} {\bibfield  {journal}
  {\bibinfo  {journal} {New J. Phys.}\ }\textbf {\bibinfo {volume} {18}},\
  \bibinfo {pages} {065001} (\bibinfo {year} {2016})}\BibitemShut {NoStop}%
\bibitem [{\citenamefont {Wang}\ and\ \citenamefont {Clerk}(2019)}]{Wang2019}%
  \BibitemOpen
  \bibfield  {author} {\bibinfo {author} {\bibfnamefont {Y.-X.}\ \bibnamefont
  {Wang}}\ and\ \bibinfo {author} {\bibfnamefont {A.~A.}\ \bibnamefont
  {Clerk}},\ }\bibfield  {title} {\enquote {\bibinfo {title} {Non-{H}ermitian
  dynamics without dissipation in quantum systems},}\ }\href {\doibase
  10.1103/PhysRevA.99.063834} {\bibfield  {journal} {\bibinfo  {journal} {Phys.
  Rev. A}\ }\textbf {\bibinfo {volume} {99}},\ \bibinfo {pages} {063834}
  (\bibinfo {year} {2019})}\BibitemShut {NoStop}%
\bibitem [{\citenamefont {Neudecker}(1969)}]{Neudecker1969}%
  \BibitemOpen
  \bibfield  {author} {\bibinfo {author} {\bibfnamefont {H.}~\bibnamefont
  {Neudecker}},\ }\bibfield  {title} {\enquote {\bibinfo {title} {A note on
  {K}ronecker matrix products and matrix equation systems},}\ }\href {\doibase
  10.1137/0117057} {\bibfield  {journal} {\bibinfo  {journal} {SIAM J. Appl.
  Math.}\ }\textbf {\bibinfo {volume} {17}},\ \bibinfo {pages} {603--606}
  (\bibinfo {year} {1969})}\BibitemShut {NoStop}%
\bibitem [{\citenamefont {Muhi\v{c}}\ and\ \citenamefont
  {Plestenjak}(2014)}]{Muhic2014}%
  \BibitemOpen
  \bibfield  {author} {\bibinfo {author} {\bibfnamefont {A.}~\bibnamefont
  {Muhi\v{c}}}\ and\ \bibinfo {author} {\bibfnamefont {B.}~\bibnamefont
  {Plestenjak}},\ }\bibfield  {title} {\enquote {\bibinfo {title} {A method for
  computing all values $\lambda$ such that ${A}+\lambda{B}$ has a multiple
  eigenvalue},}\ }\href {\doibase https://doi.org/10.1016/j.laa.2013.10.015}
  {\bibfield  {journal} {\bibinfo  {journal} {Lin. Alg. Appl.}\ }\textbf
  {\bibinfo {volume} {440}},\ \bibinfo {pages} {345} (\bibinfo {year}
  {2014})}\BibitemShut {NoStop}%
\bibitem [{\citenamefont {Hackenbroich}\ \emph {et~al.}(2003)\citenamefont
  {Hackenbroich}, \citenamefont {Viviescas},\ and\ \citenamefont
  {Haake}}]{Haake2003}%
  \BibitemOpen
  \bibfield  {author} {\bibinfo {author} {\bibfnamefont {G.}~\bibnamefont
  {Hackenbroich}}, \bibinfo {author} {\bibfnamefont {C.}~\bibnamefont
  {Viviescas}}, \ and\ \bibinfo {author} {\bibfnamefont {F.}~\bibnamefont
  {Haake}},\ }\bibfield  {title} {\enquote {\bibinfo {title} {Quantum
  statistics of overlapping modes in open resonators},}\ }\href {\doibase
  10.1103/PhysRevA.68.063805} {\bibfield  {journal} {\bibinfo  {journal} {Phys.
  Rev. A}\ }\textbf {\bibinfo {volume} {68}},\ \bibinfo {pages} {063805}
  (\bibinfo {year} {2003})}\BibitemShut {NoStop}%
\bibitem [{\citenamefont {Franke}\ \emph {et~al.}(2019)\citenamefont {Franke},
  \citenamefont {Hughes}, \citenamefont {Dezfouli}, \citenamefont {Kristensen},
  \citenamefont {Busch}, \citenamefont {Knorr},\ and\ \citenamefont
  {Richter}}]{Franke19}%
  \BibitemOpen
  \bibfield  {author} {\bibinfo {author} {\bibfnamefont {S.}~\bibnamefont
  {Franke}}, \bibinfo {author} {\bibfnamefont {S.}~\bibnamefont {Hughes}},
  \bibinfo {author} {\bibfnamefont {M.~K.}\ \bibnamefont {Dezfouli}}, \bibinfo
  {author} {\bibfnamefont {P.~T.}\ \bibnamefont {Kristensen}}, \bibinfo
  {author} {\bibfnamefont {K.}~\bibnamefont {Busch}}, \bibinfo {author}
  {\bibfnamefont {A.}~\bibnamefont {Knorr}}, \ and\ \bibinfo {author}
  {\bibfnamefont {M.}~\bibnamefont {Richter}},\ }\bibfield  {title} {\enquote
  {\bibinfo {title} {Quantization of quasinormal modes for open cavities and
  plasmonic cavity quantum electrodynamics},}\ }\href {\doibase
  10.1103/PhysRevLett.122.213901} {\bibfield  {journal} {\bibinfo  {journal}
  {Phys. Rev. Lett.}\ }\textbf {\bibinfo {volume} {122}},\ \bibinfo {pages}
  {213901} (\bibinfo {year} {2019})}\BibitemShut {NoStop}%
\bibitem [{\citenamefont {Franke}\ \emph {et~al.}(2020)\citenamefont {Franke},
  \citenamefont {Richter}, \citenamefont {Ren}, \citenamefont {Knorr},\ and\
  \citenamefont {Hughes}}]{Franke2020}%
  \BibitemOpen
  \bibfield  {author} {\bibinfo {author} {\bibfnamefont {S.}~\bibnamefont
  {Franke}}, \bibinfo {author} {\bibfnamefont {M.}~\bibnamefont {Richter}},
  \bibinfo {author} {\bibfnamefont {J.}~\bibnamefont {Ren}}, \bibinfo {author}
  {\bibfnamefont {A.}~\bibnamefont {Knorr}}, \ and\ \bibinfo {author}
  {\bibfnamefont {S.}~\bibnamefont {Hughes}},\ }\bibfield  {title} {\enquote
  {\bibinfo {title} {Quantized quasinormal-mode description of nonlinear
  cavity-{QED} effects from coupled resonators with a {F}ano-like resonance},}\
  }\href {\doibase 10.1103/PhysRevResearch.2.033456} {\bibfield  {journal}
  {\bibinfo  {journal} {Phys. Rev. Research}\ }\textbf {\bibinfo {volume}
  {2}},\ \bibinfo {pages} {033456} (\bibinfo {year} {2020})}\BibitemShut
  {NoStop}%
\bibitem [{\citenamefont {Khandelwal}\ \emph {et~al.}(2021)\citenamefont
  {Khandelwal}, \citenamefont {Brunner},\ and\ \citenamefont
  {Haack}}]{Khandelwal2021}%
  \BibitemOpen
  \bibfield  {author} {\bibinfo {author} {\bibfnamefont {S.}~\bibnamefont
  {Khandelwal}}, \bibinfo {author} {\bibfnamefont {N.}~\bibnamefont {Brunner}},
  \ and\ \bibinfo {author} {\bibfnamefont {G.}~\bibnamefont {Haack}},\
  }\bibfield  {title} {\enquote {\bibinfo {title} {Signatures of exceptional
  points in a quantum thermal machine},}\ }\href@noop {} {\  (\bibinfo {year}
  {2021})},\ \Eprint {http://arxiv.org/abs/arXiv:2101.11553} {arXiv:2101.11553}
  \BibitemShut {NoStop}%
\bibitem [{\citenamefont {Vivieskas}\ and\ \citenamefont
  {Hackenbroich}(2004)}]{Vivieskas2004}%
  \BibitemOpen
  \bibfield  {author} {\bibinfo {author} {\bibfnamefont {G.}~\bibnamefont
  {Vivieskas}}\ and\ \bibinfo {author} {\bibfnamefont {J.}~\bibnamefont
  {Hackenbroich}},\ }\bibfield  {title} {\enquote {\bibinfo {title} {Quantum
  theory of multimode fields: {A}pplications to optical resonators},}\ }\href
  {\doibase 10.1088/1464-4266/6/4/004} {\bibfield  {journal} {\bibinfo
  {journal} {J. Opt. B}\ }\textbf {\bibinfo {volume} {6}},\ \bibinfo {pages}
  {211} (\bibinfo {year} {2004})}\BibitemShut {NoStop}%
\bibitem [{\citenamefont {Horn}\ and\ \citenamefont
  {Johnson}(2012)}]{HornBook}%
  \BibitemOpen
  \bibfield  {author} {\bibinfo {author} {\bibfnamefont {R.~A.}\ \bibnamefont
  {Horn}}\ and\ \bibinfo {author} {\bibfnamefont {C.~R.}\ \bibnamefont
  {Johnson}},\ }\href@noop {} {\emph {\bibinfo {title} {Matrix Analysis}}},\
  \bibinfo {edition} {2nd}\ ed.\ (\bibinfo  {publisher} {Cambridge University
  Press},\ \bibinfo {address} {USA},\ \bibinfo {year} {2012})\BibitemShut
  {NoStop}%
\bibitem [{\citenamefont {Lin}\ and\ \citenamefont {Nori}(1996)}]{Lin1996}%
  \BibitemOpen
  \bibfield  {author} {\bibinfo {author} {\bibfnamefont {Y.-L.}\ \bibnamefont
  {Lin}}\ and\ \bibinfo {author} {\bibfnamefont {F.}~\bibnamefont {Nori}},\
  }\bibfield  {title} {\enquote {\bibinfo {title} {Quantum interference from
  sums over closed paths for electrons on a three-dimensional lattice in a
  magnetic field: {T}otal energy, magnetic moment, and orbital
  susceptibility},}\ }\href {\doibase 10.1103/PhysRevB.53.13374} {\bibfield
  {journal} {\bibinfo  {journal} {Phys. Rev. B}\ }\textbf {\bibinfo {volume}
  {53}},\ \bibinfo {pages} {13374--13385} (\bibinfo {year} {1996})}\BibitemShut
  {NoStop}%
\bibitem [{\citenamefont {Lin}\ and\ \citenamefont {Nori}(2002)}]{Lin2002}%
  \BibitemOpen
  \bibfield  {author} {\bibinfo {author} {\bibfnamefont {Y.-L.}\ \bibnamefont
  {Lin}}\ and\ \bibinfo {author} {\bibfnamefont {F.}~\bibnamefont {Nori}},\
  }\bibfield  {title} {\enquote {\bibinfo {title} {Quantum interference in
  superconducting wire networks and {J}osephson junction arrays: {A}n
  analytical approach based on multiple-loop {A}haronov-{B}ohm {F}eynman path
  integrals},}\ }\href {\doibase 10.1103/PhysRevB.65.214504} {\bibfield
  {journal} {\bibinfo  {journal} {Phys. Rev. B}\ }\textbf {\bibinfo {volume}
  {65}},\ \bibinfo {pages} {214504} (\bibinfo {year} {2002})}\BibitemShut
  {NoStop}%
\bibitem [{\citenamefont {Shchukin}\ and\ \citenamefont
  {Vogel}(2005)}]{Shchukin2005}%
  \BibitemOpen
  \bibfield  {author} {\bibinfo {author} {\bibfnamefont {E.~V.}\ \bibnamefont
  {Shchukin}}\ and\ \bibinfo {author} {\bibfnamefont {W.}~\bibnamefont
  {Vogel}},\ }\bibfield  {title} {\enquote {\bibinfo {title} {Nonclassical
  moments and their measurement},}\ }\href {\doibase
  10.1103/PhysRevA.72.043808} {\bibfield  {journal} {\bibinfo  {journal} {Phys.
  Rev. A}\ }\textbf {\bibinfo {volume} {72}},\ \bibinfo {pages} {043808}
  (\bibinfo {year} {2005})}\BibitemShut {NoStop}%
\bibitem [{\citenamefont {Miranowicz}\ and\ \citenamefont
  {Piani}(2006)}]{Miranowicz2006}%
  \BibitemOpen
  \bibfield  {author} {\bibinfo {author} {\bibfnamefont {A.}~\bibnamefont
  {Miranowicz}}\ and\ \bibinfo {author} {\bibfnamefont {M.}~\bibnamefont
  {Piani}},\ }\bibfield  {title} {\enquote {\bibinfo {title} {Comment on
  {\textquotedblleft}inseparability criteria for continuous bipartite quantum
  states{\textquotedblright}},}\ }\href
  {https://doi.org/10.1103/physrevlett.97.058901} {\bibfield  {journal}
  {\bibinfo  {journal} {Phys, Rev. Lett.}\ }\textbf {\bibinfo {volume} {97}},\
  \bibinfo {pages} {058901} (\bibinfo {year} {2006})}\BibitemShut {NoStop}%
\bibitem [{\citenamefont {Miranowicz}\ \emph {et~al.}(2009)\citenamefont
  {Miranowicz}, \citenamefont {Piani}, \citenamefont {Horodecki},\ and\
  \citenamefont {Horodecki}}]{Miranowicz2009}%
  \BibitemOpen
  \bibfield  {author} {\bibinfo {author} {\bibfnamefont {A.}~\bibnamefont
  {Miranowicz}}, \bibinfo {author} {\bibfnamefont {M.}~\bibnamefont {Piani}},
  \bibinfo {author} {\bibfnamefont {P.}~\bibnamefont {Horodecki}}, \ and\
  \bibinfo {author} {\bibfnamefont {R.}~\bibnamefont {Horodecki}},\ }\bibfield
  {title} {\enquote {\bibinfo {title} {Inseparability criteria based on
  matrices of moments},}\ }\href {https://doi.org/10.1103/physreva.80.052303}
  {\bibfield  {journal} {\bibinfo  {journal} {Phys. Rev. A}\ }\textbf {\bibinfo
  {volume} {80}},\ \bibinfo {pages} {052303} (\bibinfo {year}
  {2009})}\BibitemShut {NoStop}%
\bibitem [{\citenamefont {Kogias}\ \emph {et~al.}(2015)\citenamefont {Kogias},
  \citenamefont {Skrzypczyk}, \citenamefont {Cavalcanti}, \citenamefont
  {Ac{\'{\i}}n},\ and\ \citenamefont {Adesso}}]{Kogias2015}%
  \BibitemOpen
  \bibfield  {author} {\bibinfo {author} {\bibfnamefont {I.}~\bibnamefont
  {Kogias}}, \bibinfo {author} {\bibfnamefont {P.}~\bibnamefont {Skrzypczyk}},
  \bibinfo {author} {\bibfnamefont {D.}~\bibnamefont {Cavalcanti}}, \bibinfo
  {author} {\bibfnamefont {A.}~\bibnamefont {Ac{\'{\i}}n}}, \ and\ \bibinfo
  {author} {\bibfnamefont {G.}~\bibnamefont {Adesso}},\ }\bibfield  {title}
  {\enquote {\bibinfo {title} {Hierarchy of steering criteria based on moments
  for all bipartite quantum systems},}\ }\href
  {https://doi.org/10.1103/physrevlett.115.210401} {\bibfield  {journal}
  {\bibinfo  {journal} {Phys. Rev. Lett.}\ }\textbf {\bibinfo {volume} {115}},\
  \bibinfo {pages} {210401} (\bibinfo {year} {2015})}\BibitemShut {NoStop}%
\bibitem [{\citenamefont {Navascu{\'{e}}s}\ \emph {et~al.}(2007)\citenamefont
  {Navascu{\'{e}}s}, \citenamefont {Pironio},\ and\ \citenamefont
  {Ac{\'{\i}}n}}]{Navascus2007}%
  \BibitemOpen
  \bibfield  {author} {\bibinfo {author} {\bibfnamefont {M.}~\bibnamefont
  {Navascu{\'{e}}s}}, \bibinfo {author} {\bibfnamefont {S.}~\bibnamefont
  {Pironio}}, \ and\ \bibinfo {author} {\bibfnamefont {A.}~\bibnamefont
  {Ac{\'{\i}}n}},\ }\bibfield  {title} {\enquote {\bibinfo {title} {Bounding
  the set of quantum correlations},}\ }\href
  {https://doi.org/10.1103/physrevlett.98.010401} {\bibfield  {journal}
  {\bibinfo  {journal} {Phys. Rev. Lett.}\ }\textbf {\bibinfo {volume} {98}},\
  \bibinfo {pages} {010401} (\bibinfo {year} {2007})}\BibitemShut {NoStop}%
\bibitem [{\citenamefont {Richter}\ and\ \citenamefont
  {Vogel}(2002)}]{Richter2002}%
  \BibitemOpen
  \bibfield  {author} {\bibinfo {author} {\bibfnamefont {T.}~\bibnamefont
  {Richter}}\ and\ \bibinfo {author} {\bibfnamefont {W.}~\bibnamefont
  {Vogel}},\ }\bibfield  {title} {\enquote {\bibinfo {title} {Nonclassicality
  of quantum states: {A} hierarchy of observable conditions},}\ }\href
  {https://doi.org/10.1103/PhysRevLett.89.283601} {\bibfield  {journal}
  {\bibinfo  {journal} {Phys. Rev. Lett.}\ }\textbf {\bibinfo {volume} {89}},\
  \bibinfo {pages} {283601} (\bibinfo {year} {2002})}\BibitemShut {NoStop}%
\bibitem [{\citenamefont {Vogel}(2008)}]{Vogel2008}%
  \BibitemOpen
  \bibfield  {author} {\bibinfo {author} {\bibfnamefont {W.}~\bibnamefont
  {Vogel}},\ }\bibfield  {title} {\enquote {\bibinfo {title} {Nonclassical
  correlation properties of radiation fields},}\ }\href
  {https://doi.org/10.1103/physrevlett.100.013605} {\bibfield  {journal}
  {\bibinfo  {journal} {Phys. Rev. Lett.}\ }\textbf {\bibinfo {volume} {100}},\
  \bibinfo {pages} {013605} (\bibinfo {year} {2008})}\BibitemShut {NoStop}%
\bibitem [{\citenamefont {Miranowicz}\ \emph {et~al.}(2010)\citenamefont
  {Miranowicz}, \citenamefont {Bartkowiak}, \citenamefont {Wang}, \citenamefont
  {Liu},\ and\ \citenamefont {Nori}}]{Miranowicz2010}%
  \BibitemOpen
  \bibfield  {author} {\bibinfo {author} {\bibfnamefont {A.}~\bibnamefont
  {Miranowicz}}, \bibinfo {author} {\bibfnamefont {M.}~\bibnamefont
  {Bartkowiak}}, \bibinfo {author} {\bibfnamefont {X.}~\bibnamefont {Wang}},
  \bibinfo {author} {\bibfnamefont {Y.-X.}\ \bibnamefont {Liu}}, \ and\
  \bibinfo {author} {\bibfnamefont {F.}~\bibnamefont {Nori}},\ }\bibfield
  {title} {\enquote {\bibinfo {title} {Testing nonclassicality in multimode
  fields: A unified derivation of classical inequalities},}\ }\href
  {https://doi.org/10.1103/physreva.82.013824} {\bibfield  {journal} {\bibinfo
  {journal} {Phys. Rev. A}\ }\textbf {\bibinfo {volume} {82}},\ \bibinfo
  {pages} {013824} (\bibinfo {year} {2010})}\BibitemShut {NoStop}%
\end{thebibliography}%

\end{document}